\documentclass[aps,pre,twocolumn,showpacs,superscriptaddress,final,floatfix]{revtex4-1}

\usepackage{amsmath,amssymb}
\usepackage[pdftex]{graphicx}
\usepackage[update]{epstopdf}
\usepackage{color}
\usepackage{textcomp}
\usepackage[utf8]{inputenc}

\begin{document}

\def\be{\begin{equation}}
\def\ee{\end{equation}}
\def\bfi{\begin{figure}}
\def\efi{\end{figure}}
\def\bea{\begin{eqnarray}}
\def\eea{\end{eqnarray}}

\title{Growth Kinetics and Aging Phenomena in a Frustrated System}

\author{Manoj Kumar}
\email{manojkmr8788@gmail.com}
\affiliation{Centre for Fluid and Complex Systems, Coventry University, CV1 5FB, United Kingdom.}

\author{Federico Corberi}
\email{corberi@sa.infn.it}
\affiliation {Dipartimento di Fisica ``E.~R. Caianiello'', and INFN, Gruppo Collegato di Salerno, and CNISM, Unit\`a di Salerno,Universit\`a  di Salerno, 
via Giovanni Paolo II 132, 84084 Fisciano (SA), Italy.}

\author{Eugenio Lippiello}
\email{eugenio.lippiello@unina2.it}
\affiliation{Department of Mathematics and Physics, University of Campania L. Vanvitelli, Viale Lincoln 5, 81100 Caserta, Italy.}

\author{Sanjay Puri}
\email{purijnu@gmail.com}
\affiliation{School of Physical Sciences, Jawaharlal Nehru University, New Delhi 110067, India.}
\date{\today}

\begin{abstract}
We study numerically the ordering kinetics in a two-dimensional Ising model with random coupling 
where the fraction of antiferromagnetic links $a$ can be gradually tuned. We show that, upon increasing such fraction, the behavior changes in a radical way. Small $a$ does not prevent the system from a complete ordering, but this occurs in an extremely (logarithmically) slow manner. However, larger values of this parameter destroy complete ordering, due to frustration, and the evolution is comparatively faster (algebraic). Our study shows a precise correspondence between the kind of developing order, ferromagnetic versus frustrated, and the speed of evolution. The aging properties of the system are studied by focusing on the scaling properties of two-time quantities, the autocorrelation and linear response functions. We find that the contribution of an equilibrium and an aging part to these functions occurs differently in the various regions of the phase diagram of the model. When quenching inside the ferromagnetic phase, the two-time quantities are obtained by the addition of these parts. Instead, in the paramagnetic phase, these two contributions enter multiplicatively. Both of the scaling forms are shown with excellent accuracy, and the corresponding scaling functions and exponents have been  determined and discussed. 
\end{abstract}
\maketitle

\section{Introduction}
\label{intro}

The {\it kinetics of phase ordering} remains an interesting problem in non-equilibrium statistical mechanics. This term refers to the nonequilibrium evolution of a system when it is rendered thermodynamically unstable by an instantaneous quench from a high-temperature phase to a low-temperature phase \cite{bray2002theory,puri2009kinetics}. The system develops ordering among local regions known as domains, which grow in time $t$ with a characteristic length scale $L(t)$ until the system reaches the equilibrium state. A fundamental quantity of interest is the domain growth-law, i.e., how $L(t)$ depends on  the time $t$ elapsed after the quench. For pure systems, i.e., in the absence of quenched disorder, it is given by the power-law behavior $L(t)\sim t^{1/z}$ with an universal exponent $z$ that depends only on few relevant parameters such as the dimension of order parameter and the presence of conservation laws. In the presence of quenched disorder, when frustration is absent or irrelevant, the kinetics is slowed down due to presence of energy barriers \cite{lai1988classes}. Many efforts 
have been made~\cite{fisher1988nonequilibrium,*huse1989remanent,corberi2015, corberi2015phase, paul2004domain, *paul2005domain,*paul2007superaging,rieger2005growing, henkel2006ageing,*henkel2008,baumann2007phase,burioni2007phase,*burioni2013topological,lippiello2010scaling,corberi2011growth,corberi2012crossover,puri2004ordering,puri1991non,*puri1992non,puri1993non,oh1986monte,oguz1990domain,*oguz1994domain,rao1993kinetics,*rao1995slow,aron2008scaling,cugliandolo2010topics,corberi2015coarsening,huse1985pinning,mandal2014characterization,park2010aging,park2012domain,kumar2017ordering,kumar2017random,corberi2017equilibrium} to understand if this leads to a {\it logarithmic} or to an algebraic growth law.  Various classes of disordered systems have been investigated including, for example, Ising models with random bonds~\cite{corberi2015phase, paul2004domain, *paul2005domain,*paul2007superaging, henkel2006ageing,*henkel2008,baumann2007phase,lippiello2010scaling,corberi2011growth,puri1991non,park2012domain,oh1986monte,biswal1996domain,bray1991universality,corberi2017}, random-fields \cite{grant1984domain,anderson1987growth,corberi2012crossover,puri1993non,oguz1990domain,oguz1994domain,rao1993kinetics,aron2008scaling,mandal2014characterization}, and  with site or bond dilution \cite{grest1985impurity,castellano1998coarsening,corberi2015,corberi2015phase,park2010aging,corberi2013scaling,corberi2019}.

Huse and Henley \cite{huse1985pinning}, under the assumption that the energy-barriers in disordered systems $E_B(L)$ scales as a power-law in $L$, i.e, $E_B(L)\sim L^{\psi}$, predicted a logarithmic growth, $L(t)\sim (\ln t)^{1/\psi}$ with the barrier exponent $\psi>0$. Nevertheless, finding logarithmic growth numerically has remained a challenge due to a slow evolution. For example, several numerical studies \cite{paul2004domain,paul2005domain,paul2007superaging,rieger2005growing,henkel2006ageing,henkel2008,baumann2007phase,lippiello2010scaling,park2012domain,oh1986monte} have found an algebraic growth with a disorder-dependent exponent. But extensive simulations on the same system in Ref.~\cite{corberi2011growth} have found a signature of logarithmic growth at late times. Recently, Cugliandolo et al. \cite{cugliandolo2010topics,iguain2009growing} also have argued that an algebraic-behavior is an intermediate regime of growth.  Further, an increasing number of studies have reported a clean crossover from power-law to a logarithmic behavior in various systems such as Ising models with random fields \cite{corberi2012crossover,aron2008scaling,mandal2014characterization, kumar2017random}, random dilution \cite{grest1985impurity,corberi2015, corberi2015phase, park2010aging,ikeda1990ordering,corberi2013scaling}, and polymers (or elastic strings) in random media \cite{kolton2005nonequilibrium,noh2009relaxation,monthus2009eigenvalue,iguain2009growing}. Moreover, a number of experimental studies on random-bond \cite{shenoy1999coarsening,schins1993domain,likodimos2001thermally,*likodimos2000kinetics}  and random diluted systems \cite{ikeda1990ordering} also have reported a logarithmic growth.  Therefore, it is likely that a pre-asymptotic power-law coarsening regime exists, followed by a truly logarithmic which is sometimes hardly accessible numerically. 
 
In addition, recently it has also been found~\cite{corberi2015phase} a nontrivial dependence of the growth law $L(t)$ on the amount of randomness in diluted Ising models~\cite{corberi2015,corberi2015phase,corberi2013scaling}. 
Specifically, for a sufficiently small fraction of diluted sites (or bonds) $d$, the kinetics of growth slows down upon increasing $d$ until a certain threshold value $d^*$, after which increasing further $d$ produces a faster growth. In~\cite{corberi2015,corberi2015phase,corberi2013scaling} it is argued that $L(t)$ increases asymptotically in a  logarithmic way for any $0<d<d_c$, where $d_c=1-p_c$ is the percolation threshold above which the networks becomes disconnected, but it turns into an algebraic behaviour with a temperature-dependent exponent right at $d_c$. This is due to the fact that the fractal topology of the network at the percolation threshold  plays an important role in softening of the pinning energy barriers which cause the speeding up of the evolution.  This interplay between algebraic (at $d=0$ and $d=d_c$) and logarithmic (for $0<d<d_c$) growth law is responsible for  the non monotonous dependence of the speed of growth on the dilution strength $d$ as mentioned above.
 
All the systems discussed to date are such that the addition of quenched disorder leaves the structure of the 
equilibrium states, with two free energy minima at low temperatures, qualitatively preserved. However, there are cases in which the effect of disorder is so strong as to  alter this structure possibly. This may occur, in particular, when the disorder is associated with frustration. With adding frustration, the problem becomes more complicated as even the low-temperature equilibrium properties in these systems are still debated \cite{vincent1997slow}. It is useful to stress that studying the off equilibrium kinetics of such systems may help shedding some light on the equilibrium structure as well since the coarsening behavior discussed before is expected to be associated to systems with a ferromagnetic-like equilibrium phase space structure whereas a different dynamical evolution characterizes system with a mean-field glassy scenario.
 
This paper focuses on growth kinetics and aging in a disordered system with frustration. It provides an overview of two of our recent works \cite{corberi2017equilibrium,corberi2019effects}, where we investigated equilibrium and off-equilibrium dynamics in a frustrated Ising magnet. The frustration is tunable with one limit being the non-frustrated system, where the properties are better understood. Specifically, we considered a two-dimensional Ising model with both ferromagnetic and antiferromagnetic interactions, and  tune  the fraction $a$ of the latter gradually. In an effort to stay as close as possible to the ferromagnetic system, we set the strength $|J_+|$ of ferromagnetic couplings larger than that, $|J_-|$, of the antiferromagnetic ones.

Regarding equilibrium properties, the system exhibits different phases in the low-temperature phase diagram (see Fig.~\ref{fig_phase_diagr}). As $a$ is progressively increased, one moves from a {\it ferromagnetic} (FM) phase ($a<a_f$), where frustration plays a minor role, to a strongly frustrated {\it paramagnetic} (PM) phase ($a_f\le a \le a_a$), which at $T=0$ is a spin-glass phase  \cite{amoruso2003scalings,*jorg2006strong,mezard1987spin} (shown by a thick green line in Fig.~\ref{fig_phase_diagr}). For even larger value of $a$, we enter in an {\it antiferromagnetic} (AFM) region ($a>a_a$) . Therefore, by tuning $a$ one can study the system evolution in transition from a ferromagnetic phase to a deeply frustrated one.
\bfi
\begin{center}
\includegraphics[width=0.95\columnwidth]{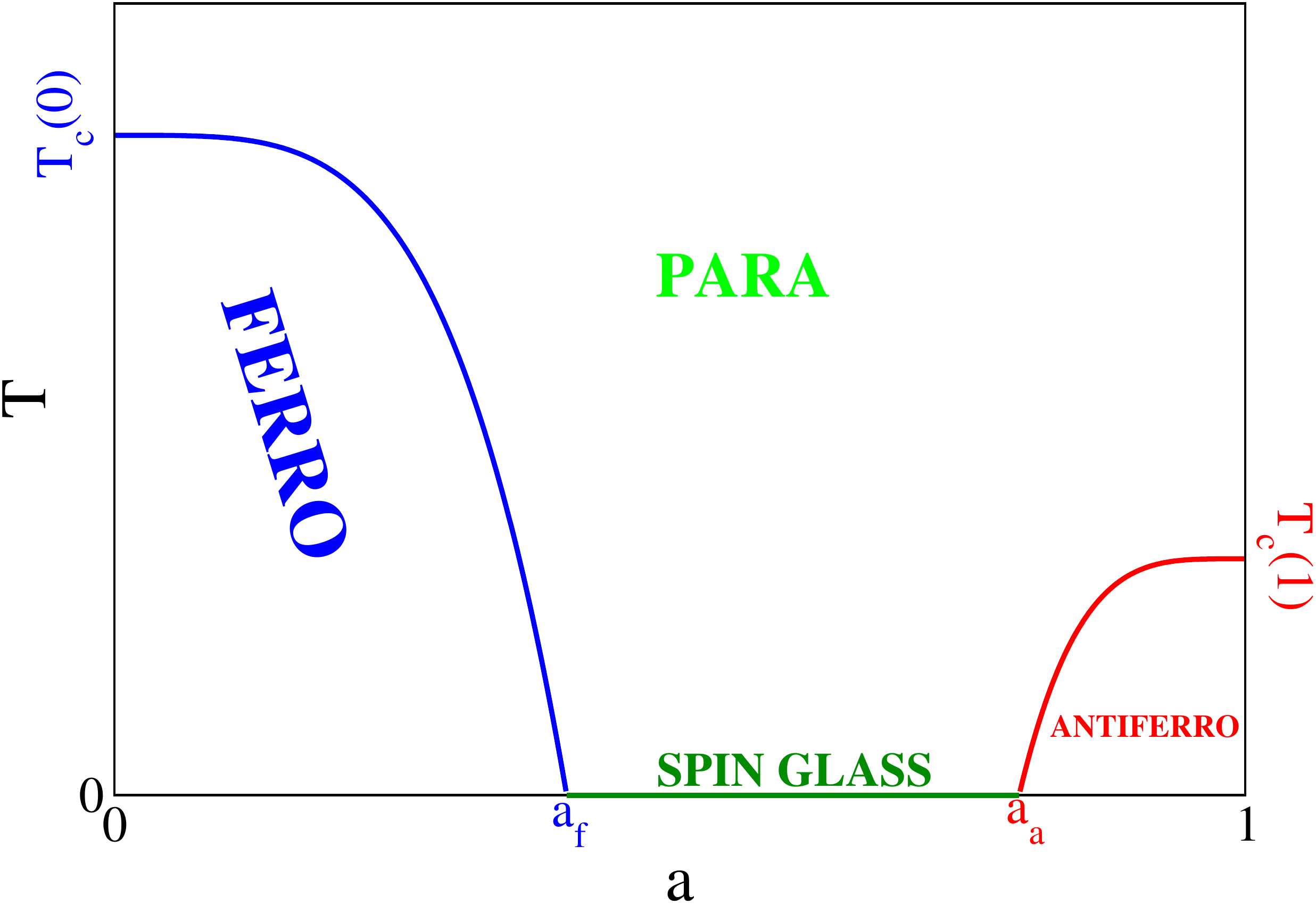}
\end{center}
\caption{Schematic representation of the phase-diagram of the model.}
\label{fig_phase_diagr}
\efi
 
Another important feature of slowly evolving systems is that they are characterized by the aging phenomenon~\cite{cugliandolo2002dynamics,calabrese2005ageing}, which is embodied by  dynamical scaling, and is well characterized by two-time quantities, such as the autocorrelation function $C(t,t_w)$ and the associated linear response function $R(t,t_w)$ \cite{bouchaud1998out}, with $t\ge t_w$. In equilibrium states these quantities are linearly related according to the usual {\it fluctuation-dissipation theorem} (FDT). On general grounds, for out-of-equilibrium states the response function cannot be written in terms of $C(t,t_w)$ alone~\cite{lippiello2005off}. However, the relation between these two quantities is expected to be meaningful and to bear important information both on the equilibrium structure and on the dynamical properties~\cite{parisi1999generalized,*franz1999response}. The nonequilibrium scaling properties of two-time quantities are quite well understood for a non-disordered system~\cite{corberi2004effective,corberi2011growing,barrat1998monte}. In particular, it is known how the contribution of degrees of freedom which are thermalized to the bath temperature and  those of a genuinely out of equilibrium nature add up to form a stationary and a non-stationary term, $C_{\rm eq}$ and $C_{\rm ag}$ respectively (and similarly for the response), in terms of which two-times quantities can be written. 
However, for disordered systems, a similar understanding in general is missing~\cite{lippiello2005off,franz1998measuring,*franz1999response,baiesi2009fluctuations,corberi2010fluctuation,lippiello2008non,lippiello2008nonlinear}, though a violation of FDT has been found \cite{barrat1998monte,crisanti2003violation}.

In Ref.~\cite{corberi2019effects}, we computed these two-time functions in the frustrated magnet and determined their scaling forms. We showed that not only $C(t,t_w)$ and $R(t,t_w)$ take different scaling forms in the various regions of the phase-diagram (i.e., FM or AFM and PM), but also that $C_{\rm eq}$ and $C_{\rm ag}$ combine differently to form the whole
correlation  $C$ (and similarly for the response function) in such different phases of the system.  
Specifically, we showed how the additive structure of the two-time quantities ($C=C_{\rm eq}+C_{\rm ag}$) in the FM phase, turns into a multiplicative one  in the 
PM region where $C(t,t_w)=C_{\rm eq}(t-t_w)\cdot C_{\rm ag}(t,t_w)$, and similarly for the response function. 

In this paper, we highlight our main results on growth kinetics and aging in a two-dimensional frustrated magnet. This paper is  organized as follows. In Sec.~\ref{model} we introduce the model and describe the structure of equilibrium states and the phase diagram.  In Sec.~\ref{sec_kin}, we present numerical results for the growth kinetics of the model in its different phases.  Sec.~\ref{aging} is devoted to the study of aging phenomena, in which we discuss the scaling properties of two-time quantities, and then present detailed numerical results for these quantities. Finally, in Sec.~\ref{concl}, we conclude the paper with a summary and a discussion of our findings.
 
\section{Model and Phase Structure}
\label{model}

We consider the random-bond spin model given by the Hamiltonian
\be
{\cal H}(\{s_i\})=-\sum _{\langle ij\rangle}J_{ij}s_is_j,
\label{ham}
\ee
where $s_i = \pm 1$ are the Ising spins, and  $\langle ij\rangle$ denotes nearest neighbors sites of a two-dimensional
square lattice. $J_{ij}$  are the uncorrelated stochastic random coupling constants, drawn from a bimodal distribution, which takes a value $J_0 - \epsilon$ with probability $a$, and $J_0 + \epsilon$ with probability $1-a$, i.e., 
\be
P(J_{ij})=a\,\delta _{J_{ij} ,J_0 - \epsilon} +(1-a)\,\delta _{J_{ij},J_0 + \epsilon}, 
\ee
where $J_0 > 0$, and $\delta $ is the Kronecker function. Clearly, $\epsilon \le J_0$ corresponds to a non-frustrated case. Here, instead, we use $ \epsilon > J_0$,
meaning that the fraction $a$ of bonds are AFM with 
$J _{ij}<0$ (which we also denote as $J_-=J_0-\epsilon$) and the remaining ones are FM with $J _{ij}>0$ (denoted as $J_+=J_0+\epsilon$).

We consider a simple case with
\begin{equation}
 J_0<\epsilon <\frac{q}{q-2}J_0,
\label{condition}
\end{equation}
where $q$ is the coordination number of the lattice.
When this condition holds,  a spin to which at least an FM bond is attached will always lower its energy by pointing along the direction of the majority
(if a majority exists) of spins to which it is connected by 
FM bonds. For instance, even if a spin has three antiferromagnetic and only one ferromagnetic bond, the energy
will lower by aligning it to the spin on the other end of the ferromagnetic bond.
Hence, Eq.~\eqref{condition} corresponds to a {\it ferromagnetic-always-wins} condition. 
This choice has been made in order to stay as close as possible to a ferromagnetic system, in order
to understand its properties more easily. However, as we will show soon, this does not prevent the system from
exhibiting a frustration dominated phase.
Notice that $J_0=0$ is the usual Edwards-Anderson spin-glass systems where both positive and negative bonds are of equal strength $(J_{\pm}=\pm \epsilon)$, and therefore do not obey Eq.~\eqref{condition}. 

Here we set $J_0=1$, and choose $\epsilon = 1.25 J_0$ (i.e. $J_+=J_0+\epsilon=2.25$ and $J_-=J_0-\epsilon=-0.25$), which obviously satisfies Eq.~\eqref{condition}. All the numerical data are presented for square lattices of size $512^2$ with periodic boundary conditions  applied on both sides. 

We start with a discussion of the $T=0$ equilibrium states of the model. The ground state problem of a frustrated Ising spin system can be exactly solved on a planar graph (without periodic boundary condition), using the {\it minimum-weight–perfect-matching} (MWPM) algorithm \cite{landry2002ground,thomas2007matching}. However, for non-planar graphs, the kind of system considered here, we have used a highly efficient iterative windowing technique developed by Khoshbakht and Weigel \cite{khoshbakht2018domain}. This algorithm is based on the MWPM approach by mapping the ground state problem on the toroidal lattices and can be used to find the exact ground state in  polynomial time up to $3000^2$ spins on a system of full periodic boundary conditions. 

In order to classify the phase structure of the model, it is useful to consider the two global order parameters, viz., the spontaneous magnetization $m$ and the staggered magnetization $M$, defined as
\begin{equation}
 m=\frac{1}{N} \sum _is_i, \quad M=\frac{1}{N} \sum _i\sigma_i, 
\end{equation}
where $\sigma _i=(-1)^i s_i$ is the staggered spin, the index $i$ runs over the lattice sites in such a way that two nearest neighbors (NNs) always have a different value of $(-1)^{i}$. In other words, the staggered magnetization $M$ is basically  the difference between the magnetization of two sub-lattices. In the following, we  describe the physical phases of the model as $a$ is varied, using the values of $m$ and $M$ measured in the ground states as a guide.
\begin{figure}
  \centering
  \rotatebox{0}{\resizebox{.96\columnwidth}{!}{\includegraphics{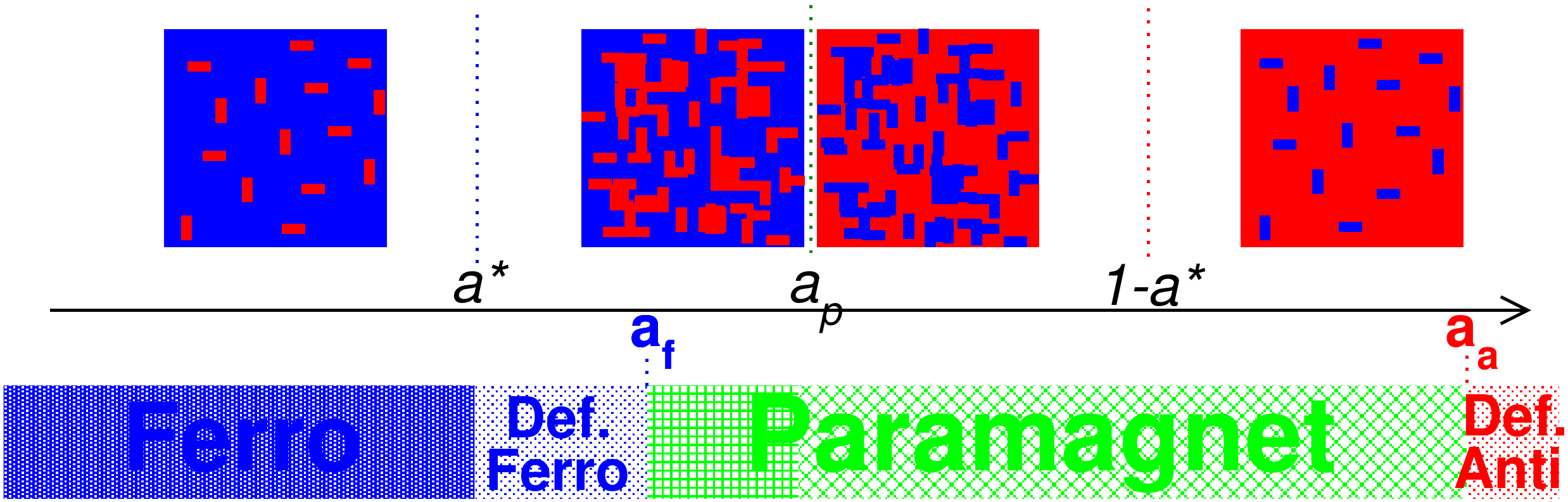}}}
  \caption{In the upper stripe, four typical bond configurations are pictorially shown,
    corresponding to $0<a<a^*$, $a^*<a\lesssim a_p$, $a_p\lesssim a<1-a^*$ and
    $1-a^*<a<1$, from left to right, respectively.
    FM bonds are drawn in blue, AFM ones in red.
    The bar below the configuration stripe describes the physical phases of the
    systems as $a$ is varied, e.g., if FM, PM, etc.
    }
\label{fig_substrate}
\end{figure}
\begin{figure}
\centering
\rotatebox{0}{\resizebox{.95\columnwidth}{!}{\includegraphics{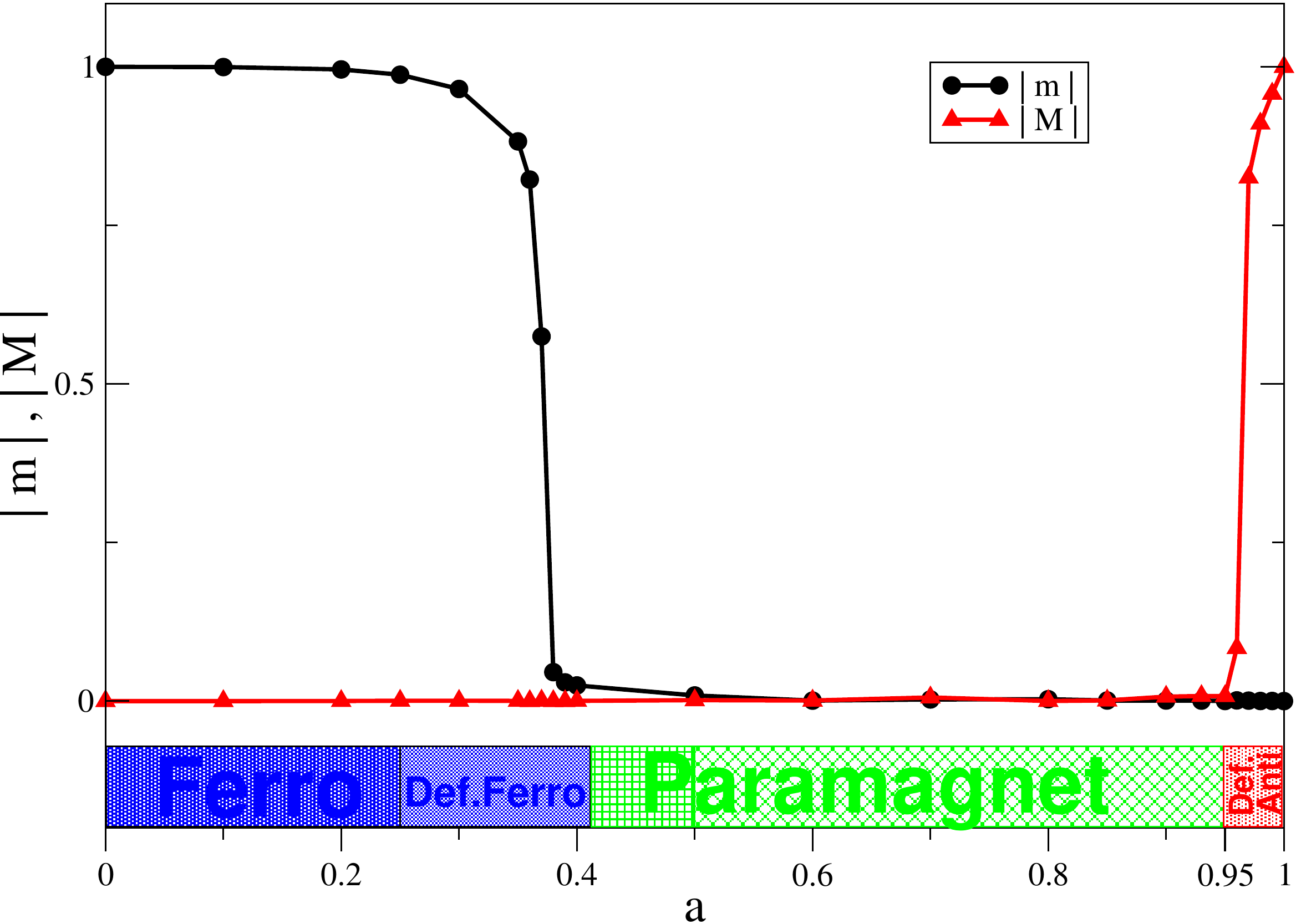}}}
\caption{Plot of the absolute value of the magnetization $|m|$ and of the staggered magnetization $|M|$ in the ground states at $T=0$ for different values of $a$. The ground states are obtained on a $512^2$ lattice with the periodic boundary condition (Fig. from Ref.~\cite{corberi2017equilibrium}).}
\label{fig_magn}
\end{figure}
\begin{figure}
 \centering
\includegraphics[width=0.95\columnwidth]{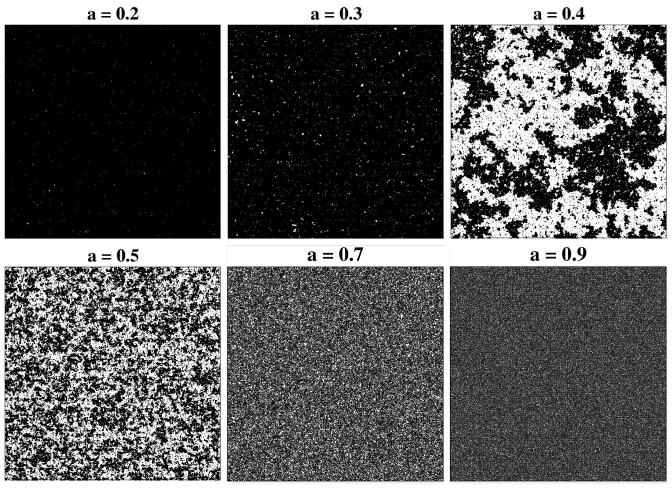}
\caption{Configurations of the GS for a $512^2$ lattice  
for different values of $a$. Spins up are marked in black and spins down are in white (Fig. from Ref.~\cite{corberi2017equilibrium}).}
\label{fig_gs}
\end{figure}

\subsection{Ferromagnetic Phase ($0\leq a<a_f$)}
\label{reg1}

This phase can be split into the two sectors with $0\le a<a^*$ and $a^*\le a <a_f$. \\
{\it Sector $0<a<a^*$}: In this region, there are very few AFM bonds, which are basically isolated in a sea of FM ones, as shown in Fig.~\ref{fig_substrate}, where a pictorial representation is provided. Therefore, also from the condition~(\ref{condition}), the ground state is akin to a usual FM system, and hence we expect $|m|\simeq 1$ and $M=0$. We can see in Fig.~\ref{fig_magn} that this is indeed the case. The value of $a^*$ at which the AFM bonds starts clustering is expected to be located between $a=0.2$ and $a=0.3$ \cite{corberi2015phase}.  A representation of a actual ground state configuration of the model for various values of $a$ is shown in Fig.~\ref{fig_gs}, and clearly the picture with $a=0.2$ (in the upper left panel) confirms the above description, showing a complete FM ordered structure. \\
{\it  Sector  $a^*\le a<a_f$}: In this  region, there is still a prevalence of FM order, which extends up to
$a_f \gtrsim 0.4$ (see Fig.~\ref{fig_magn}). This is due to the fact that the number of FM bonds is larger than that of the AFM ones and also because they are comparatively much stronger, being $J_+=9|J_-|$. However since AFM bonds can also  coalesce, regions with down spins may be found locally, as it can be seen in Fig.~\ref{fig_gs} for $a=0.3$ and $a=0.4$ (upper central and right panel). This is why we call this sector as {\it defective ferromagnet}. In this extended  FM region, the magnetization $|m|$ decreases upon raising $a$ which  vanishes at the transition point $a=a_f$, and $M=0$, as it can be observed in Fig.~\ref{fig_magn}.

\subsection{Paramagnetic phase ($a_f\leq a\leq a_a$)}
\label{regpara}

Also in this phase, we can distinguish two subregions, namely those with $ a_f<a<a_p$ and with $a_p<a<a_a$ (Fig~\ref{fig_substrate}), as we discuss below. \\
{\it Sector $ a_f<a<a_p$}: In this sector, the FM bonds still prevail to form a sea that spans the system. However, the AFM bonds also get grouped to form sufficiently connected paths so as to destroy the FM state. Therefore, one has $m=0$ and also $M=0$, since negative bonds are in a minority which is insufficient to establish an  AFM ordering in this sector. This is confirmed in Fig.~\ref{fig_magn}. Due to the fact that $m=M\simeq0$, we generically call this region as {\it paramagnetic}. Further,  $a=a_p=0.5$ is the bond percolation threshold above which FM bonds do no span the system. A ground state of the system for $a=0.5$ is shown in Fig.~\ref{fig_gs} (bottom row, left). \\
{\it Sector  $a_p<a<a_a$}: In this region, there are still FM islands in a sea of AFM bonds. Therefore,  $m=0$, as expected throughout this region, as it is observed in Fig.~\ref{fig_magn}. However, the presence of a spanning sea of AFM bonds is not sufficient to establish a global AFM order even when $a$ is so large that FM bonds are isolated. This is obviously due to the fact that AFM interactions are  weak as compared to FM ones, and, indeed,  we see in Fig.~\ref{fig_magn} that the property $M=0$ extends up to $a=a_a$, where $a_a$ is located around $a\gtrsim 0.95$.

\begin{figure}
\centering
\includegraphics[width=\columnwidth]{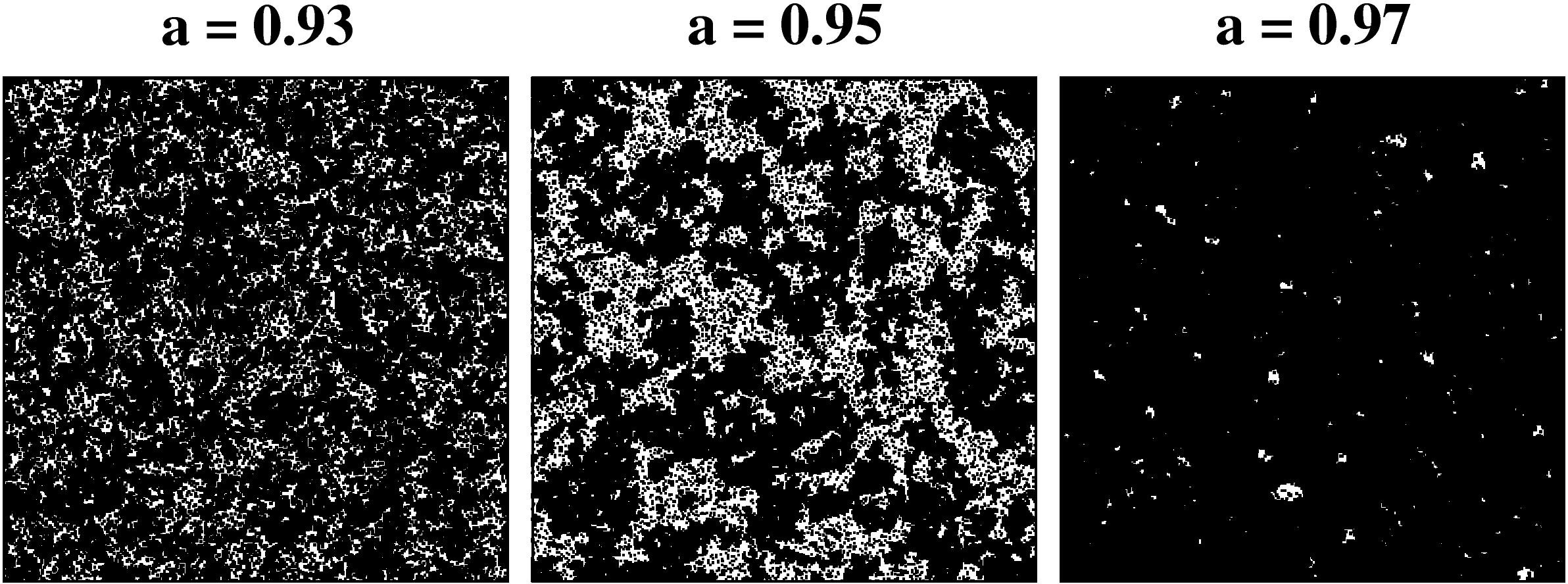}
\caption{Ground state configurations of the staggered spin $\sigma _i$ for a $512^2$ lattice for different values of $a$. $\sigma _i=1$ are marked in black and $\sigma _i=-1$ are in white.}
\label{fig_gs_anti}
\end{figure}

The development of AFM order can easily be observed  by plotting the staggered spin $\sigma _i$ instead of $s_i$, because plotting $\{s_i\}$ results in a uniform grey plot in which AFM structure cannot be seen clearly. This is done in  Fig.~\ref{fig_gs_anti}, where AFM structures can be easily spotted as  black or white regions.

\subsection{Antiferromagnetic Phase ($a_a\leq a\leq1$)}
\label{reg4}

In this region the AFM order sets in, and there are very few and far apart strong FM bonds, which represent a kind of defects in an otherwise perfectly ordered AFM state. Each of these produces a defect in an otherwise antiferromagnetically ordered system. In this region,  one has $m=0$ and $M\neq 0$, as it can be seen in Fig.~\ref{fig_magn}. An AFM structure can be clearly seen from a ground state at $a=0.97$ in Fig.~\ref{fig_gs_anti}, where staggered spins $\{\sigma _i\}$ are plotted.

\section{Numerical Simulations for Growth Kinetics}
\label{sec_kin}

\subsection{Simulation Details}

 The system is prepared in a disordered state with spins pointing randomly up or down, corresponding to $T=\infty$. It is then quenched to a low temperature $T_f$ at time $t=0$. Here, we stress that quenching to a very small $T_f$ can result in sluggish kinetics, and no substantial growth of $L(t)$ can be detected in the simulation window. On the other hand, a larger choice of the quench temperature may result in $T_f>T_c(a)$. Therefore, a reasonable choice of $T_f$ becomes necessary in order to study the off-equilibrium growth kinetics of the system.

We found, out of many $T_f$, the  two suitable choices $T_f=0.4$ and $T_f=0.75$, which represent a good compromise between the two contrasting issues discussed above. Both these temperatures are much below the critical temperature $T_c(a=0)\simeq 2.269J_+ \simeq  5.105$ of the clean ferromagnet. From the side of AFM, $T_f=0.4$ is smaller than the critical temperature of the clean antiferromagnet $T_c(a=1)\simeq 2.269J_-\simeq 0.567$ while the $T_f=0.75$ is above.  Coming to the PM region, since $T_c(a)=0$, for values of $a$ in the PM phase, the quench is necessarily made above the critical temperature.

After the quenching, which occurs at $t=0$, the system has been evolved using non-conserved dynamics~\cite{bray2002theory, puri2009kinetics} with the Glauber transition rates:
\begin{eqnarray}
\label{glauber_algo}
  W(s_i\to -s_i)&=& \frac{1}{2}\left[1-\tanh \left(\frac{\beta \varDelta E}{2}\right)\right] \nonumber \\ &=&\left[1+\exp\,(\beta \varDelta E)\right]^{-1},
\end{eqnarray}
where $\varDelta E$ is the change in energy resulting due to single spin-flip $(s_i\rightarrow -s_i)$, given as 
\begin{equation}
  \varDelta E =2 s_i\sum_{j\in n_i}J_{ij}s_j. 
\end{equation}
Here, $n_i$ refers to the set of nearest neighbors of site $i$. Using this algorithm, the system is evolved up to $t=10^7$ 
Monte-Carlo steps, each of which amounts  to attempted updates of $N(=512^2)$ spins in the lattice.  All simulation results have been averaged over $10^3$ different  realizations of disorder and initial conditions.

The main observable we are interested in is the typical size of domains which grows as a function of time, i.e., $L(t)$. We  define it as the inverse of the excess energy
\begin{equation}
L(t)=N \left[\langle E(t) \rangle-E_{\rm eq}\right]^{-1},
\label{lt}
\end{equation}
where $\langle E(t) \rangle$ is the average value of the energy at time $t$ and $E_{\rm eq}$ is the average energy of the equilibrium state at $T=T_f$. This can be obtained by evolving the ground state at $T=T_f$ until stationarity is achieved. The use of Eq.~(\ref{lt}) to determine $L(t)$ is standard in phase-ordering kinetics and, in a non-disordered system, can be easily understood from the fact that  the excess non-equilibrium energy is stored on the interfaces (e.g., domain walls),  whose density scales as the inverse of  typical domain size $L(t)$~\cite{bray2002theory}. Notice that as the present model has a frustrated (or paramagnetic) phase in the region $a_f\le a \le a_a$, $L(t)$ in Eq. (\ref{lt})  should prudentially be regarded simply as the inverse distance from the equilibrium energy.  

\subsection{Numerical Results}

\begin{figure}
\centering
\rotatebox{0}{\resizebox{.95\columnwidth}{!}{\includegraphics{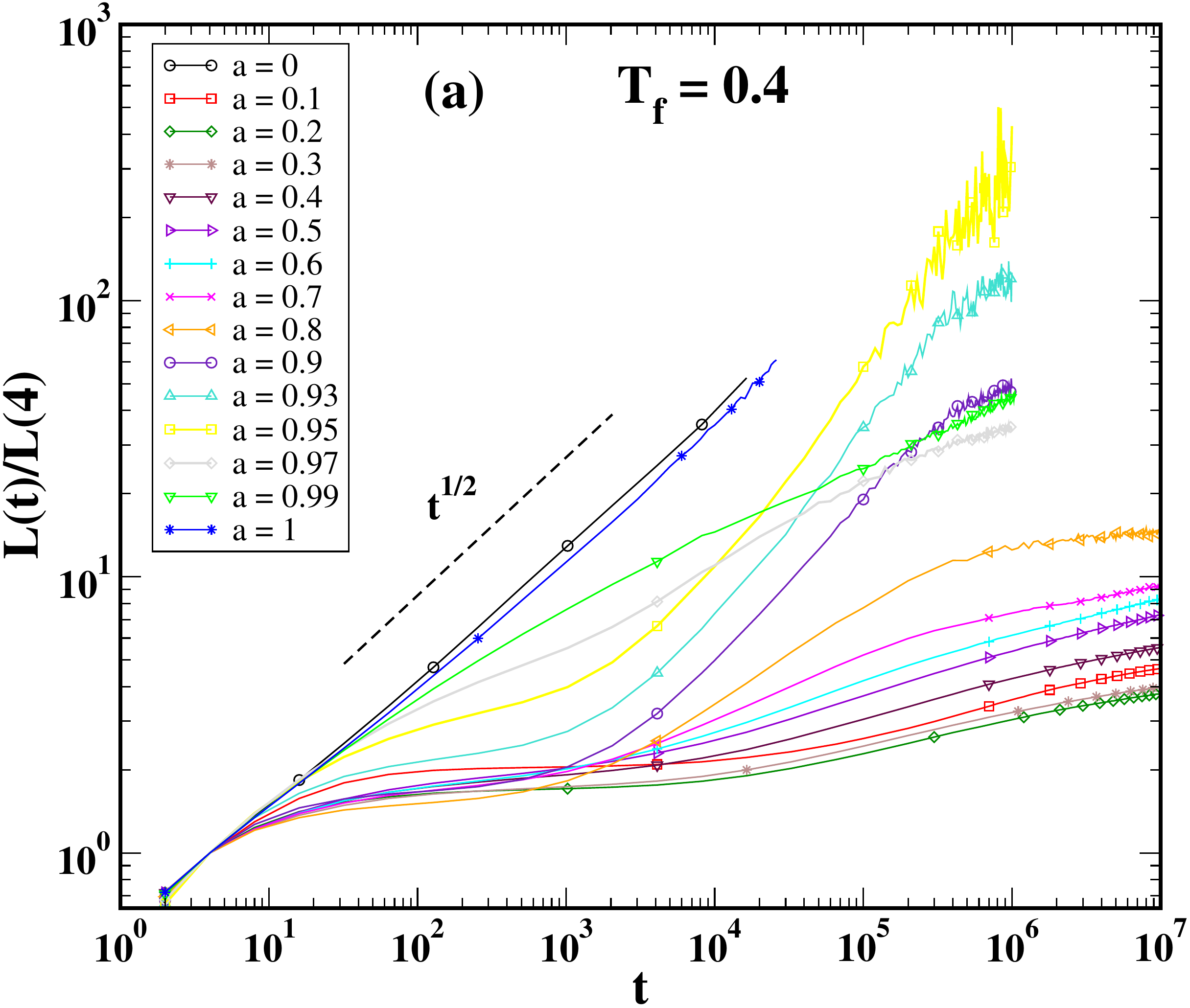}}}
\vspace{0.3cm}
\rotatebox{0}{\resizebox{.95\columnwidth}{!}{\includegraphics{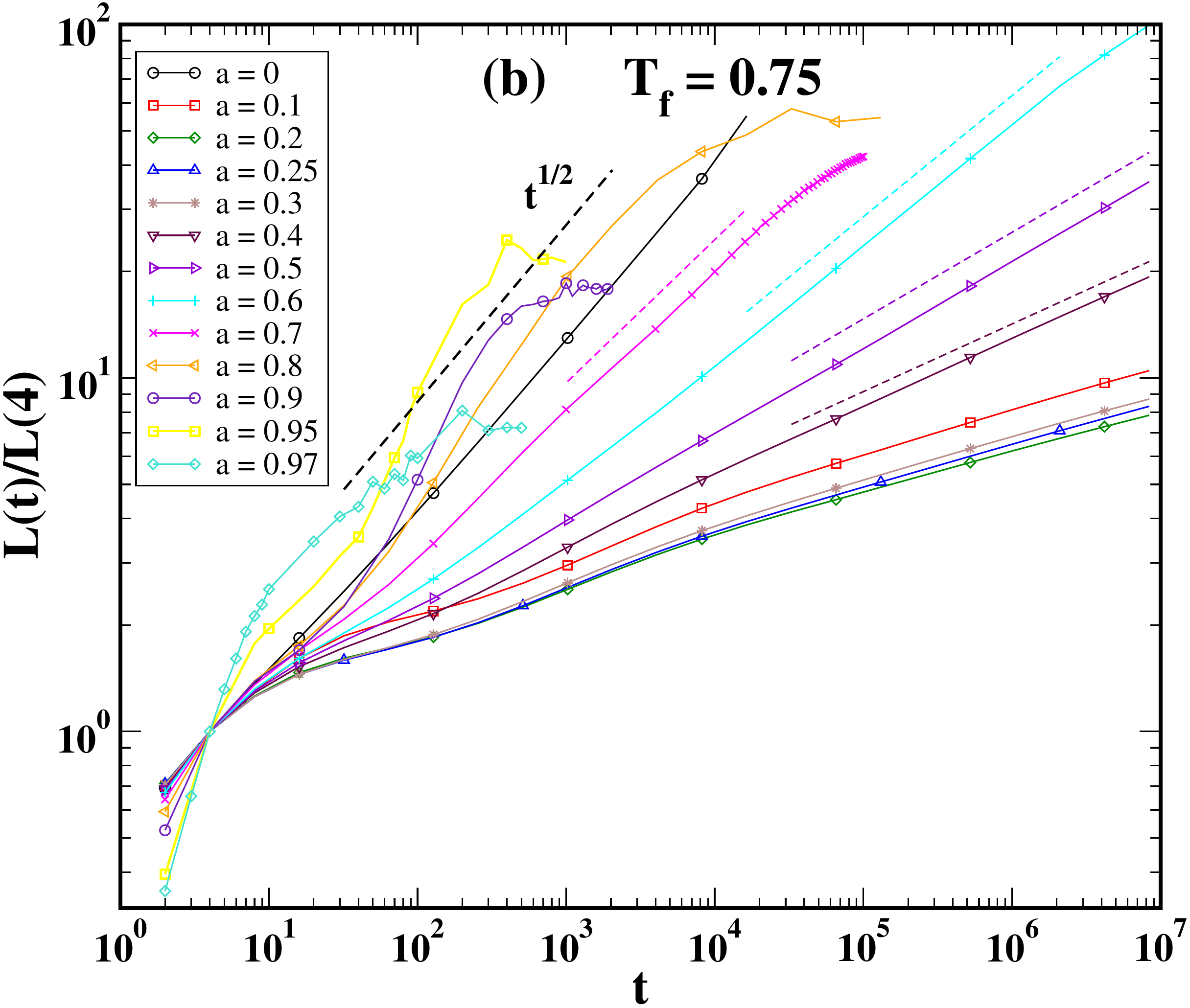}}}
\caption{Log-log plot of $L(t)$ vs $t$ for different values of $a$ (specified in the legend) for the temperature quench at (a) (upper panel) $T_f=0.4$  and (b) (lower panel) $T_f=0.75$.  The black dashed lines are the power-law $(t^{1/2})$ fits, which correspond to the pure case for $a=0$ or for $a=1$. The other dashed lines in the lower panel, near to (and of the same color of) the data for $a=0.4,0.5,0.6,0.7$, are the power-law fits of $t^{1/z}$ with $z\simeq 5.2,4.2,2.93,2.53$, respectively (Fig. from Ref.~\cite{corberi2017equilibrium}).}
\label{fig_quench}
\end{figure}

We now present our numerical simulations on the growth kinetics after the temperature quench. Figs.~\ref{fig_quench}(a) and (b) show the behavior of $L(t)$, in the whole range of values of $a$,  for $T_f=0.4$ and $T_f=0.75$, respectively. Notice that we have plotted $L(t)/L(t=4)$ to better compare  different $L(t)$ curves for different $a$. 

As a general remark, an oscillating behavior of the growth can be seen for $T_f=0.4$ in Fig~\ref{fig_quench}(a). This is usually interpreted~\cite{corberi2015,corberi2015phase,burioni2007phase,*burioni2013topological} as a stop and go mechanism when interfaces get pinned on defects such as weak AFM bonds in an FM phase or vice versa. For instance, in an FM phase for small $a$, the smallest energetic barrier $e_b$ encountered by a piece of an interface  when it crosses from a single AFM bond to FM ones, is typically given as $e_b=J_+-J_-=2\epsilon$. Then,  the associated Arrhenius time to escape the pinned state is $\tau \simeq \exp\,(e_b/k_BT)$. With  $\epsilon=1.25$ in our simulations, one has $\tau \simeq  518$ for $T_f=0.4$ and $\tau \simeq 28$ for $T_f=0.75$. We  see in Fig.~\ref{fig_quench} that, for $T_f=0.4$, this value is very well compatible with the time where $L(t)$, after becoming very slow, starts growing  faster again. As $T_f$ is raised, the stop and go mechanism, although still present, is less coherent, and the oscillations are smeared out. This is observed at $T_f = 0.75$, where the oscillatory phenomenon is hinted. Moreover, the speed of ordering increases upon raising $T_f$.

Let us now describe the nature of kinetics in the various phases of the system.

\subsubsection{Ferromagnetic Region $0\leq a\leq a_f$}
\label{ferrodyn}

\begin{figure}
\centering
\includegraphics[width=0.95\columnwidth]{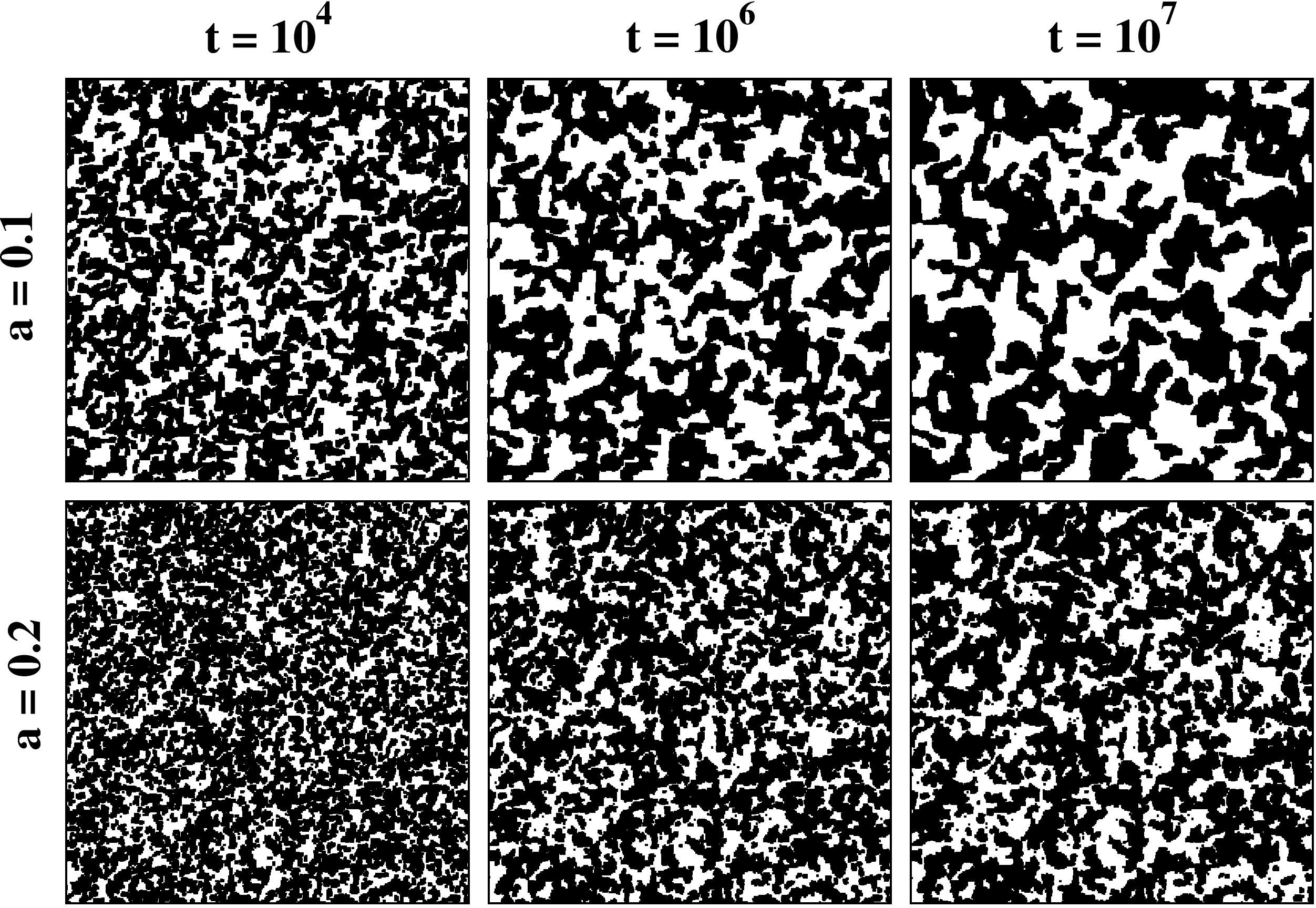}
\caption{Evolution of $\{s_i\}$-configurations, where $s_i=\pm 1$ are marked in black and white, respectively,
in the FM region at different times (see key) after a quench
to $T_f=0.4$.  The value of $a$ is  $0.1$  in the upper row and $0.2$ in the lower row.}
\label{fig_quench_ferro}
\end{figure}

In this region, starting from $a=0$ (pure case) where the expected behavior $L(t)\propto t^{1/2}$ is clearly observed (Fig.~\ref{fig_quench}), the growth slows down upon rising $a$, but this occurs only up to a certain value, which we interpret as $a=a^*$, which is located around $a=0.2$. Upon increasing $a$ beyond $a^*$, the phase-ordering process speeds up again, up to $a_f$. This is very well observed both for $T_f=0.4$ and $T_f=0.75$. The value of $a^*$, at these two temperatures is comparable, as it is expected assuming that this  behavior can be ascribed to the topology of the bond network alone. The morphology of the growing domains can be observed from Fig.~\ref{fig_quench_ferro}. Upon increasing $a$, domains become more jagged and possibly fractal. 
\begin{figure}
\centering
\rotatebox{0}{\resizebox{.95\columnwidth}{!}{\includegraphics{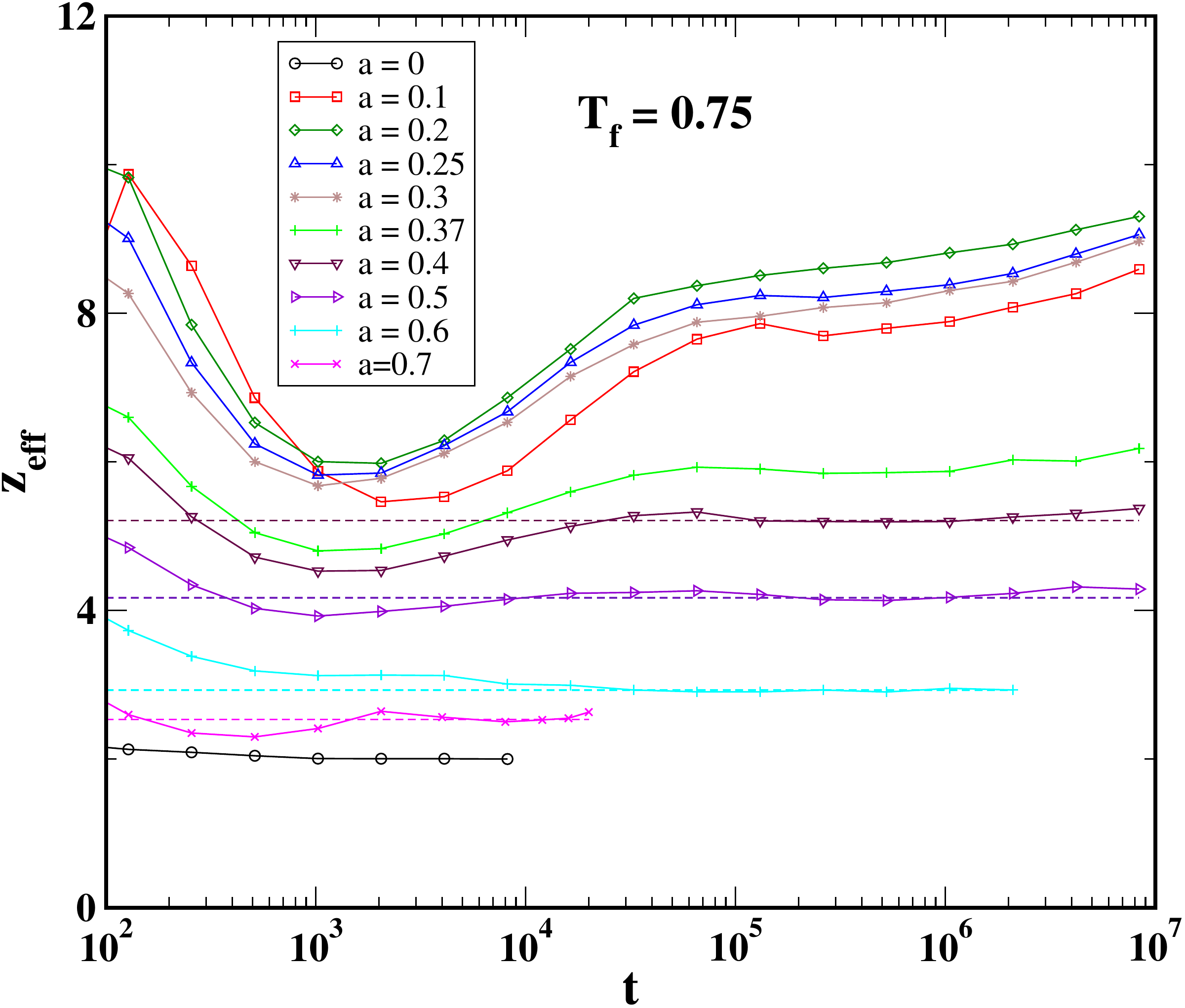}}}
\caption{Log-linear plot of effective exponent $z_{\rm eff}(t)$ as a function of time $t$ for a
quench to $T_f=0.75$, for different values of $a$ as specified in the key. The dashed lines indicate the estimated algebraic growth for $a=$ 0.4, 0.5, 0.6, and 0.7 with asymptotic values of $z_{\rm eff} \simeq$ 5.2, 4.17, 2.93, and 2.53, respectively.}
\label{fig_eff_exp}
\end{figure}

Next, to determine the nature of the growth, we compute the effective growth exponent $z_{\rm eff}$, defined as, 
\begin{equation}
 \frac{1}{z_{\rm eff}(t)}=\frac{d[\ln L(t)]}{d[\ln t]}.
\label{zeff}
\end{equation}
At $T_f=0.4$, the oscillating nature of the curves shadow the genuine growth law, and it is therefore almost impossible to come up with any quantitative statement about the neat growth, e.g., if it is consistent with a power-law or with a logarithm or something else. Therefore, we plot  $z_{\rm eff}$ for  $T_f=0.75$ in Fig.~\ref{fig_eff_exp}, which clearly depicts the non monotonous nature of growth.   For $a=0$, asymptotically $z_{\rm eff}\simeq 2$, as expected. When $a$ is progressively increased from $a=0$ to $a=0.4\simeq a_f$, looking $z_{\rm eff}$ at sufficiently large times ($t>10^4$), one sees that it initially starts rising until $a$ reaches the value $a^*\simeq 0.2$ and then decreases continuously, meaning that the growth slows down upto $a=a^*$ and then  speeds up with increasing $a$ from $a=a^*$. 
This non-monotonous fashion  of the growth mechanism has also been  found in disordered non-frustrated systems \cite{corberi2015,corberi2015phase,corberi2013scaling}. 
Furthermore, a positive slope or a curvature   in $z_{\rm eff}$ as function of  time $t$ can also be observed for all  values of $a$ in the range $0<a\le 0.3$, which implies  a logarithmic nature of the growth. Instead, for $a\gtrsim 0.4 \simeq a_f$, 
 $z_{\rm eff}(t)$ at late times  is showing an approximately constant behavior, representing an algebraic growth, $L(t)\sim t^{1/z_{\rm eff}}$. 
The dashed lines represent our best estimate of $z_{\rm eff}$ in this region.

The different asymptotic behavior\textemdash algebraic versus logarithmic\textemdash
observed at $a=a_f$ with respect to the rest of the FM region 
can also be interpreted upon thinking $a_f$ as the lower limit of the
PM region, where algebraic behaviors are observed. We will comment further on this point below.

\subsubsection{Paramagnetic Region $a_f<a<a_a$}
\label{paradyn}

In this region, we expect a kind of spin-glass order at $T_f=0$. For a two-dimensional spin glass (corresponding to $J_0=0$ in our model) it has been shown \cite{rieger2005growing,rieger1994aging,kisker1996off,*franz2003quasiequilibrium,chamon2011fluctuations} 
that the existence of a spin-glass phase at $T=0$ rules the kinetic in a long lasting pre-asymptotic regime in which an algebraic behavior of a growing length scale has been identified \cite{rieger2005growing}. Interestingly, we see from Fig.~\ref{fig_quench}(b), the data is consistent with an algebraic growth $L(t)\propto t^{1/z}$, with an $a$-dependent exponent, in agreement with Ref.~\cite{rieger2005growing}.

The algebraic increase of $L(t)$  can also be confirmed from inspection of the effective exponent $z_{\rm eff}$ in Fig.~\ref{fig_eff_exp}, where at the late time regime $t\gtrsim 10^3$-$10^4$, this quantity stays basically constant except for some noisy behavior. Notice also that $1/z_{\rm eff}$ raises as $a$ is increased, which, at least partly, can be ascribed to the fact that the {\it number} of the largest barriers, which are associated to the fraction of $J_+$, keep decreasing with increasing $a$. A power-law for $L(t)$ in this PM region, as opposed to the logarithmic one in the  FM region, can perhaps be determined by the spin-glass structure, which has many quasi-equivalent low-energy states that can speed up of the evolution from logarithmic to algebraic.

\subsubsection{Antiferromagnetic Region $a\geq a_a$}
\label{antidyn}

\begin{figure} [tb]
\centering
\includegraphics[width=0.95\columnwidth]{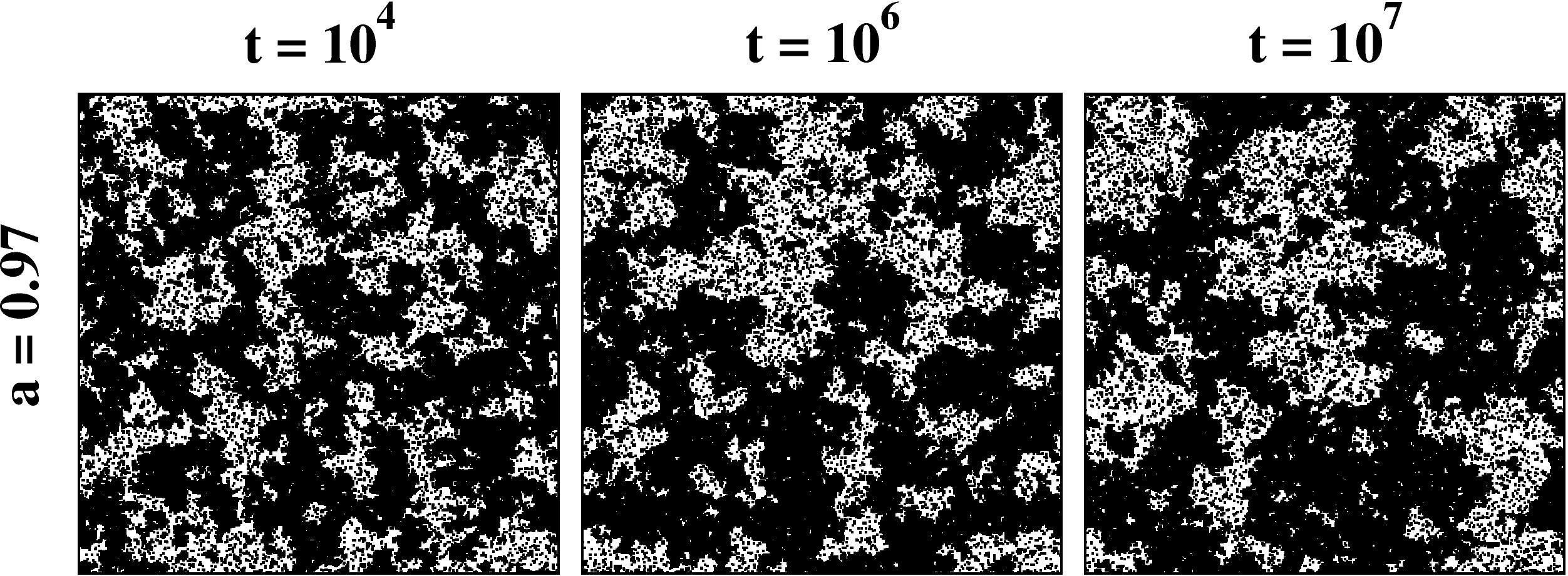}
\caption{Evolving morphologies at different times of the staggered spins in the AFM region after a quench
to $T_f=0.4$ with $a=0.97$.  $\sigma_i=\pm 1$ are marked in black and white, respectively.}
\label{fig_quench_anti}
\end{figure}

In this region with AFM order we expect a situation mirroring the one discussed in the FM region, with the obvious
correspondences $a=0 \leftrightarrow a=1$, and $a=a_f \leftrightarrow a=a_a$. This picture agrees with what is observed in our simulation. Considering the data in Fig.~\ref{fig_quench}(a), one clearly observes the non-monotonic behavior of growth in the same way as observed  in the FM phase. Upon decreasing $a$ from the pure antiferromagnet value $a=1$, the growth kinetics quickly becomes much slower in going to $a=0.97$, and then increases again until the upper  limit of the PM phase is achieved at $a=a_a\gtrsim 0.95$. The evolution of the staggered spin $\sigma _i$ is visualized in Fig.~\ref{fig_quench_anti}.  

\section{Aging Phenomena}
\label{aging}

To study the aging phenomena of a system, it is useful to compute the two-time quantities, namely, the autocorrelation and the response function. The autocorrelation function is defined as
\be
C(t,t_w)=\frac{1}{N}\sum _{i=1}^N[ \langle s_i(t)s_i(t_w)\rangle- \langle s_i(t)\rangle \langle s_i(t_w)\rangle],
\label{defC}
\ee
where $t_w$ and $t$  are called waiting and observation time, respectively, with $t\ge t_w$. $\langle \dots \rangle $ denotes average over initial conditions (i.e, thermal average) as well as  over the realizations of the quenched disorder. The impulsive auto-response function is defined as
\be
R(t,t_w)=\frac{1}{N}\sum _{i=1}^N \left . \frac{\delta \langle s_i(t)\rangle _h}
{\delta h_i(t_w)}\right \vert _{h=0},
\ee
where $h_i(t)$ is a time-dependent magnetic field and $\langle \dots \rangle_h$ means an average in the presence of such field. This quantity describe the (spatially average) linear response of a spin $s_i$ at time $t>t_w$ when a sufficiently weak magnetic field $h_i(t_w)$ is applied at the same site $i$ at a previous time $t_w$. Since  this quantity is very noisy one usually measures the so called {\it integrated} auto-response function, also defined as the {\it zero-field cooled susceptibility}
\be
\chi(t,t_w)=\int _{t_w}^t dt'\, R(t,t') .
\label{defchi}
\ee
This quantity has an enhanced signal/noise ratio and is therefore more suitable to numerical investigations~\cite{corberi2003scaling}. Let us now discuss the known scaling behavior of these two time quantities $C(t,t_w)$ and $R(t,t_w)$ (or $\chi(t,t_w)$) in different systems. 

\subsection{Scaling Behaviors of Two-time Quantities} \label{secscalA}

\noindent {\it Non-disordered systems}: For systems without quenched disorder, the scaling behavior of the two-time quantities depends on the kind of quench. More precisely, as discussed in Ref.~\cite{corberi2003scaling}, one has three different behaviors corresponding to i) a sub-critical quench to  $T_f< T_c$, ii) a critical quench to $T_f=T_c>0$ for $d>d_L$ (lower critical dimension) and iii) a critical quench to $T_f=T_c=0$ for $d=d_L$ .
We discuss them separately below.

\subsubsection{Sub-critical Quench to $T_f<T_c$ ($d>d_L$)}
\label{additive}

In this case, for large $t_w$, $C(t,t_w)$ and $R(t,t_w)$ are represented by the sum of two contributions~\cite{corberi2001connection}, given as
\be
C(t,t_w)=C_{\rm eq}(t-t_w)+C_{\rm ag}(t,t_w),
\label{Cadd}
\ee
\be
R(t,t_w)=R_{\rm eq}(t-t_w)+R_{\rm ag}(t,t_w),
\label{Radd}
\ee
where $C_{\rm eq}$ and $C_{\rm ag}$ (and similarly for $R$) are an equilibrium and an aging term, respectively.
The first term describe the equilibrium (or stationary) contribution formed in the interior of domains which reach  
equilibrium very fast. The second contribution is the remaining out-of-equilibrium (or aging) contribution due to the interfaces which are very slow in evolution. The equilibrium part is time-translational invariant (TTI), hence it depend only on the time difference $t-t_w$, and obeys the fluctuation-dissipation theorem
\be
TR_{\rm eq}(t-t_w)=-dC_{\rm eq}(t-t_w)/dt.
\label{fdtR}
\ee
The aging parts obey the  scaling forms
\be
C_{\rm ag}(t,t_w)=\mathcal{C}\left ( \frac{L(t)}{L(t_w)}\right ),
\label{Cagscalbelow}
\ee
and
\be
R_{\rm ag}(t,t_w)=L(t_w)^{-(z+\alpha)}\mathcal{R}\left ( \frac{L(t)}{L(t_w)}\right ),
\label{Rscaladd}
\ee
where $\mathcal{C}$ and $\mathcal R$ are scaling functions and $\alpha $ is an aging  
exponent~\cite{hinrichsen2008dynamical,henkel2008,henkel2003local,henkel2003scaling,henkel2001aging,*corberi2003comment,baumann2007phase,barrat1998monte,corberi2001interface,corberi2001connection,corberi2003scaling,corberi2004generic,corberi2005correction,henkel2004scaling,*corberi2005comment,*henkel2005reply,lippiello2005off,lippiello2006test}.

Similarly, from Eqs.~(\ref{Radd},\ref{Rscaladd}), the integrated auto-response can also be written in an additive structure, $\chi(t,t_w)=\chi_{\rm eq}(t-t_w)+\chi_{\rm ag}(t,t_w)$, with 
\be
\chi_{\rm ag}(t,t_w)=L(t_w)^{-\alpha} \mathcal F \left  ( \frac{L(t)}{L(t_w)}\right ).
\label{chiagferro}
\ee

Additionally, $C_{\rm eq}$ and  $C_{\rm ag}$ in Eq.~\eqref{Cadd} have some limiting properties given as follows. 
For short-time differences $t-t_w$, $C_{\rm ag}\cong \mathcal{C} (1) = q_{\rm EA}$, where $q_{\rm EA}$ is the so called Edwards-Anderson order parameter, which for a ferromagnet is simply the squared spontaneous magnetization, $q_{\rm EA}=m^2$. Instead, in the large time-difference regime, namely with $t-t_w\to \infty$ with fixed $L(t)/L(t_w)$, $C_{\rm eq}(t-t_w)=0$ and only the aging part contributes to $C(t,t_w)$. Concerning the  scaling function $\mathcal C (y)$, it is expected to behave as 
\be
\mathcal C (y) \sim y^{-\lambda_C}
\label{lambdac}
\ee
for $y\gg1$ \cite{henkel2011non}, where $\lambda_C$ is an autocorrelation exponent.

On the other hand, for the response function, $R_{\rm eq}$ obeys the FDT in Eq.~(\ref{fdtR}) and vanishes in the large time-difference regime while, conversely, $R_{\rm ag}$ vanish in the short time-difference regime.
The above features affect the behavior of the integrated autoresponse which in the aging regime at the late times, exhibits a power-law decay of the scaling function $\mathcal{F}(y)$ in Eq.~\eqref{chiagferro} as
\begin{equation}
\mathcal F (y) \sim y^{-\lambda_\chi}
\label{lambdachi}
\end{equation}
 at large $y$, with the exponent $\lambda_\chi$  
expected to be the same as the exponent $\alpha$ in the FM phase ($a<a_f \simeq 0.4$) \cite{corberi2013scaling,corberi2003comment}.

Let us mention that the above features are independent of $T_f$ and hence apply down to $T_f=0$, as the temperature is an irrelevant parameter in the renormalization group sense~\cite{mazenko1985kinetics,bray1990renormalization}.

\subsubsection{Critical Quench to $T_f=T_c>0$ ($d>d_L$)}
\label{multiplicative}

In this case the equilibrium and aging parts of both $C(t,t_w)$ and $R(t,t_w)$ take the multiplicative  forms~\cite{godreche2002nonequilibrium,calabrese2005ageing,corberi2004effective}
\be
C(t,t_w)=C_{\rm eq}(t-t_w)C_{\rm ag}(t,t_w),
\label{Cmult}
\ee
\be
R(t,t_w)=R_{\rm eq}(t-t_w)R_{\rm ag}(t,t_w),
\label{Rmult} 
\ee
where 
\be
C_{\rm ag}(t,t_w)= \widetilde {\rm C} \left (\frac{L(t)}{L(t_w)}\right ),
\label{scalCcrit}
\ee
\be
R_{\rm ag}(t,t_w)=\widetilde {\rm R} \left (\frac{L(t)}{L(t_w)}\right ),
\label{scalRmult}
\ee
with $\widetilde {\rm C}(x)$ and $\widetilde {\rm R}(x)$ scaling functions (different from the ones of Eqs.~(\ref{Cagscalbelow},\ref{Rscaladd})), whereas the equilibrium contributions obey
\be
C_{\rm eq}(t-t_w)=(t-t_w+t_0)^{-B}
\label{Ceqmult}
\ee
and
\be
R_{\rm eq}(t-t_w)=(t-t_w+t_0)^{-(1+A)}.
\ee
where $t_0$ is a microscopic time. The FDT theorem in Eq.~(\ref{fdtR}) implies that $A=B$, which are further related as $A=B=(d-2+\eta)/z_c$ to the static critical exponent $\eta$ and to the dynamic critical exponent $z_c$.   

Notice that, from Eqs.~(\ref{Rmult},\ref{scalRmult}), unlike the previous case of sub-critical quenches, here it is not possible to decompose the integrated response $\chi$ into its equilibrium and aging parts. However, it has been shown~\cite{corberi2006scaling} that the quantity $1-T_f\chi(t,t_w)$, which represents the distance from the equilibrium static value, scales as
\be
1-T_f\chi(t,t_w)=L(t_w)^{-zB} \mathcal G \left (\frac{L(t)}{L(t_w)}\right ),
\label{distance}
\ee
where $\mathcal G$ is a scaling function, and the exponent $zB$ is a scaling exponent. Moreover, in the short time difference regime, namely letting $t_w$ become large while keeping $t-t_w$ fixed, from Eqs.~(\ref{Cmult},\ref{Rmult}) one gets $C(t,t_w)\propto C_{\rm eq}(t-t_w)$, and $R(t,t_w)\propto R_{\rm eq}(t-t_w)$.

\subsubsection{Quenches to $T_f=T_c=0$ ($d=d_L$)}

At $d_L$, $T_c=0$ and hence $T_f=0$ can be viewed also as a limiting case of a critical quench. This raises a question about which structure (additive or multiplicative) of two-time quantities applies in this case. In other words, whether we consider this as the limit for $d\to d_L$ of a quench at the critical temperature (hence the multiplicative structure would apply), or the limit for $d\to d_L$ of a sub-critical quench at $T_f=0$ (hence the additive decomposition would hold). The correct picture is the latter, since an equilibrium system without quenched disorder is perfectly ordered at $T_f=0$, and therefore  the scaling structure of two-time quantities must be akin to one of the sub-critical quenches. In addition, since the equilibrium state at $T=0$ has no dynamics, then $C_{\rm eq}(t-t_w) \equiv 0$ and $C(t,t_w)=C_{\rm ag}(t,t_w)$. The same property is shared by $\chi _{\rm eq}$ in scalar  systems with a discrete (up-down) symmetry, while in vectorial systems with continuous symmetry $\chi _{\rm eq} \neq 0$ due to the presence of Goldstone modes. The distinguishing feature of the quench at $T_f=0$ with $d=d_L$ (with respect to those at $T_f=0$ with $d>d_L$) is the value of the response function exponent, which turns out to be $\alpha=0$ in this case \cite{corberi2001interface,corberi2001connection,corberi2003scaling,burioni2007phase,lippiello2000fluctuation,godreche2000response,corberi2002universality,corberi2002slow,corberi2002off,burioni2006aging}.

\subsubsection{Disordered Systems}

In disordered systems, it is expected that the presence of quenched disordered can modify the scaling behavior with respect to what is known for the non-disordered case. In spin-glass models with $p$-spins~\cite{cugliandolo1993analytical} or mean-field interactions~\cite{cugliandolo1995weak,*cugliandolo1994out,mezard1987spin}, when quenches to a phase with $q_{\rm EA}>0$ is made, symmetry breaking occurs and, as expected, one has an additive structure but with a response function exponent $\alpha=0$, in contrast to $\alpha \neq 0$ found in the case of quenches below the critical point in the absence of disorder. 
 
\subsection{Numerical Results}

In the following, we present numerical simulations for the scaling behavior of $\chi(t,t_w)$ and $C(t,t_w)$  in the different phases of the present model and analyze their scaling behavior. The numerical results  are presented for a temperature quench to $T_f=0.75$. To compute $\chi$ numerically without applying the small perturbation we use the generalization of the fluctuation-dissipation theorem to non-equilibrium states derived in Refs.~\cite{lippiello2005off,franz1998measuring,*franz1999response,baiesi2009fluctuations,corberi2010fluctuation,lippiello2008non,lippiello2008nonlinear}. To determine the equilibrium parts of two-time quantities $C_{\rm eq}$ and $\chi_{\rm eq}$, we use the ground state of system at $T=0$ as an initial condition, and equilibrate it at the final temperature of quench $T_f=0.75$, and then compute the above mentioned two-time quantities.
In studying the scaling behavior of two-time quantities we will also determine the value of the associated exponents,  introduced 
in Sec.~\ref{secscalA}, whose estimated values are collected in a
table~\ref{exp_values}.

\subsubsection{Quenches with $a<a_f$}

In this  FM phase symmetry breaking takes place and hence an additive structure is expected for the two-time quantities. This means that  for the autocorrelation, according to Eqs.~(\ref{Cadd},\ref{Cagscalbelow}), the plot of $C(t,t_w)-C_{\rm eq}(t-t_w)$ against $L(t)/L(t_w)$ for any given value of $a<a_f$,   should show a data collapse. Fig.~\ref{compare_C} is the plot which confirms this structure. Here, in panels (a) and (b), we plot $C(t,t_w)$ versus the time difference $(t-t_w)$ at $a=0.1$ and $a=0.3$, respectively, for different values of $t_w$ (see the figure caption). In these two panels, it is clear that the longer the waiting time $t_w$, the later  $C(t,t_w)$ decays, which is the hallmark of aging.
In the main frame of panel (c), we plot the aging quantity $C_{\rm ag}(t,t_w) \equiv C(t,t_w)-C_{\rm eq}(t-t_w)$ against $L(t)/L(t_w)-1$ which shows a very neat data collapse both for $a=0.1$ and for $a=0.3$. We point out that a good data collapse is also obtained when $C_{\rm eq}$ is not subtracted as in this case $C_{\rm eq} (t-t_w) \approx0$. Notice that we have used $L(t)/L(t_w)-1$ on the $x$-axis in order to better show the small time-difference regime.
Thus  this plot proves quite convincingly that the scaling structure described in Sec.~\ref{additive} for a non-disordered case also applies to the present disordered case. However, the scaling function $\mathcal{C}$  in Eq.~(\ref{Cagscalbelow}) changes with varying $a$. This is a clear indication of a violation of the so-called superuniversality hypothesis~\cite{cugliandolo2010topics, fisher1988nonequilibrium,*huse1989remanent}, which basically amounts to the fact that scaling functions should be independent on the presence/amount of disorder, in the present system.

In the inset of Fig.~\ref{compare_C}(c) we plot $C_{\rm ag} (t, t_w)$ against $L(t)/L(t_w)$ (on a double-log scale) to show the scaling function $\mathcal C (y)$ which is expected to decay as $\mathcal C (y) \sim y^{-\lambda_C}$ for $y\gg1$ (Eq.~\ref{lambdac}). More precisely, we consider different $a=0,0.1,0.3$, but at a fix $t_w=1$ in order to reach the asymptotic regime  of power-law decay. For $a=0$, $\lambda_C\simeq 1.25$, as expected for the pure Ising case in two-dimension \cite{fisher1988nonequilibrium,*huse1989remanent,humayun1991non,liu1991nonequilibrium,lorenz2007numerical}. For $a=0.1$, $\lambda_C \simeq 1.07$, and for $a=0.3$, $\lambda_C \simeq 1$. Clearly, the value of $\lambda_C$ is consistent with the Yeung-Rao-Desai inequality $\lambda_C \ge d/2$ \cite{yeung1996bounds}.  
\bfi
\begin{center}
\includegraphics*[width=0.96\columnwidth]{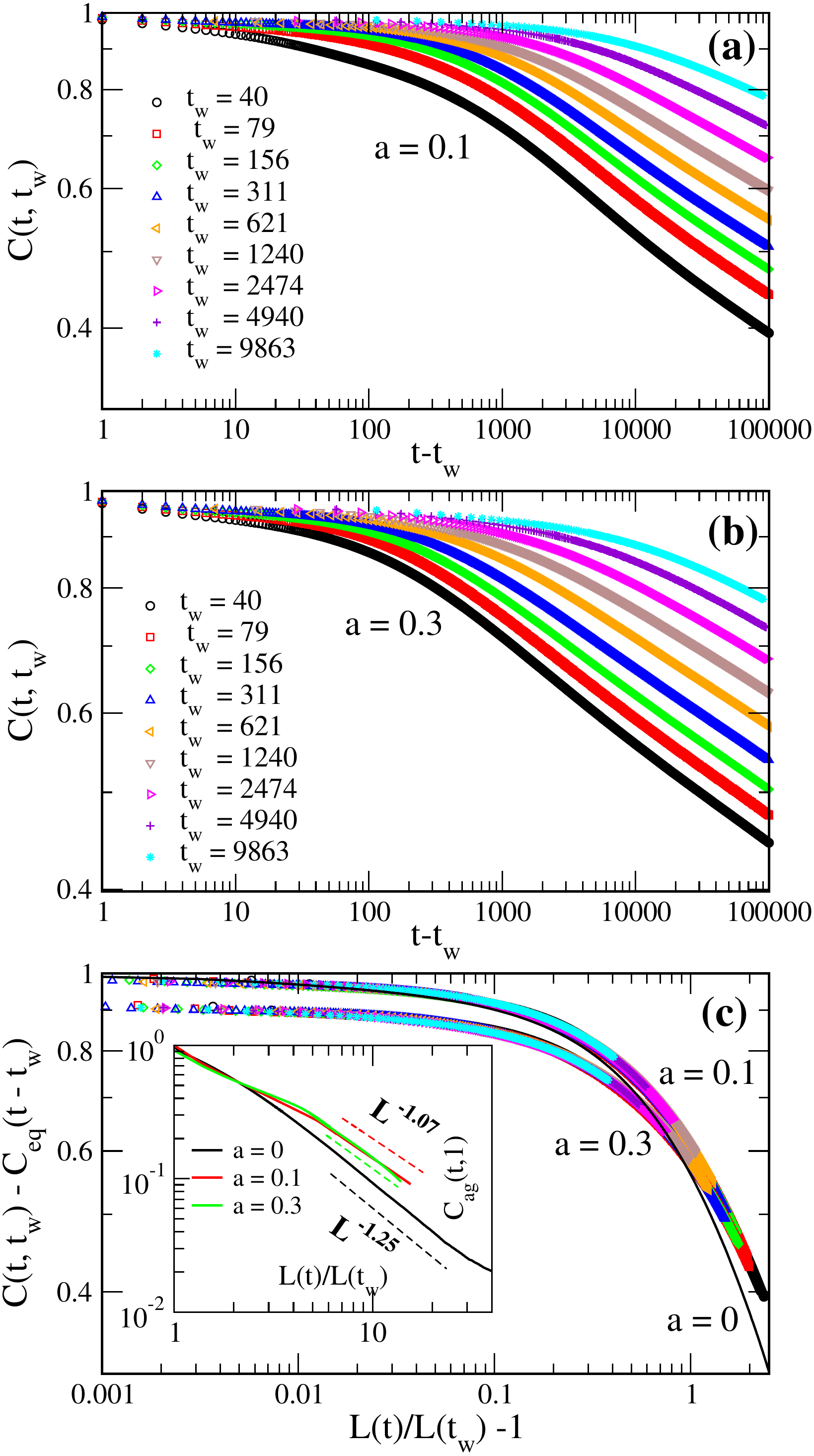}
\end{center}
\caption{(a) Plot of $C(t,t_w)$ versus time difference $(t-t_w)$ at $a=0.1$ for different values of $t_w$, drawn with
different colors (see key). (b) As in panel (a) but for $a=0.3$. (c) In the main frame, $C(t,t_w)-C_{\rm eq}(t-t_w)$ is plotted against $L(t)/L(t_w)-1$ (on a log-log scale) for $a=0.1$ (upper set of curves), and
$a=0.3$ (lower set of curves). Data for the same value of $t_w$ and for different $a$ are plotted with the same colors.   The solid back curve 
curve is the scaling function $\mathcal{C}$ of the pure case with $a=0$. 
The inset in panel (c) is the plot of $C_{\rm {ag}} (t, t_w=1)\equiv C(t,1)-C_{\rm eq}(t-1)$ vs. $L(t)/L(t_w)$  for different $a=0,0.1,0.3$. The dashed lines indicate the power-law decay at large $t$,  $C_{\rm ag}(t) \sim L(t)^{-\lambda_C}$, with $\lambda_C \simeq$ 1.25, 1.07, 1 for $a=$ 0, 0.1, 0.3, respectively.}
\label{compare_C}
\efi
\bfi
\begin{center}
\includegraphics[width=0.96\columnwidth]{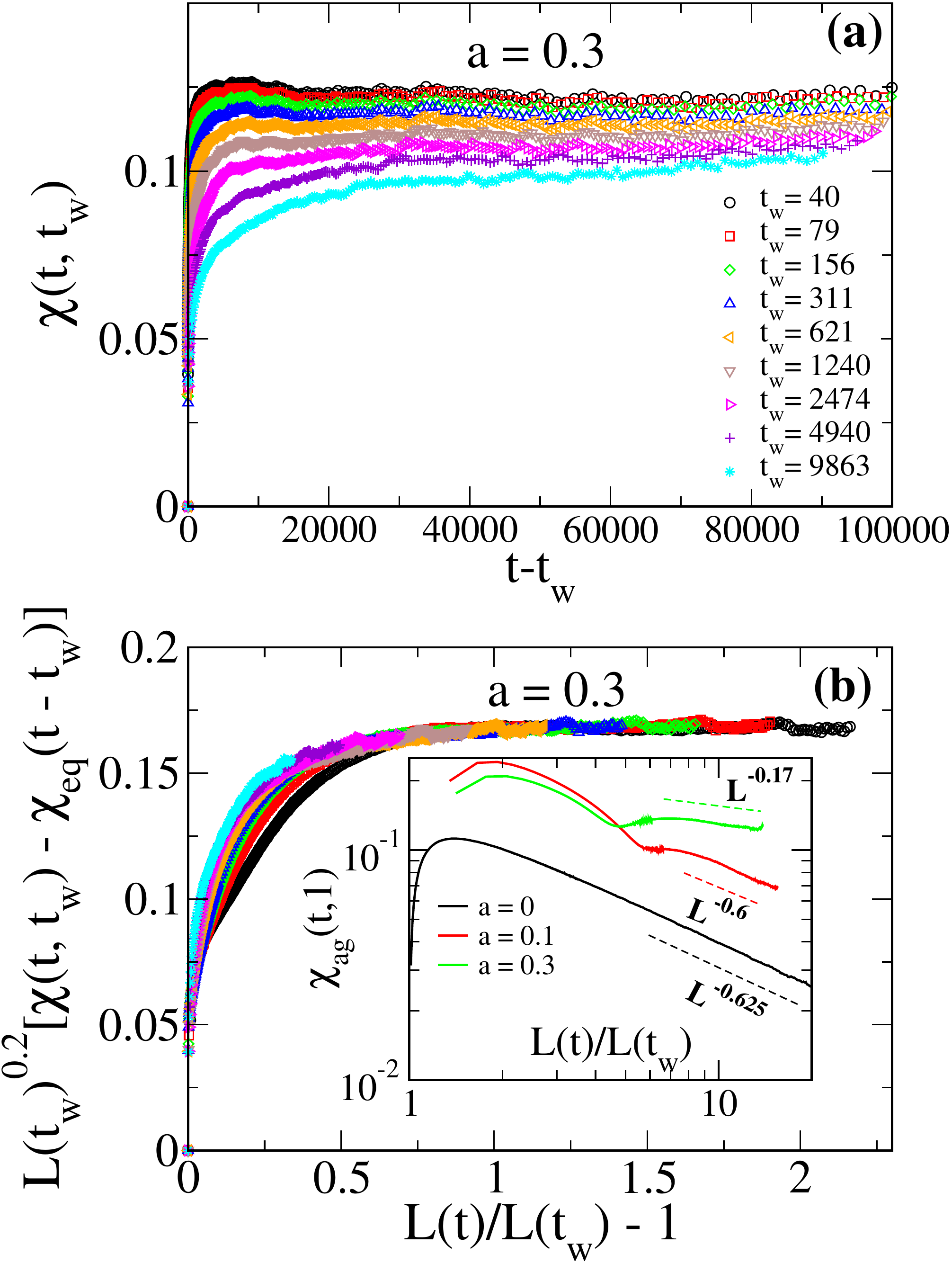}
\end{center}
\caption{ (a) Plot of $\chi(t,t_w)$ versus time difference $(t-t_w)$ at $a=0.3$ for different values of $t_w$, drawn with
different colors (see key). (b) In the main frame, $L(t_w)^\alpha [\chi(t,t_w)-\chi_{\rm eq}(t-t_w)]$ is plotted against $L(t)/L(t_w)-1$ for $a=0.3$. Curves for different values of $t_w$ are drawn with
different colors (see key in the upper panel). The values of $\alpha$  for different $a<a_f \simeq 0.4$ are $\alpha=0.625,0.6,0.4,0.2$ for $a=0,0.1,0.2,0.3$, respectively. The inset is the plot of the aging quantity $\chi_{\rm {ag}} (t, t_w)\equiv \chi(t,t_w)-\chi_{\rm eq}(t-t_w)$ vs. $L(t)/L(t_w)$ at $t_w=1$ and for different $a=0,0.1,0.3$. The dashed lines indicate the power-law decay of  $\chi_{\rm ag}(t)\sim L(t)^{-\lambda_\chi}$ at large $t$, with  $\lambda_\chi \simeq$ 0.625, 0.62, 0.17 for $a=$ 0, 0.1, 0.3, respectively. The exponent $\lambda_\chi$ is of the same order as the exponent $\alpha$, as expected.}
\label{compare_Chi}
\efi

Let us now analyze the response function.  In panel (a) of Fig.~\ref{compare_Chi}, we plot $\chi(t,t_w)$ versus the time difference $(t-t_w)$ at $a=0.3$ for different values of $t_w$ (see the figure caption). In panel (b), we show a plot of $L(t_w)^\alpha [\chi (t,t_w) -\chi_{\rm eq}(t-t_w)]$ against $L(t)/L(t_w)-1$,  where the aging exponent $\alpha>0$ is determined from the data collapse.  Notice that the data collapses only in the region of large time separation where the system show aging behavior. The good collapse shows that the additive combination of Eq.~(\ref{chiagferro}) also holds here. Here also, we  stress that the scaling function $\mathcal{F}$ depends quite strongly on $a$, a fact that invalidates superuniversality, as already noticed for the autocorrelation function.

To explore the behavior of $\alpha$ as function of $a$ we examine the autoresponse integrated function in the aging regime at the late times, where we expect the power-law decay [Eq.~\eqref{lambdachi}] at large $y$.
In the inset of the lower panel of Fig.~\ref{compare_Chi}(c), $\chi_{\rm ag} (t, t_w)$ is plotted against $L(t)/L(t_w)$ (on a double-log scale) for different $a=0,0.1,0.3$ at a fix $t_w=1$ in order to reach the asymptotic regime  of power-law decay.  The dashed lines show the behavior $\chi_{\rm ag}(t)\sim L(t)^{-\lambda_\chi}$ for $L(t)\gg1$. The estimated values are 
$\lambda_\chi \simeq$ 0.625, 0.6, 0.17 for $a=$ 0, 0.1, 0.3, respectively, which turn out to be comparable with the aging exponent $\alpha$, as expected. For $a=0$, $\lambda_\chi=\lambda_C/z$, as expected for the pure Ising case in the two dimensions~\cite{bray2002theory,picone2004local}.  

\subsubsection{Quenches with $a= a_f$}

A quench is made at $T_f=0$ with $a=a_f$ is a case of a critical quench and hence one expects a multiplicative scaling structure as the one discussed in Sec.~\ref{multiplicative}. This is further due to the fact that the equilibrium magnetization vanishes at $a_f$ (Fig.~\ref{fig_magn}) and hence $q_{\rm EA}=0$ in this case. This same structure of the two-time quantities is expected  to be observed also in our numerical quenches to finite temperatures, provided they are sufficiently deep in temperature and limited in time in order not to exceed the long-lasting aging stage. We already know  that this is the case with our choice of $T_f$ as the evolution of $L(t)$ in Fig.~\ref{fig_quench}(b) shows no signs of convergence to an equilibrium value. 
\bfi
\begin{center}
\includegraphics*[width=0.97\columnwidth]{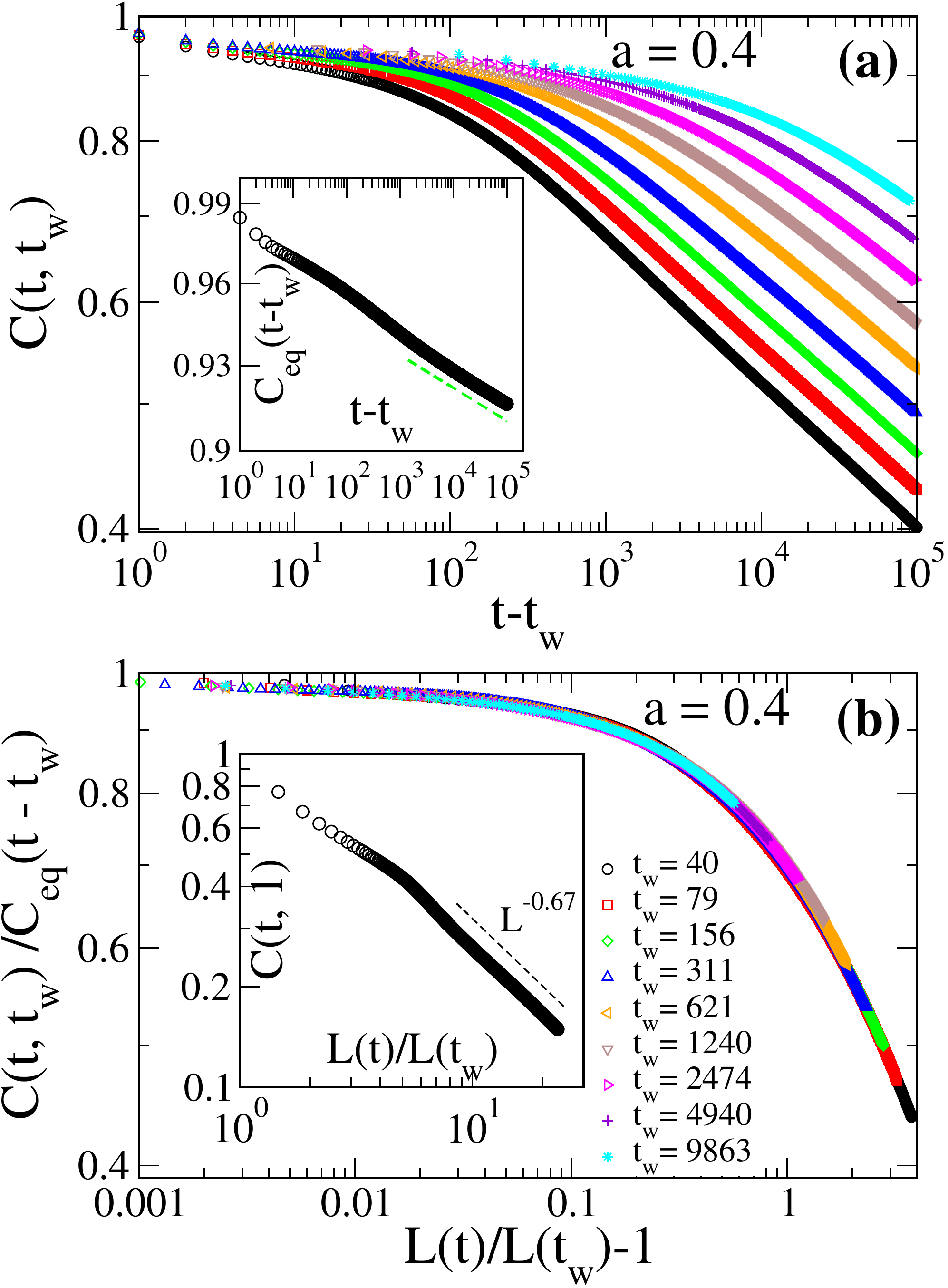}
\end{center}
\caption{(a) In the main frame, $C(t,t_w)$ is plotted against the time delay $(t-t_w)$ at $a=0.4$ for different values of $t_w$, drawn with
different colors (see key in lower panel). In the inset the equilibrium correlation $C_{\rm eq}(t-t_w)$ is plotted (on a log-log scale) against
$t-t_w$. The green dashed line is the algebraic form $(t-t_w)^{-0.005}$. (b) $C(t,t_w)/C_{\rm eq}(t-t_w)$ is plotted (on a log-log scale) against  $L(t)/L(t_w)-1$ for $a=0.4$. Curves for different values of $t_w$ are drawn with
different colors, see key. The inset of this panel is the plot of $C_(t, t_w)$ vs. $L(t)/L(t_w)$ at $t_w=1$ for $a=0.4$ and the dashed lines indicate the asymptotic behavior $C(t) \sim L(t)^{-\lambda_C}$ for $t\gg1$ with $\lambda_C \simeq 0.67$.}
\label{autocdivceqL_T075_a04}
\efi

Let us now focus on the numerical data for this case.  The main panel of Fig.~\ref{autocdivceqL_T075_a04}(a) is the unscaled plot of $C(t,t_w)$ vs the time-delay $t-t_w$, for different values of $t_w$, at $a=0.4$ (see the
figure caption). Scaling this data according to Eqs.~(\ref{Cmult},\ref{scalCcrit}) would require us to plot $C(t,t_w)/C_{\rm eq}(t-t_w)$ versus $L(t)/L(t_w)$. That is what we present in the main panel of Fig.~\ref{autocdivceqL_T075_a04}(b), where a nice data collapse of the curves at different $t_w$ can clearly be seen. This is a clearcut confirmation of the multiplicative structure in the quench of the present model at $a=0.4 (\simeq a_f)$.
The inset of Fig.~\ref{autocdivceqL_T075_a04}(a) shows the behavior of $C_{\rm eq}(t-t_w)$ on a double logarithmic scale. From Eq.~\eqref{Ceqmult},  by neglecting $t_0$ for large $t-t_w$, this quantity should obey $C_{\rm eq}(t-t_w)\simeq (t-t_w)^{-B}$. This is shown by a dashed green line. Notice that $C_{\rm eq}$ decays with a very small exponent $B\simeq 0.005$.

Coming to the asymptotic behavior of $C(t,t_w)$, from Eqs.~(\ref{Cmult},~\ref{scalCcrit},~\ref{Ceqmult}) it is expected
\begin{equation}
 C(t,t_w)=t_w^{-B}(t/t_w-1)^{-B}\widetilde {\rm C} \left (\frac{L(t)}{L(t_w)}\right ),
\end{equation}
i.e, $C(t,t_w)=t_w^{-B}\widetilde {\mathcal{C}}(y)$, with $\widetilde {\mathcal{C}}(y) \sim y^{-\lambda_C}$ for $y\gg1$. In order to reliably measure $\lambda_C$, we look at $C(t,t_w=1)$, namely we focus on the shortest possible waiting time. The inset of Fig.~\ref{autocdivceqL_T075_a04}(b)  is the plot of this quantity vs. $L(t)/L(t_w)$  for $a=0.4$. The dashed lines indicate the large $t$ behavior $C(t) \sim L(t)^{-\lambda_C}$, with $\lambda_C \simeq 0.67$.
\bfi
\begin{center}
\includegraphics*[width=0.95\columnwidth]{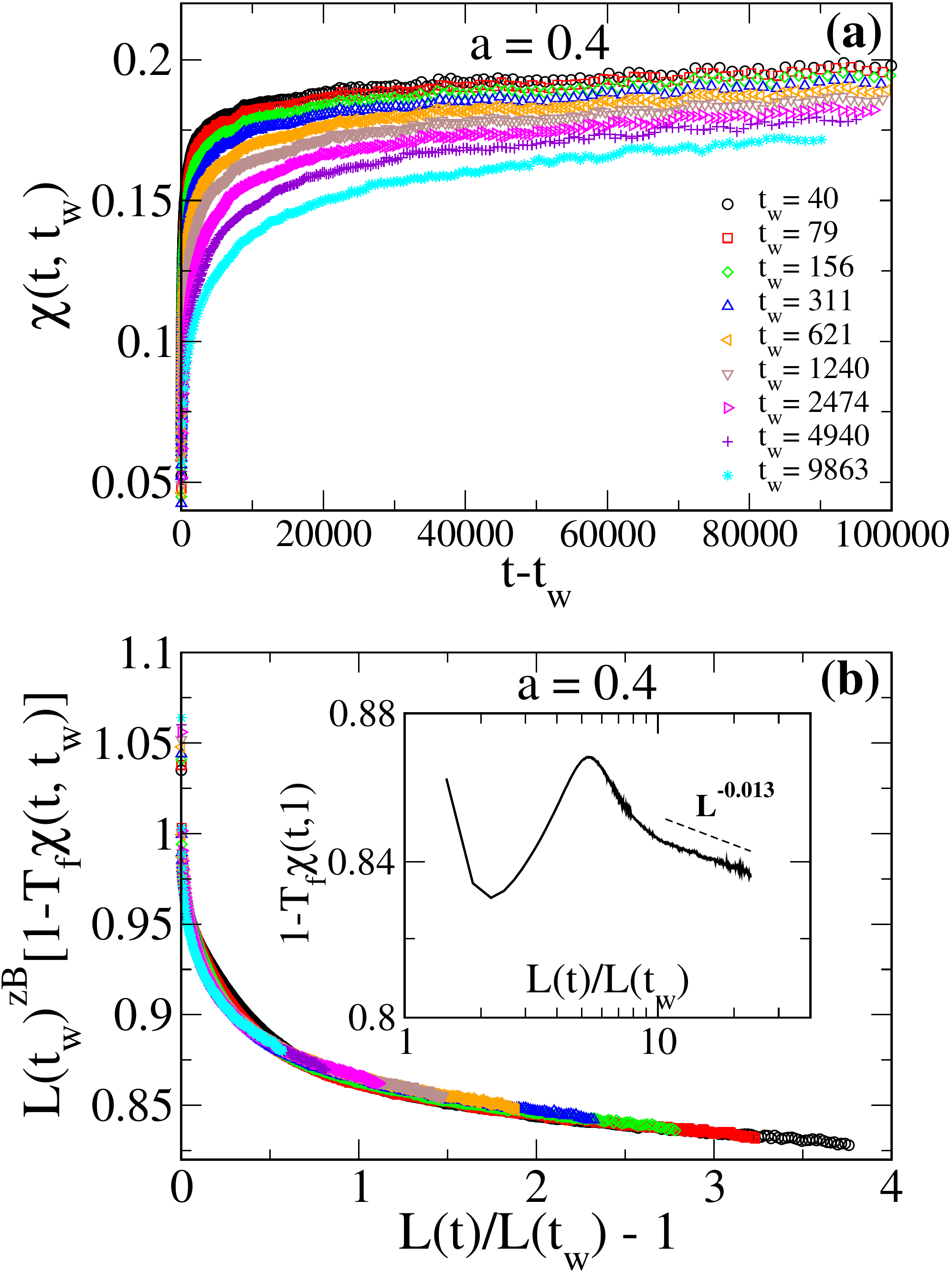}
\end{center}
\caption{ (a) Plot of $\chi(t,t_w)$ versus time difference $(t-t_w)$ at $a=0.4$ for different values of $t_w$, drawn with
different colors (see key). (b) In the main frame, $L(t_w)^{zB}[1-T_f\chi(t,t_w)]$ is plotted against $L(t)/L(t_w)-1$ with $zB=0.025$ for $a=0.4$. Curves for different values of $t_w$ are drawn with
different colors (see key in the upper panel). The inset shows the plot of $1-T_f\chi(t,t_w)$  against $L(t)/L(t_w)$ (on a double-log scale) at a fix $t_w=1$ for $a=0.4$ and a dashed line represent the large $t$ behavior $1-T_f\chi(t) \sim L(t)^{-\lambda_\chi}$ with $\lambda_\chi \simeq 0.013$.}
\label{chi_a04_risc}
\efi

Next, we look at $\chi (t,t_w)$, which is plotted for $a = 0.4$ in Fig.~\ref{chi_a04_risc}(a) as a function of the time delay $t-t_w$,  for different values of $t_w$ (see the figure caption).  The scaling of $\chi (t,t_w)$ is expected to be of the form of Eq.~(\ref{distance}). The exponent $B$ appearing in this scaling relation is already determined in the inset of Fig.~\ref{autocdivceqL_T075_a04}(a), where $C_{\rm eq}$ decays with a very small exponent $B\simeq 0.005$ for sufficiently large $(t-t_w)$ as shown by a dashed green line. Now recall Fig.~\ref{fig_quench}(b), where the asymptotic growth exponent $z\simeq 5.2$ for $a=0.4$ (shown by a maroon dashed line). Using these exponents $z$ and $B$, we plot $L(t_w)^{zB} [1-T_f\chi (t,t_w)]$ against $L(t)/L(t_w)-1$ in Fig.~\ref{chi_a04_risc}(b). According to Eq.~(\ref{distance}), we should observe data collapse of this data for values of $t_w$. This is indeed what one sees, which confirms the multiplicative scaling also for $\chi(t,t_w)$ in this case. 

Finally, considering the large $y$ behavior, the response scaling function $\mathcal{G}(y)$ in Eq.~(\ref{distance}) should scale asymptotically as $y^{-\lambda_\chi}$. As already discussed for the autocorrelation function, in order to enter the asymptotic regime of the power-law decay, we look at the data for $t_w=1$, and plot $1-T_f\chi (t,1)$ against $L(t)/L(t_w)$ on a double-log scale for $a=0.4$ in the inset of Fig.~\ref{chi_a04_risc}(b). The dashed line is the asymptotic behavior $1-T_f\chi (t,1)\sim L(t)^{-\lambda_\chi}$ with the exponent $\lambda_\chi \simeq 0.013$.

\subsubsection{Quenches with $a> a_f$}

As mentioned in Sec.~\ref{paradyn}, an algebraic behavior of $L(t)$ is not only observed at the critical point $(a=a_f)$, but also in the whole region $a_f<a<a_a$,  presumably due to the proximity of the spin-glass phase at $T=0$. Therefore, also in the case of quench with $a> a_f$, the two-time quantities are expected to show scaling. However, it is not obvious which scaling structure should apply to this case.
On the one hand, the system could feel the critical point at $(a=a_f,T=0)$, in that case one would expect basically the same multiplicative structure observed in the quench at $a=a_f$ and with the same exponents. On the other hand, the system can be influenced by the spin-glass phase with $q_{\rm EA}>0$, by virtue of which one could guess an additive structure. 
\bfi
\begin{center}
\includegraphics*[width=0.95\columnwidth]{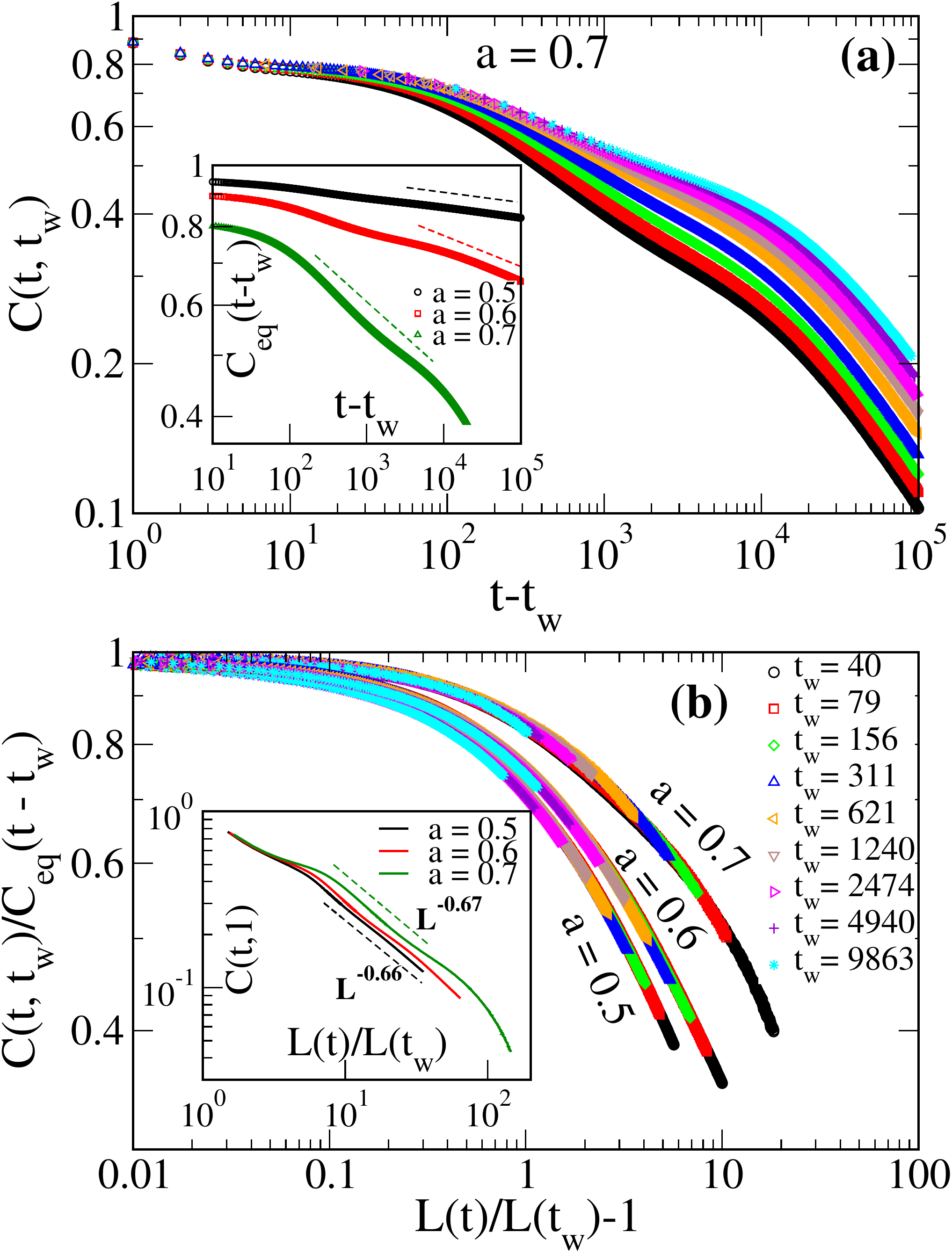}
\end{center}
\caption{(a) In the main frame, $C(t,t_w)$ is plotted against the time delay $(t-t_w)$ at $a=0.7$ for different values of $t_w$, drawn with
different colors and symbols (see key in lower panel). In the inset we have plotted the equilibrium correlation $C_{\rm eq}(t-t_w)$ (on a log-log scale) for different $a=0.5,0.6,0.7$ (see key in the inset). The dashed lines are the power-law fits of $(t-t_w)^{-B}$ for determining the exponent $B$: $B=0.016,0.05,0.12$ for $a=0.5,0.6,0.7$, respectively. (b) In the main frame, $C(t,t_w)/C_{\rm eq}(t-t_w)$ is plotted (on a log-log scale) against $L(t)/L(t_w)-1$ for $a=0.5,0.6$, and $0.7$.
Curves for different values of $t_w$ are drawn with
different colors (see key),  and for same $t_w$ but for different $a$'s are plotted with the same colors and same symbols. In the inset of this panel, $C_(t, t_w)$ vs. $L(t)/L(t_w)$ is plotted at a fix $t_w=1$ and for $a=0.5,0.6,0.7$ (see key in the inset).   The dashed lines are the asymptotic behavior $C(t) \sim L(t)^{-\lambda_C}$ for $t\gg1$ with $\lambda_C \simeq$ 0.66, 0.67, 0.67 for $a=$ 0.5, 0.6, 0.7, respectively. 
}
\label{compare_C_para}
\efi
\bfi
\begin{center}
\includegraphics*[width=0.95\columnwidth]{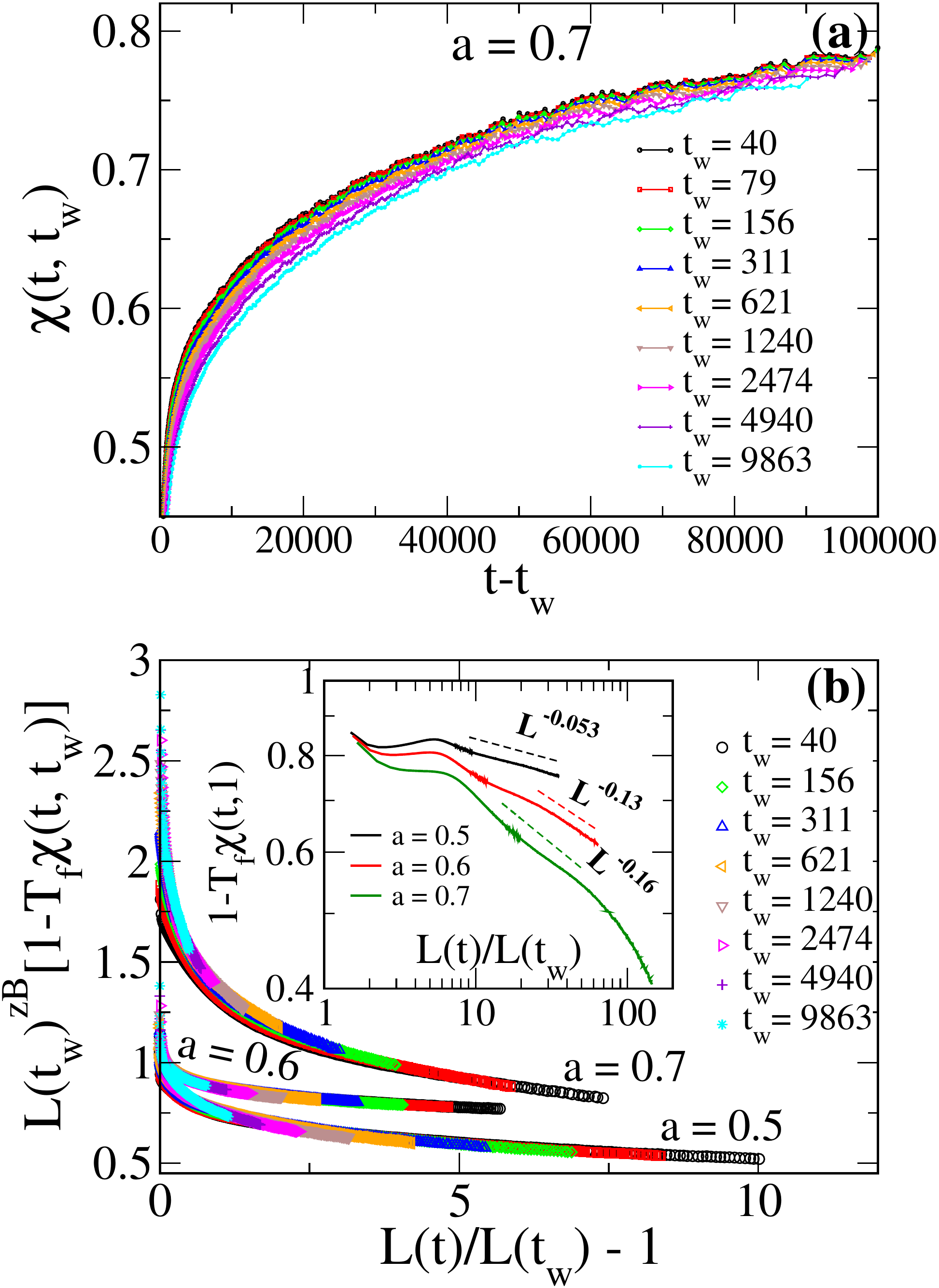}
\end{center}
\caption{ (a) Plot of $\chi(t,t_w)$ versus time difference $(t-t_w)$ for $a=0.7$ and different values of $t_w$, drawn with
different colors (see key). (b) In the main frame, $L(t_w)^{zB}[1-T_f\chi(t,t_w)]$ is plotted against $L(t)/L(t_w)-1$ with the exponent $zB=$ 0.067, 0.15, 0.3 for $a=$ 0.5, 0.6, and 0.7, respectively. Curves for different values of $t_w$ are drawn with different colors (see key). In the inset we plot $1-T_f\chi(t,t_w)$  against $L(t)/L(t_w)$ (on a double-log scale), for $t_w=1$ and $a=0.5,0.6,0.7$ (see key in the inset). The dashed lines are the large $t$ behavior $1-T_f\chi(t) \sim L(t)^{-\lambda_\chi}$ with $\lambda_\chi \simeq 0.053$, 0.13, 0.16 for $a=0.5$, 0.6, 0.7, respectively.}
\label{chiL_T075_a05_resc}
\efi

From our numerical data, we found that for all $a>a_f$, $C(t,t_w)<C_{\rm eq}(t-t_w)$ for any $t>t_w$ which implies $C_{\rm ag}=C-C_{\rm eq}<0$, a fact that rules out the additive scheme, as $C_{\rm ag}<0$ is unphysical.  Indeed, form our simulation data, a multiplicative structure is very well verified as we found the data collapse of $C(t,t_w)/C_{\rm eq}(t-t_w)$ for various values of $t_w$. This data is shown in the main frame of Fig.~\ref{compare_C_para}(b) for  different values of $a=0.5$, 0.6, and 0.7. An excellent data collapse can be clearly seen for all $a>a_f$.  The upper panel Fig.~\ref{compare_C_para}(a) is the unscaled plot of $C(t,t_w)$ against the time difference $(t-t_w)$, for $a=0.7$. In an inset of this figure,  we show $C_{\rm eq}(t-t_w)$ on a double logarithmic scale for $a=0.5,0.6,0.7$. From this data, we determine the
exponent $B$ from the power-law fit of $C_{\rm eq}\sim (t-t_w)^{-B}$ for large $t-t_w$ as shown by the dashed lines. The exponent $B$ comes out as  $B=0.016,0.05,0.12$ for $a=0.5,0.6,0.7$, respectively, which will be needed to scale $\chi(t,t_w)$.  Next, having understood how to extract the autocorrelation exponent $\lambda_C$, we plot $C(t,1)$ vs $L(t)/L(t_w)$ (on a double-log scale) in the inset of Fig.~\ref{compare_C_para}(b), for $a=0.5,0.6,0.7$. The dashed lines give the asymptotic behavior $L(t)^{-\lambda_C}$ with $\lambda_C \simeq$ 0.66, 0.67, 0.67 for $a=$ 0.5, 0.6, 0.7, respectively. Notice that the exponent $\lambda_C$ takes a value around $\lambda _C\simeq 0.67$ for all values of $a$ in the paramagnetic region, suggesting a unique value. 

Now we move to analyze the response function $\chi (t,t_w)$, which is shown for $a=0.7$ in Fig.~\ref{chiL_T075_a05_resc}(a).
Recalling the exponent $B$ determined in inset of Fig.~\ref{compare_C_para}(a) and the dynamic exponent $z$ from Fig.~\ref{fig_quench}(b), we next plotted $L(t_w)^{zB} [1-T_f\chi(t,t_w)]$ versus $L(t)/L(t_w)-1$ in the lower panel Fig.~\ref{chiL_T075_a05_resc}(b) for $a=0.5,0.6,0.7$ and for different $t_w$-values (see the figure caption). Clearly, from this figure, our determined values of static and dynamic exponents $z$ and $B$ produce a nice data collapse for all different $a=0.5$, 0.6, and 0.7. As before, given Eq.~\eqref{distance},  this confirms the multiplicative structure also for this case of quench with $a> a_f$.  The inset of Fig.~\ref{chiL_T075_a05_resc}(b) displays the asymptotic behavior given by the power-law decay $1-T_f \chi(t,t_w) \sim L(t)^{-\lambda_\chi}$, with the exponent $\lambda_\chi \simeq$ 0.053, 0.13, and 0.16 for $a=$ 0.5, 0.6, and 0.7, respectively. Notice that in this case the exponent $\lambda_\chi \ne zB$, as expected, 
unlike in the previous case of a quench with $a\lesssim a_f$. Furthermore, notice that both the scaling functions  $\widetilde{C}(y)$ and $\mathcal{G}(y)$ for $C(t,t_w)$ and $\chi(t,t_w)$, respectively, in Figs.~\ref{compare_C_para}(b) and \ref{chiL_T075_a05_resc}(b) are different for different $a$, which again consistently showing a violation of superuniversality. 

Finally, we find it useful to list in a Table~\ref{exp_values} the measured values of all the exponents: the autocorrelation exponent $\lambda_C$, the auto-response exponent $\lambda_\chi$, and the scaling exponents $\alpha$ and $zB$.
\begin{table}

\centering
\begin{tabular}{| c | c | c |c|c|}
 \hline 
  $a$&$\alpha$&$zB$&$\lambda_C$&$\lambda_\chi$\\ \hline
  0&0.625&-&1.25&0.625\\ \hline
  0.1&0.62&-&1.07&0.6\\ \hline
  0.2&0.4&-&1.05&0.42\\ \hline
  0.3&0.2&-&1.0&0.17\\ \hline
  0.4&-&0.025&0.67&0.013\\ \hline
  0.5&-&0.067&0.66&0.053\\ \hline
  0.6&-&0.15&0.67&0.13\\ \hline
 0.7&-&0.28&0.67&0.16\\ \hline
 \end{tabular}
\caption{Values of  the autocorrelation exponent $\lambda_C$, the auto-response exponent $\lambda_\chi$, and the scaling exponents $\alpha$, $zB$ for all considered values of $a$ at the temperature quench $T_f=0.75$.}
\label{exp_values}

\end{table}

\section{Summary and Discussion}
\label{concl}

This paper was a review of our results on growth kinetics and aging in a two-dimensional frustrated system. After a quenching has been made to a low-temperature $T_f$, the system show different phases with varying the frustration strength namely the fraction of negative couplings $a$. These phases are ferromagnetic for $0\le a<a_f$, paramagnetic with a zero-temperature spin-glass phase for $a_f\le a<a_a$, and an antiferromagnetic for $a_a \le a <1$. We have found that the system shows a nontrivial behavior of both the growth and the aging properties for quenches operated in different phases. Our major findings are summarized as follows.

In the FM and AFM phases, the speed of growth varies in a non-monotonic way as the amount of disorder $a$ is increased. Specifically, speaking of the FM phase for definiteness, there exists a value $a^*\sim 0.2$ where the kinetics is slower than for any other value of $a$. This non-monotonous type of the growth behavior can be related to the topology of the bond network (see Sec.~\ref{model}). This result is consistent with what has already been found in models of disordered magnets without frustration~\cite{corberi2015,corberi2015phase,corberi2013scaling}, and hence extends its validity to the realm of frustrated systems, pointing toward a general robustness. We showed that the growth law $L(t)$ is of a logarithmic type in the whole FM ($0<a<a_f$) as well as AFM region $a_a<a<1$. Instead, in the PM region we find that $L(t)$ grows algebraically in the whole phase, irrespective of the fact that the geometrical properties of the bond network greatly change as $a$ is varied in $[a_f,a_a]$. 

In the whole ferromagnetic region, for $0\le a\le a_f$, we observed that the equilibrium and the aging parts of 
the two time quantities (autocorrelation and response function) combine in an additive way to form the complete correlation and response , i.e., $C(t,t_w)=C_{\rm eq}(t-t_w)+C_{\rm ag}(t,t_w)$ (and similarly for the response function). However, the scaling functions for the aging parts of both quantities strongly depend on the amount $a$ of frustration. This shows that the property of superuniversality is not obeyed. Further, the aging  exponent $\alpha$ depend on $a$, and vanishes in the limit $a\to a_f^-$. This can be interpreted due to the fact that, in this limit, the system has fractal structure, and therefore interfaces are free to move without experiencing any restoring force. The same fact, indeed, can be considered the origin of the speeding up of the growth for $a\gtrsim a_f$. 

At $a=a_f$, $T_c=0$, which is in fact a critical point of the FM phase, the additive structure breaks down, and a multiplicative structure emerges where the equilibrium and the aging parts of two-time quantities multiply, i.e., $C=C_{\rm eq}\cdot C_{\rm ag}$. This is consistent with  other spin-glass systems~\cite{berthier2002geometrical,manssen2015aging}, where a similar behavior has been observed. 
The same multiplicative structure is found even when the system is quenched in the PM region with $a>a_f$. 
Further, the scaling functions of both $C$ and $\chi$, and the exponent $zB$  turns out to be strongly dependent on the amount $a$. This is further showing a violation of superuniversality. 

 We also discussed the large-$t$ behavior where the scaling functions for both the quantities $C(t,t_w)$ and $\chi(t,t_w)$ are   governed by power-law decay with exponents $\lambda_C$ and $\lambda_\chi$. The values of these exponents are reported in Table~\ref{exp_values}. It is worth mentioning that in 
the FM phase the values of $\lambda_C$ are  in agreement with the Yeung-Rao-Desai inequality $\lambda_C \ge d/2$ \cite{yeung1996bounds}, and approach the lower-bound $d/2$ as $a\to a_f^-$. For the response function, $\lambda_\chi \simeq \alpha$ is found to be well obeyed.

We conclude this paper with a final remark that our results for the growth kinetics and the aging properties clearly distinguish the different phases of a $d=2$ frustrated system. In this respect, this motivates a further interest in investigating the three-dimensional case of the frustrated system where its phase-structure  can be elucidated from a detailed nonequilibrium study of the growth and the aging quantities.

\acknowledgments
MK would like to acknowledge the support of the Royal Society - SERB Newton International fellowship (NIF$\backslash$R1$\backslash$180386). The numerical simulations were performed on High-performance computing clusters at SPS, JNU, New Delhi, and at IUAC, New Delhi. SP is grateful to the Department of Science and Technology (India) for support through a J.C. Bose fellowship. \\
\ \\
{\bf Author contribution}: FC, EL and SP conceived, designed and supervised the project. MK developed the codes, obtained the data, and performed the analysis. All the authors contributed to manuscript preparation.

\bibliography{dgdiso.bib}

\begin{thebibliography}{118}%
\makeatletter
\providecommand \@ifxundefined [1]{%
 \@ifx{#1\undefined}
}%
\providecommand \@ifnum [1]{%
 \ifnum #1\expandafter \@firstoftwo
 \else \expandafter \@secondoftwo
 \fi
}%
\providecommand \@ifx [1]{%
 \ifx #1\expandafter \@firstoftwo
 \else \expandafter \@secondoftwo
 \fi
}%
\providecommand \natexlab [1]{#1}%
\providecommand \enquote  [1]{``#1''}%
\providecommand \bibnamefont  [1]{#1}%
\providecommand \bibfnamefont [1]{#1}%
\providecommand \citenamefont [1]{#1}%
\providecommand \href@noop [0]{\@secondoftwo}%
\providecommand \href [0]{\begingroup \@sanitize@url \@href}%
\providecommand \@href[1]{\@@startlink{#1}\@@href}%
\providecommand \@@href[1]{\endgroup#1\@@endlink}%
\providecommand \@sanitize@url [0]{\catcode `\\12\catcode `\$12\catcode
  `\&12\catcode `\#12\catcode `\^12\catcode `\_12\catcode `\%12\relax}%
\providecommand \@@startlink[1]{}%
\providecommand \@@endlink[0]{}%
\providecommand \url  [0]{\begingroup\@sanitize@url \@url }%
\providecommand \@url [1]{\endgroup\@href {#1}{\urlprefix }}%
\providecommand \urlprefix  [0]{URL }%
\providecommand \Eprint [0]{\href }%
\providecommand \doibase [0]{http://dx.doi.org/}%
\providecommand \selectlanguage [0]{\@gobble}%
\providecommand \bibinfo  [0]{\@secondoftwo}%
\providecommand \bibfield  [0]{\@secondoftwo}%
\providecommand \translation [1]{[#1]}%
\providecommand \BibitemOpen [0]{}%
\providecommand \bibitemStop [0]{}%
\providecommand \bibitemNoStop [0]{.\EOS\space}%
\providecommand \EOS [0]{\spacefactor3000\relax}%
\providecommand \BibitemShut  [1]{\csname bibitem#1\endcsname}%
\let\auto@bib@innerbib\@empty
\bibitem [{\citenamefont {Bray}(2002)}]{bray2002theory}%
  \BibitemOpen
  \bibfield  {author} {\bibinfo {author} {\bibfnamefont {A.~J.}\ \bibnamefont
  {Bray}},\ }\href@noop {} {\bibfield  {journal} {\bibinfo  {journal} {Advances
  in Physics}\ }\textbf {\bibinfo {volume} {51}},\ \bibinfo {pages} {481}
  (\bibinfo {year} {2002})}\BibitemShut {NoStop}%
\bibitem [{\citenamefont {Puri}(2009)}]{puri2009kinetics}%
  \BibitemOpen
  \bibfield  {author} {\bibinfo {author} {\bibfnamefont {S.}~\bibnamefont
  {Puri}},\ }in\ \href@noop {} {\emph {\bibinfo {booktitle} {Kinetics of Phase
  Transitions}}}\ (\bibinfo  {publisher} {CRC Press},\ \bibinfo {year} {2009})\
  pp.\ \bibinfo {pages} {13--74}\BibitemShut {NoStop}%
\bibitem [{\citenamefont {Lai}\ \emph {et~al.}(1988)\citenamefont {Lai},
  \citenamefont {Mazenko},\ and\ \citenamefont {Valls}}]{lai1988classes}%
  \BibitemOpen
  \bibfield  {author} {\bibinfo {author} {\bibfnamefont {Z.}~\bibnamefont
  {Lai}}, \bibinfo {author} {\bibfnamefont {G.~F.}\ \bibnamefont {Mazenko}}, \
  and\ \bibinfo {author} {\bibfnamefont {O.~T.}\ \bibnamefont {Valls}},\
  }\href@noop {} {\bibfield  {journal} {\bibinfo  {journal} {Physical Review
  B}\ }\textbf {\bibinfo {volume} {37}},\ \bibinfo {pages} {9481} (\bibinfo
  {year} {1988})}\BibitemShut {NoStop}%
\bibitem [{\citenamefont {Fisher}\ and\ \citenamefont
  {Huse}(1988)}]{fisher1988nonequilibrium}%
  \BibitemOpen
  \bibfield  {author} {\bibinfo {author} {\bibfnamefont {D.~S.}\ \bibnamefont
  {Fisher}}\ and\ \bibinfo {author} {\bibfnamefont {D.~A.}\ \bibnamefont
  {Huse}},\ }\href@noop {} {\bibfield  {journal} {\bibinfo  {journal} {Physical
  Review B}\ }\textbf {\bibinfo {volume} {38}},\ \bibinfo {pages} {373}
  (\bibinfo {year} {1988})}\BibitemShut {NoStop}%
\bibitem [{\citenamefont {Huse}(1989)}]{huse1989remanent}%
  \BibitemOpen
  \bibfield  {author} {\bibinfo {author} {\bibfnamefont {D.~A.}\ \bibnamefont
  {Huse}},\ }\href@noop {} {\bibfield  {journal} {\bibinfo  {journal} {Physical
  Review B}\ }\textbf {\bibinfo {volume} {40}},\ \bibinfo {pages} {304}
  (\bibinfo {year} {1989})}\BibitemShut {NoStop}%
\bibitem [{\citenamefont {Corberi}\ \emph
  {et~al.}(2015{\natexlab{a}})\citenamefont {Corberi}, \citenamefont
  {Lippiello},\ and\ \citenamefont {Zannetti}}]{corberi2015}%
  \BibitemOpen
  \bibfield  {author} {\bibinfo {author} {\bibfnamefont {F.}~\bibnamefont
  {Corberi}}, \bibinfo {author} {\bibfnamefont {E.}~\bibnamefont {Lippiello}},
  \ and\ \bibinfo {author} {\bibfnamefont {M.}~\bibnamefont {Zannetti}},\
  }\href@noop {} {\bibfield  {journal} {\bibinfo  {journal} {Journal of
  Statistical Mechanics: Theory and Experiment}\ }\textbf {\bibinfo {volume}
  {2015}},\ \bibinfo {pages} {P10001} (\bibinfo {year}
  {2015}{\natexlab{a}})}\BibitemShut {NoStop}%
\bibitem [{\citenamefont {Corberi}\ \emph
  {et~al.}(2015{\natexlab{b}})\citenamefont {Corberi}, \citenamefont
  {Zannetti}, \citenamefont {Lippiello}, \citenamefont {Burioni},\ and\
  \citenamefont {Vezzani}}]{corberi2015phase}%
  \BibitemOpen
  \bibfield  {author} {\bibinfo {author} {\bibfnamefont {F.}~\bibnamefont
  {Corberi}}, \bibinfo {author} {\bibfnamefont {M.}~\bibnamefont {Zannetti}},
  \bibinfo {author} {\bibfnamefont {E.}~\bibnamefont {Lippiello}}, \bibinfo
  {author} {\bibfnamefont {R.}~\bibnamefont {Burioni}}, \ and\ \bibinfo
  {author} {\bibfnamefont {A.}~\bibnamefont {Vezzani}},\ }\href@noop {}
  {\bibfield  {journal} {\bibinfo  {journal} {Physical Review E}\ }\textbf
  {\bibinfo {volume} {91}},\ \bibinfo {pages} {062122} (\bibinfo {year}
  {2015}{\natexlab{b}})}\BibitemShut {NoStop}%
\bibitem [{\citenamefont {Paul}\ \emph {et~al.}(2004)\citenamefont {Paul},
  \citenamefont {Puri},\ and\ \citenamefont {Rieger}}]{paul2004domain}%
  \BibitemOpen
  \bibfield  {author} {\bibinfo {author} {\bibfnamefont {R.}~\bibnamefont
  {Paul}}, \bibinfo {author} {\bibfnamefont {S.}~\bibnamefont {Puri}}, \ and\
  \bibinfo {author} {\bibfnamefont {H.}~\bibnamefont {Rieger}},\ }\href@noop {}
  {\bibfield  {journal} {\bibinfo  {journal} {EPL (Europhysics Letters)}\
  }\textbf {\bibinfo {volume} {68}},\ \bibinfo {pages} {881} (\bibinfo {year}
  {2004})}\BibitemShut {NoStop}%
\bibitem [{\citenamefont {Paul}\ \emph {et~al.}(2005)\citenamefont {Paul},
  \citenamefont {Puri},\ and\ \citenamefont {Rieger}}]{paul2005domain}%
  \BibitemOpen
  \bibfield  {author} {\bibinfo {author} {\bibfnamefont {R.}~\bibnamefont
  {Paul}}, \bibinfo {author} {\bibfnamefont {S.}~\bibnamefont {Puri}}, \ and\
  \bibinfo {author} {\bibfnamefont {H.}~\bibnamefont {Rieger}},\ }\href@noop {}
  {\bibfield  {journal} {\bibinfo  {journal} {Physical Review E}\ }\textbf
  {\bibinfo {volume} {71}},\ \bibinfo {pages} {061109} (\bibinfo {year}
  {2005})}\BibitemShut {NoStop}%
\bibitem [{\citenamefont {Paul}\ \emph {et~al.}(2007)\citenamefont {Paul},
  \citenamefont {Schehr},\ and\ \citenamefont {Rieger}}]{paul2007superaging}%
  \BibitemOpen
  \bibfield  {author} {\bibinfo {author} {\bibfnamefont {R.}~\bibnamefont
  {Paul}}, \bibinfo {author} {\bibfnamefont {G.}~\bibnamefont {Schehr}}, \ and\
  \bibinfo {author} {\bibfnamefont {H.}~\bibnamefont {Rieger}},\ }\href@noop {}
  {\bibfield  {journal} {\bibinfo  {journal} {Physical Review E}\ }\textbf
  {\bibinfo {volume} {75}},\ \bibinfo {pages} {030104} (\bibinfo {year}
  {2007})}\BibitemShut {NoStop}%
\bibitem [{\citenamefont {Rieger}\ \emph {et~al.}(2005)\citenamefont {Rieger},
  \citenamefont {Schehr},\ and\ \citenamefont {Paul}}]{rieger2005growing}%
  \BibitemOpen
  \bibfield  {author} {\bibinfo {author} {\bibfnamefont {H.}~\bibnamefont
  {Rieger}}, \bibinfo {author} {\bibfnamefont {G.}~\bibnamefont {Schehr}}, \
  and\ \bibinfo {author} {\bibfnamefont {R.}~\bibnamefont {Paul}},\ }\href@noop
  {} {\bibfield  {journal} {\bibinfo  {journal} {Progress of Theoretical
  Physics Supplement}\ }\textbf {\bibinfo {volume} {157}},\ \bibinfo {pages}
  {111} (\bibinfo {year} {2005})}\BibitemShut {NoStop}%
\bibitem [{\citenamefont {Henkel}\ and\ \citenamefont
  {Pleimling}(2006)}]{henkel2006ageing}%
  \BibitemOpen
  \bibfield  {author} {\bibinfo {author} {\bibfnamefont {M.}~\bibnamefont
  {Henkel}}\ and\ \bibinfo {author} {\bibfnamefont {M.}~\bibnamefont
  {Pleimling}},\ }\href@noop {} {\bibfield  {journal} {\bibinfo  {journal} {EPL
  (Europhysics Letters)}\ }\textbf {\bibinfo {volume} {76}},\ \bibinfo {pages}
  {561} (\bibinfo {year} {2006})}\BibitemShut {NoStop}%
\bibitem [{\citenamefont {Henkel}\ and\ \citenamefont
  {Pleimling}(2008)}]{henkel2008}%
  \BibitemOpen
  \bibfield  {author} {\bibinfo {author} {\bibfnamefont {M.}~\bibnamefont
  {Henkel}}\ and\ \bibinfo {author} {\bibfnamefont {M.}~\bibnamefont
  {Pleimling}},\ }\href@noop {} {\bibfield  {journal} {\bibinfo  {journal}
  {Physical Review B}\ }\textbf {\bibinfo {volume} {78}},\ \bibinfo {pages}
  {224419} (\bibinfo {year} {2008})}\BibitemShut {NoStop}%
\bibitem [{\citenamefont {Baumann}\ \emph {et~al.}(2007)\citenamefont
  {Baumann}, \citenamefont {Henkel},\ and\ \citenamefont
  {Pleimling}}]{baumann2007phase}%
  \BibitemOpen
  \bibfield  {author} {\bibinfo {author} {\bibfnamefont {F.}~\bibnamefont
  {Baumann}}, \bibinfo {author} {\bibfnamefont {M.}~\bibnamefont {Henkel}}, \
  and\ \bibinfo {author} {\bibfnamefont {M.}~\bibnamefont {Pleimling}},\
  }\href@noop {} {\bibfield  {journal} {\bibinfo  {journal} {arXiv preprint
  arXiv:0709.3228}\ } (\bibinfo {year} {2007})}\BibitemShut {NoStop}%
\bibitem [{\citenamefont {Burioni}\ \emph {et~al.}(2007)\citenamefont
  {Burioni}, \citenamefont {Cassi}, \citenamefont {Corberi},\ and\
  \citenamefont {Vezzani}}]{burioni2007phase}%
  \BibitemOpen
  \bibfield  {author} {\bibinfo {author} {\bibfnamefont {R.}~\bibnamefont
  {Burioni}}, \bibinfo {author} {\bibfnamefont {D.}~\bibnamefont {Cassi}},
  \bibinfo {author} {\bibfnamefont {F.}~\bibnamefont {Corberi}}, \ and\
  \bibinfo {author} {\bibfnamefont {A.}~\bibnamefont {Vezzani}},\ }\href@noop
  {} {\bibfield  {journal} {\bibinfo  {journal} {Physical Review E}\ }\textbf
  {\bibinfo {volume} {75}},\ \bibinfo {pages} {011113} (\bibinfo {year}
  {2007})}\BibitemShut {NoStop}%
\bibitem [{\citenamefont {Burioni}\ \emph {et~al.}(2013)\citenamefont
  {Burioni}, \citenamefont {Corberi},\ and\ \citenamefont
  {Vezzani}}]{burioni2013topological}%
  \BibitemOpen
  \bibfield  {author} {\bibinfo {author} {\bibfnamefont {R.}~\bibnamefont
  {Burioni}}, \bibinfo {author} {\bibfnamefont {F.}~\bibnamefont {Corberi}}, \
  and\ \bibinfo {author} {\bibfnamefont {A.}~\bibnamefont {Vezzani}},\
  }\href@noop {} {\bibfield  {journal} {\bibinfo  {journal} {Physical Review
  E}\ }\textbf {\bibinfo {volume} {87}},\ \bibinfo {pages} {032160} (\bibinfo
  {year} {2013})}\BibitemShut {NoStop}%
\bibitem [{\citenamefont {Lippiello}\ \emph {et~al.}(2010)\citenamefont
  {Lippiello}, \citenamefont {Mukherjee}, \citenamefont {Puri},\ and\
  \citenamefont {Zannetti}}]{lippiello2010scaling}%
  \BibitemOpen
  \bibfield  {author} {\bibinfo {author} {\bibfnamefont {E.}~\bibnamefont
  {Lippiello}}, \bibinfo {author} {\bibfnamefont {A.}~\bibnamefont
  {Mukherjee}}, \bibinfo {author} {\bibfnamefont {S.}~\bibnamefont {Puri}}, \
  and\ \bibinfo {author} {\bibfnamefont {M.}~\bibnamefont {Zannetti}},\
  }\href@noop {} {\bibfield  {journal} {\bibinfo  {journal} {EPL (Europhysics
  Letters)}\ }\textbf {\bibinfo {volume} {90}},\ \bibinfo {pages} {46006}
  (\bibinfo {year} {2010})}\BibitemShut {NoStop}%
\bibitem [{\citenamefont {Corberi}\ \emph
  {et~al.}(2011{\natexlab{a}})\citenamefont {Corberi}, \citenamefont
  {Lippiello}, \citenamefont {Mukherjee}, \citenamefont {Puri},\ and\
  \citenamefont {Zannetti}}]{corberi2011growth}%
  \BibitemOpen
  \bibfield  {author} {\bibinfo {author} {\bibfnamefont {F.}~\bibnamefont
  {Corberi}}, \bibinfo {author} {\bibfnamefont {E.}~\bibnamefont {Lippiello}},
  \bibinfo {author} {\bibfnamefont {A.}~\bibnamefont {Mukherjee}}, \bibinfo
  {author} {\bibfnamefont {S.}~\bibnamefont {Puri}}, \ and\ \bibinfo {author}
  {\bibfnamefont {M.}~\bibnamefont {Zannetti}},\ }\href@noop {} {\bibfield
  {journal} {\bibinfo  {journal} {Journal of Statistical Mechanics: Theory and
  Experiment}\ }\textbf {\bibinfo {volume} {2011}},\ \bibinfo {pages} {P03016}
  (\bibinfo {year} {2011}{\natexlab{a}})}\BibitemShut {NoStop}%
\bibitem [{\citenamefont {Corberi}\ \emph {et~al.}(2012)\citenamefont
  {Corberi}, \citenamefont {Lippiello}, \citenamefont {Mukherjee},
  \citenamefont {Puri},\ and\ \citenamefont {Zannetti}}]{corberi2012crossover}%
  \BibitemOpen
  \bibfield  {author} {\bibinfo {author} {\bibfnamefont {F.}~\bibnamefont
  {Corberi}}, \bibinfo {author} {\bibfnamefont {E.}~\bibnamefont {Lippiello}},
  \bibinfo {author} {\bibfnamefont {A.}~\bibnamefont {Mukherjee}}, \bibinfo
  {author} {\bibfnamefont {S.}~\bibnamefont {Puri}}, \ and\ \bibinfo {author}
  {\bibfnamefont {M.}~\bibnamefont {Zannetti}},\ }\href@noop {} {\bibfield
  {journal} {\bibinfo  {journal} {Physical Review E}\ }\textbf {\bibinfo
  {volume} {85}},\ \bibinfo {pages} {021141} (\bibinfo {year}
  {2012})}\BibitemShut {NoStop}%
\bibitem [{\citenamefont {Puri}(2004)}]{puri2004ordering}%
  \BibitemOpen
  \bibfield  {author} {\bibinfo {author} {\bibfnamefont {S.}~\bibnamefont
  {Puri}},\ }\href@noop {} {\bibfield  {journal} {\bibinfo  {journal} {Phase
  Transitions}\ }\textbf {\bibinfo {volume} {77}},\ \bibinfo {pages} {469}
  (\bibinfo {year} {2004})}\BibitemShut {NoStop}%
\bibitem [{\citenamefont {Puri}\ \emph {et~al.}(1991)\citenamefont {Puri},
  \citenamefont {Chowdhury},\ and\ \citenamefont {Parekh}}]{puri1991non}%
  \BibitemOpen
  \bibfield  {author} {\bibinfo {author} {\bibfnamefont {S.}~\bibnamefont
  {Puri}}, \bibinfo {author} {\bibfnamefont {D.}~\bibnamefont {Chowdhury}}, \
  and\ \bibinfo {author} {\bibfnamefont {N.}~\bibnamefont {Parekh}},\
  }\href@noop {} {\bibfield  {journal} {\bibinfo  {journal} {Journal of Physics
  A: Mathematical and General}\ }\textbf {\bibinfo {volume} {24}},\ \bibinfo
  {pages} {L1087} (\bibinfo {year} {1991})}\BibitemShut {NoStop}%
\bibitem [{\citenamefont {Puri}\ and\ \citenamefont
  {Parekh}(1992)}]{puri1992non}%
  \BibitemOpen
  \bibfield  {author} {\bibinfo {author} {\bibfnamefont {S.}~\bibnamefont
  {Puri}}\ and\ \bibinfo {author} {\bibfnamefont {N.}~\bibnamefont {Parekh}},\
  }\href@noop {} {\bibfield  {journal} {\bibinfo  {journal} {Journal of Physics
  A: Mathematical and General}\ }\textbf {\bibinfo {volume} {25}},\ \bibinfo
  {pages} {4127} (\bibinfo {year} {1992})}\BibitemShut {NoStop}%
\bibitem [{\citenamefont {Puri}\ and\ \citenamefont
  {Parekh}(1993)}]{puri1993non}%
  \BibitemOpen
  \bibfield  {author} {\bibinfo {author} {\bibfnamefont {S.}~\bibnamefont
  {Puri}}\ and\ \bibinfo {author} {\bibfnamefont {N.}~\bibnamefont {Parekh}},\
  }\href@noop {} {\bibfield  {journal} {\bibinfo  {journal} {Journal of Physics
  A: Mathematical and General}\ }\textbf {\bibinfo {volume} {26}},\ \bibinfo
  {pages} {2777} (\bibinfo {year} {1993})}\BibitemShut {NoStop}%
\bibitem [{\citenamefont {Oh}\ and\ \citenamefont {Choi}(1986)}]{oh1986monte}%
  \BibitemOpen
  \bibfield  {author} {\bibinfo {author} {\bibfnamefont {J.~H.}\ \bibnamefont
  {Oh}}\ and\ \bibinfo {author} {\bibfnamefont {D.-I.}\ \bibnamefont {Choi}},\
  }\href@noop {} {\bibfield  {journal} {\bibinfo  {journal} {Physical Review
  B}\ }\textbf {\bibinfo {volume} {33}},\ \bibinfo {pages} {3448} (\bibinfo
  {year} {1986})}\BibitemShut {NoStop}%
\bibitem [{\citenamefont {Oguz}\ \emph {et~al.}(1990)\citenamefont {Oguz},
  \citenamefont {Chakrabarti}, \citenamefont {Toral},\ and\ \citenamefont
  {Gunton}}]{oguz1990domain}%
  \BibitemOpen
  \bibfield  {author} {\bibinfo {author} {\bibfnamefont {E.}~\bibnamefont
  {Oguz}}, \bibinfo {author} {\bibfnamefont {A.}~\bibnamefont {Chakrabarti}},
  \bibinfo {author} {\bibfnamefont {R.}~\bibnamefont {Toral}}, \ and\ \bibinfo
  {author} {\bibfnamefont {J.~D.}\ \bibnamefont {Gunton}},\ }\href@noop {}
  {\bibfield  {journal} {\bibinfo  {journal} {Physical Review B}\ }\textbf
  {\bibinfo {volume} {42}},\ \bibinfo {pages} {704} (\bibinfo {year}
  {1990})}\BibitemShut {NoStop}%
\bibitem [{\citenamefont {Oguz}(1994)}]{oguz1994domain}%
  \BibitemOpen
  \bibfield  {author} {\bibinfo {author} {\bibfnamefont {E.}~\bibnamefont
  {Oguz}},\ }\href@noop {} {\bibfield  {journal} {\bibinfo  {journal} {Journal
  of Physics A: Mathematical and General}\ }\textbf {\bibinfo {volume} {27}},\
  \bibinfo {pages} {2985} (\bibinfo {year} {1994})}\BibitemShut {NoStop}%
\bibitem [{\citenamefont {Rao}\ and\ \citenamefont
  {Chakrabarti}(1993)}]{rao1993kinetics}%
  \BibitemOpen
  \bibfield  {author} {\bibinfo {author} {\bibfnamefont {M.}~\bibnamefont
  {Rao}}\ and\ \bibinfo {author} {\bibfnamefont {A.}~\bibnamefont
  {Chakrabarti}},\ }\href@noop {} {\bibfield  {journal} {\bibinfo  {journal}
  {Physical review letters}\ }\textbf {\bibinfo {volume} {71}},\ \bibinfo
  {pages} {3501} (\bibinfo {year} {1993})}\BibitemShut {NoStop}%
\bibitem [{\citenamefont {Rao}\ and\ \citenamefont
  {Chakrabarti}(1995)}]{rao1995slow}%
  \BibitemOpen
  \bibfield  {author} {\bibinfo {author} {\bibfnamefont {M.}~\bibnamefont
  {Rao}}\ and\ \bibinfo {author} {\bibfnamefont {A.}~\bibnamefont
  {Chakrabarti}},\ }\href@noop {} {\bibfield  {journal} {\bibinfo  {journal}
  {Physical Review E}\ }\textbf {\bibinfo {volume} {52}},\ \bibinfo {pages}
  {R13} (\bibinfo {year} {1995})}\BibitemShut {NoStop}%
\bibitem [{\citenamefont {Aron}\ \emph {et~al.}(2008)\citenamefont {Aron},
  \citenamefont {Chamon}, \citenamefont {Cugliandolo},\ and\ \citenamefont
  {Picco}}]{aron2008scaling}%
  \BibitemOpen
  \bibfield  {author} {\bibinfo {author} {\bibfnamefont {C.}~\bibnamefont
  {Aron}}, \bibinfo {author} {\bibfnamefont {C.}~\bibnamefont {Chamon}},
  \bibinfo {author} {\bibfnamefont {L.~F.}\ \bibnamefont {Cugliandolo}}, \ and\
  \bibinfo {author} {\bibfnamefont {M.}~\bibnamefont {Picco}},\ }\href@noop {}
  {\bibfield  {journal} {\bibinfo  {journal} {Journal of Statistical Mechanics:
  Theory and Experiment}\ }\textbf {\bibinfo {volume} {2008}},\ \bibinfo
  {pages} {P05016} (\bibinfo {year} {2008})}\BibitemShut {NoStop}%
\bibitem [{\citenamefont {Cugliandolo}(2010)}]{cugliandolo2010topics}%
  \BibitemOpen
  \bibfield  {author} {\bibinfo {author} {\bibfnamefont {L.~F.}\ \bibnamefont
  {Cugliandolo}},\ }\href@noop {} {\bibfield  {journal} {\bibinfo  {journal}
  {Physica A: Statistical Mechanics and its Applications}\ }\textbf {\bibinfo
  {volume} {389}},\ \bibinfo {pages} {4360} (\bibinfo {year}
  {2010})}\BibitemShut {NoStop}%
\bibitem [{\citenamefont {Corberi}(2015)}]{corberi2015coarsening}%
  \BibitemOpen
  \bibfield  {author} {\bibinfo {author} {\bibfnamefont {F.}~\bibnamefont
  {Corberi}},\ }\href@noop {} {\bibfield  {journal} {\bibinfo  {journal}
  {Comptes Rendus Physique}\ }\textbf {\bibinfo {volume} {16}},\ \bibinfo
  {pages} {332} (\bibinfo {year} {2015})}\BibitemShut {NoStop}%
\bibitem [{\citenamefont {Huse}\ and\ \citenamefont
  {Henley}(1985)}]{huse1985pinning}%
  \BibitemOpen
  \bibfield  {author} {\bibinfo {author} {\bibfnamefont {D.~A.}\ \bibnamefont
  {Huse}}\ and\ \bibinfo {author} {\bibfnamefont {C.~L.}\ \bibnamefont
  {Henley}},\ }\href@noop {} {\bibfield  {journal} {\bibinfo  {journal}
  {Physical review letters}\ }\textbf {\bibinfo {volume} {54}},\ \bibinfo
  {pages} {2708} (\bibinfo {year} {1985})}\BibitemShut {NoStop}%
\bibitem [{\citenamefont {Mandal}\ and\ \citenamefont
  {Sinha}(2014)}]{mandal2014characterization}%
  \BibitemOpen
  \bibfield  {author} {\bibinfo {author} {\bibfnamefont {P.~K.}\ \bibnamefont
  {Mandal}}\ and\ \bibinfo {author} {\bibfnamefont {S.}~\bibnamefont {Sinha}},\
  }\href@noop {} {\bibfield  {journal} {\bibinfo  {journal} {Physical Review
  E}\ }\textbf {\bibinfo {volume} {89}},\ \bibinfo {pages} {042144} (\bibinfo
  {year} {2014})}\BibitemShut {NoStop}%
\bibitem [{\citenamefont {Park}\ and\ \citenamefont
  {Pleimling}(2010)}]{park2010aging}%
  \BibitemOpen
  \bibfield  {author} {\bibinfo {author} {\bibfnamefont {H.}~\bibnamefont
  {Park}}\ and\ \bibinfo {author} {\bibfnamefont {M.}~\bibnamefont
  {Pleimling}},\ }\href@noop {} {\bibfield  {journal} {\bibinfo  {journal}
  {Physical Review B}\ }\textbf {\bibinfo {volume} {82}},\ \bibinfo {pages}
  {144406} (\bibinfo {year} {2010})}\BibitemShut {NoStop}%
\bibitem [{\citenamefont {Park}\ and\ \citenamefont
  {Pleimling}(2012)}]{park2012domain}%
  \BibitemOpen
  \bibfield  {author} {\bibinfo {author} {\bibfnamefont {H.}~\bibnamefont
  {Park}}\ and\ \bibinfo {author} {\bibfnamefont {M.}~\bibnamefont
  {Pleimling}},\ }\href@noop {} {\bibfield  {journal} {\bibinfo  {journal} {The
  European Physical Journal B}\ }\textbf {\bibinfo {volume} {85}},\ \bibinfo
  {pages} {300} (\bibinfo {year} {2012})}\BibitemShut {NoStop}%
\bibitem [{\citenamefont {Kumar}\ \emph
  {et~al.}(2017{\natexlab{a}})\citenamefont {Kumar}, \citenamefont
  {Chatterjee}, \citenamefont {Paul},\ and\ \citenamefont
  {Puri}}]{kumar2017ordering}%
  \BibitemOpen
  \bibfield  {author} {\bibinfo {author} {\bibfnamefont {M.}~\bibnamefont
  {Kumar}}, \bibinfo {author} {\bibfnamefont {S.}~\bibnamefont {Chatterjee}},
  \bibinfo {author} {\bibfnamefont {R.}~\bibnamefont {Paul}}, \ and\ \bibinfo
  {author} {\bibfnamefont {S.}~\bibnamefont {Puri}},\ }\href@noop {} {\bibfield
   {journal} {\bibinfo  {journal} {Physical Review E}\ }\textbf {\bibinfo
  {volume} {96}},\ \bibinfo {pages} {042127} (\bibinfo {year}
  {2017}{\natexlab{a}})}\BibitemShut {NoStop}%
\bibitem [{\citenamefont {Kumar}\ \emph
  {et~al.}(2017{\natexlab{b}})\citenamefont {Kumar}, \citenamefont {Banerjee},\
  and\ \citenamefont {Puri}}]{kumar2017random}%
  \BibitemOpen
  \bibfield  {author} {\bibinfo {author} {\bibfnamefont {M.}~\bibnamefont
  {Kumar}}, \bibinfo {author} {\bibfnamefont {V.}~\bibnamefont {Banerjee}}, \
  and\ \bibinfo {author} {\bibfnamefont {S.}~\bibnamefont {Puri}},\ }\href@noop
  {} {\bibfield  {journal} {\bibinfo  {journal} {EPL (Europhysics Letters)}\
  }\textbf {\bibinfo {volume} {117}},\ \bibinfo {pages} {10012} (\bibinfo
  {year} {2017}{\natexlab{b}})}\BibitemShut {NoStop}%
\bibitem [{\citenamefont {Corberi}\ \emph {et~al.}(2017)\citenamefont
  {Corberi}, \citenamefont {Kumar}, \citenamefont {Puri},\ and\ \citenamefont
  {Lippiello}}]{corberi2017equilibrium}%
  \BibitemOpen
  \bibfield  {author} {\bibinfo {author} {\bibfnamefont {F.}~\bibnamefont
  {Corberi}}, \bibinfo {author} {\bibfnamefont {M.}~\bibnamefont {Kumar}},
  \bibinfo {author} {\bibfnamefont {S.}~\bibnamefont {Puri}}, \ and\ \bibinfo
  {author} {\bibfnamefont {E.}~\bibnamefont {Lippiello}},\ }\href@noop {}
  {\bibfield  {journal} {\bibinfo  {journal} {Physical Review E}\ }\textbf
  {\bibinfo {volume} {95}},\ \bibinfo {pages} {062136} (\bibinfo {year}
  {2017})}\BibitemShut {NoStop}%
\bibitem [{\citenamefont {Biswal}\ \emph {et~al.}(1996)\citenamefont {Biswal},
  \citenamefont {Puri},\ and\ \citenamefont {Chowdhury}}]{biswal1996domain}%
  \BibitemOpen
  \bibfield  {author} {\bibinfo {author} {\bibfnamefont {B.}~\bibnamefont
  {Biswal}}, \bibinfo {author} {\bibfnamefont {S.}~\bibnamefont {Puri}}, \ and\
  \bibinfo {author} {\bibfnamefont {D.}~\bibnamefont {Chowdhury}},\ }\href@noop
  {} {\bibfield  {journal} {\bibinfo  {journal} {Physica A: Statistical
  Mechanics and its Applications}\ }\textbf {\bibinfo {volume} {229}},\
  \bibinfo {pages} {72} (\bibinfo {year} {1996})}\BibitemShut {NoStop}%
\bibitem [{\citenamefont {Bray}\ and\ \citenamefont
  {Humayun}(1991)}]{bray1991universality}%
  \BibitemOpen
  \bibfield  {author} {\bibinfo {author} {\bibfnamefont {A.}~\bibnamefont
  {Bray}}\ and\ \bibinfo {author} {\bibfnamefont {K.}~\bibnamefont {Humayun}},\
  }\href@noop {} {\bibfield  {journal} {\bibinfo  {journal} {Journal of Physics
  A: Mathematical and General}\ }\textbf {\bibinfo {volume} {24}},\ \bibinfo
  {pages} {L1185} (\bibinfo {year} {1991})}\BibitemShut {NoStop}%
\bibitem [{\citenamefont {Corberi}\ \emph
  {et~al.}(2015{\natexlab{c}})\citenamefont {Corberi}, \citenamefont
  {Cugliandolo}, \citenamefont {Insalata},\ and\ \citenamefont
  {Picco}}]{corberi2017}%
  \BibitemOpen
  \bibfield  {author} {\bibinfo {author} {\bibfnamefont {F.}~\bibnamefont
  {Corberi}}, \bibinfo {author} {\bibfnamefont {L.~F.}\ \bibnamefont
  {Cugliandolo}}, \bibinfo {author} {\bibfnamefont {F.}~\bibnamefont
  {Insalata}}, \ and\ \bibinfo {author} {\bibfnamefont {M.}~\bibnamefont
  {Picco}},\ }\href@noop {} {\bibfield  {journal} {\bibinfo  {journal}
  {Physical Review E}\ }\textbf {\bibinfo {volume} {95}},\ \bibinfo {pages}
  {022101} (\bibinfo {year} {2015}{\natexlab{c}})}\BibitemShut {NoStop}%
\bibitem [{\citenamefont {Grant}\ and\ \citenamefont
  {Gunton}(1984)}]{grant1984domain}%
  \BibitemOpen
  \bibfield  {author} {\bibinfo {author} {\bibfnamefont {M.}~\bibnamefont
  {Grant}}\ and\ \bibinfo {author} {\bibfnamefont {J.}~\bibnamefont {Gunton}},\
  }\href@noop {} {\bibfield  {journal} {\bibinfo  {journal} {Physical Review
  B}\ }\textbf {\bibinfo {volume} {29}},\ \bibinfo {pages} {1521} (\bibinfo
  {year} {1984})}\BibitemShut {NoStop}%
\bibitem [{\citenamefont {Anderson}(1987)}]{anderson1987growth}%
  \BibitemOpen
  \bibfield  {author} {\bibinfo {author} {\bibfnamefont {S.~R.}\ \bibnamefont
  {Anderson}},\ }\href@noop {} {\bibfield  {journal} {\bibinfo  {journal}
  {Physical Review B}\ }\textbf {\bibinfo {volume} {36}},\ \bibinfo {pages}
  {8435} (\bibinfo {year} {1987})}\BibitemShut {NoStop}%
\bibitem [{\citenamefont {Grest}\ and\ \citenamefont
  {Srolovitz}(1985)}]{grest1985impurity}%
  \BibitemOpen
  \bibfield  {author} {\bibinfo {author} {\bibfnamefont {G.~S.}\ \bibnamefont
  {Grest}}\ and\ \bibinfo {author} {\bibfnamefont {D.~J.}\ \bibnamefont
  {Srolovitz}},\ }\href@noop {} {\bibfield  {journal} {\bibinfo  {journal}
  {Physical review B}\ }\textbf {\bibinfo {volume} {32}},\ \bibinfo {pages}
  {3014} (\bibinfo {year} {1985})}\BibitemShut {NoStop}%
\bibitem [{\citenamefont {Castellano}\ \emph {et~al.}(1998)\citenamefont
  {Castellano}, \citenamefont {Corberi}, \citenamefont {Marconi},\ and\
  \citenamefont {Petri}}]{castellano1998coarsening}%
  \BibitemOpen
  \bibfield  {author} {\bibinfo {author} {\bibfnamefont {C.}~\bibnamefont
  {Castellano}}, \bibinfo {author} {\bibfnamefont {F.}~\bibnamefont {Corberi}},
  \bibinfo {author} {\bibfnamefont {U.~M.~B.}\ \bibnamefont {Marconi}}, \ and\
  \bibinfo {author} {\bibfnamefont {A.}~\bibnamefont {Petri}},\ }\href@noop {}
  {\bibfield  {journal} {\bibinfo  {journal} {Le Journal de Physique IV}\
  }\textbf {\bibinfo {volume} {8}},\ \bibinfo {pages} {Pr6} (\bibinfo {year}
  {1998})}\BibitemShut {NoStop}%
\bibitem [{\citenamefont {Corberi}\ \emph {et~al.}(2013)\citenamefont
  {Corberi}, \citenamefont {Lippiello}, \citenamefont {Mukherjee},
  \citenamefont {Puri},\ and\ \citenamefont {Zannetti}}]{corberi2013scaling}%
  \BibitemOpen
  \bibfield  {author} {\bibinfo {author} {\bibfnamefont {F.}~\bibnamefont
  {Corberi}}, \bibinfo {author} {\bibfnamefont {E.}~\bibnamefont {Lippiello}},
  \bibinfo {author} {\bibfnamefont {A.}~\bibnamefont {Mukherjee}}, \bibinfo
  {author} {\bibfnamefont {S.}~\bibnamefont {Puri}}, \ and\ \bibinfo {author}
  {\bibfnamefont {M.}~\bibnamefont {Zannetti}},\ }\href@noop {} {\bibfield
  {journal} {\bibinfo  {journal} {Physical Review E}\ }\textbf {\bibinfo
  {volume} {88}},\ \bibinfo {pages} {042129} (\bibinfo {year}
  {2013})}\BibitemShut {NoStop}%
\bibitem [{\citenamefont {Corberi}\ \emph
  {et~al.}(2019{\natexlab{a}})\citenamefont {Corberi}, \citenamefont
  {Cugliandolo}, \citenamefont {Insalata},\ and\ \citenamefont
  {Picco}}]{corberi2019}%
  \BibitemOpen
  \bibfield  {author} {\bibinfo {author} {\bibfnamefont {F.}~\bibnamefont
  {Corberi}}, \bibinfo {author} {\bibfnamefont {L.~F.}\ \bibnamefont
  {Cugliandolo}}, \bibinfo {author} {\bibfnamefont {F.}~\bibnamefont
  {Insalata}}, \ and\ \bibinfo {author} {\bibfnamefont {M.}~\bibnamefont
  {Picco}},\ }\href@noop {} {\bibfield  {journal} {\bibinfo  {journal} {Journal
  of Statistical Mechanics: Theory and Experiment}\ }\textbf {\bibinfo {volume}
  {2019}},\ \bibinfo {pages} {P043203} (\bibinfo {year}
  {2019}{\natexlab{a}})}\BibitemShut {NoStop}%
\bibitem [{\citenamefont {Iguain}\ \emph {et~al.}(2009)\citenamefont {Iguain},
  \citenamefont {Bustingorry}, \citenamefont {Kolton},\ and\ \citenamefont
  {Cugliandolo}}]{iguain2009growing}%
  \BibitemOpen
  \bibfield  {author} {\bibinfo {author} {\bibfnamefont {J.~L.}\ \bibnamefont
  {Iguain}}, \bibinfo {author} {\bibfnamefont {S.}~\bibnamefont {Bustingorry}},
  \bibinfo {author} {\bibfnamefont {A.~B.}\ \bibnamefont {Kolton}}, \ and\
  \bibinfo {author} {\bibfnamefont {L.~F.}\ \bibnamefont {Cugliandolo}},\
  }\href@noop {} {\bibfield  {journal} {\bibinfo  {journal} {Physical Review
  B}\ }\textbf {\bibinfo {volume} {80}},\ \bibinfo {pages} {094201} (\bibinfo
  {year} {2009})}\BibitemShut {NoStop}%
\bibitem [{\citenamefont {Ikeda}\ \emph {et~al.}(1990)\citenamefont {Ikeda},
  \citenamefont {Endoh},\ and\ \citenamefont {Itoh}}]{ikeda1990ordering}%
  \BibitemOpen
  \bibfield  {author} {\bibinfo {author} {\bibfnamefont {H.}~\bibnamefont
  {Ikeda}}, \bibinfo {author} {\bibfnamefont {Y.}~\bibnamefont {Endoh}}, \ and\
  \bibinfo {author} {\bibfnamefont {S.}~\bibnamefont {Itoh}},\ }\href@noop {}
  {\bibfield  {journal} {\bibinfo  {journal} {Physical review letters}\
  }\textbf {\bibinfo {volume} {64}},\ \bibinfo {pages} {1266} (\bibinfo {year}
  {1990})}\BibitemShut {NoStop}%
\bibitem [{\citenamefont {Kolton}\ \emph {et~al.}(2005)\citenamefont {Kolton},
  \citenamefont {Rosso},\ and\ \citenamefont
  {Giamarchi}}]{kolton2005nonequilibrium}%
  \BibitemOpen
  \bibfield  {author} {\bibinfo {author} {\bibfnamefont {A.~B.}\ \bibnamefont
  {Kolton}}, \bibinfo {author} {\bibfnamefont {A.}~\bibnamefont {Rosso}}, \
  and\ \bibinfo {author} {\bibfnamefont {T.}~\bibnamefont {Giamarchi}},\
  }\href@noop {} {\bibfield  {journal} {\bibinfo  {journal} {Physical review
  letters}\ }\textbf {\bibinfo {volume} {95}},\ \bibinfo {pages} {180604}
  (\bibinfo {year} {2005})}\BibitemShut {NoStop}%
\bibitem [{\citenamefont {Noh}\ \emph {et~al.}(2009)\citenamefont {Noh},
  \citenamefont {Park} \emph {et~al.}}]{noh2009relaxation}%
  \BibitemOpen
  \bibfield  {author} {\bibinfo {author} {\bibfnamefont {J.~D.}\ \bibnamefont
  {Noh}}, \bibinfo {author} {\bibfnamefont {H.}~\bibnamefont {Park}},  \emph
  {et~al.},\ }\href@noop {} {\bibfield  {journal} {\bibinfo  {journal}
  {Physical Review E}\ }\textbf {\bibinfo {volume} {80}},\ \bibinfo {pages}
  {040102} (\bibinfo {year} {2009})}\BibitemShut {NoStop}%
\bibitem [{\citenamefont {Monthus}\ and\ \citenamefont
  {Garel}(2009)}]{monthus2009eigenvalue}%
  \BibitemOpen
  \bibfield  {author} {\bibinfo {author} {\bibfnamefont {C.}~\bibnamefont
  {Monthus}}\ and\ \bibinfo {author} {\bibfnamefont {T.}~\bibnamefont
  {Garel}},\ }\href@noop {} {\bibfield  {journal} {\bibinfo  {journal} {Journal
  of Statistical Mechanics: Theory and Experiment}\ }\textbf {\bibinfo {volume}
  {2009}},\ \bibinfo {pages} {P12017} (\bibinfo {year} {2009})}\BibitemShut
  {NoStop}%
\bibitem [{\citenamefont {Shenoy}\ \emph {et~al.}(1999)\citenamefont {Shenoy},
  \citenamefont {Selinger}, \citenamefont {Gr{\"u}neberg}, \citenamefont
  {Naciri},\ and\ \citenamefont {Shashidhar}}]{shenoy1999coarsening}%
  \BibitemOpen
  \bibfield  {author} {\bibinfo {author} {\bibfnamefont {D.}~\bibnamefont
  {Shenoy}}, \bibinfo {author} {\bibfnamefont {J.}~\bibnamefont {Selinger}},
  \bibinfo {author} {\bibfnamefont {K.}~\bibnamefont {Gr{\"u}neberg}}, \bibinfo
  {author} {\bibfnamefont {J.}~\bibnamefont {Naciri}}, \ and\ \bibinfo {author}
  {\bibfnamefont {R.}~\bibnamefont {Shashidhar}},\ }\href@noop {} {\bibfield
  {journal} {\bibinfo  {journal} {Physical review letters}\ }\textbf {\bibinfo
  {volume} {82}},\ \bibinfo {pages} {1716} (\bibinfo {year}
  {1999})}\BibitemShut {NoStop}%
\bibitem [{\citenamefont {Schins}\ \emph {et~al.}(1993)\citenamefont {Schins},
  \citenamefont {Arts},\ and\ \citenamefont {De~Wijn}}]{schins1993domain}%
  \BibitemOpen
  \bibfield  {author} {\bibinfo {author} {\bibfnamefont {A.}~\bibnamefont
  {Schins}}, \bibinfo {author} {\bibfnamefont {A.}~\bibnamefont {Arts}}, \ and\
  \bibinfo {author} {\bibfnamefont {H.}~\bibnamefont {De~Wijn}},\ }\href@noop
  {} {\bibfield  {journal} {\bibinfo  {journal} {Physical review letters}\
  }\textbf {\bibinfo {volume} {70}},\ \bibinfo {pages} {2340} (\bibinfo {year}
  {1993})}\BibitemShut {NoStop}%
\bibitem [{\citenamefont {Likodimos}\ \emph {et~al.}(2001)\citenamefont
  {Likodimos}, \citenamefont {Labardi}, \citenamefont {Orlik}, \citenamefont
  {Pardi}, \citenamefont {Allegrini}, \citenamefont {Emonin},\ and\
  \citenamefont {Marti}}]{likodimos2001thermally}%
  \BibitemOpen
  \bibfield  {author} {\bibinfo {author} {\bibfnamefont {V.}~\bibnamefont
  {Likodimos}}, \bibinfo {author} {\bibfnamefont {M.}~\bibnamefont {Labardi}},
  \bibinfo {author} {\bibfnamefont {X.}~\bibnamefont {Orlik}}, \bibinfo
  {author} {\bibfnamefont {L.}~\bibnamefont {Pardi}}, \bibinfo {author}
  {\bibfnamefont {M.}~\bibnamefont {Allegrini}}, \bibinfo {author}
  {\bibfnamefont {S.}~\bibnamefont {Emonin}}, \ and\ \bibinfo {author}
  {\bibfnamefont {O.}~\bibnamefont {Marti}},\ }\href@noop {} {\bibfield
  {journal} {\bibinfo  {journal} {Physical Review B}\ }\textbf {\bibinfo
  {volume} {63}},\ \bibinfo {pages} {064104} (\bibinfo {year}
  {2001})}\BibitemShut {NoStop}%
\bibitem [{\citenamefont {Likodimos}\ \emph {et~al.}(2000)\citenamefont
  {Likodimos}, \citenamefont {Labardi},\ and\ \citenamefont
  {Allegrini}}]{likodimos2000kinetics}%
  \BibitemOpen
  \bibfield  {author} {\bibinfo {author} {\bibfnamefont {V.}~\bibnamefont
  {Likodimos}}, \bibinfo {author} {\bibfnamefont {M.}~\bibnamefont {Labardi}},
  \ and\ \bibinfo {author} {\bibfnamefont {M.}~\bibnamefont {Allegrini}},\
  }\href@noop {} {\bibfield  {journal} {\bibinfo  {journal} {Physical Review
  B}\ }\textbf {\bibinfo {volume} {61}},\ \bibinfo {pages} {14440} (\bibinfo
  {year} {2000})}\BibitemShut {NoStop}%
\bibitem [{\citenamefont {Vincent}\ \emph {et~al.}(1997)\citenamefont
  {Vincent}, \citenamefont {Hammann}, \citenamefont {Ocio}, \citenamefont
  {Bouchaud},\ and\ \citenamefont {Cugliandolo}}]{vincent1997slow}%
  \BibitemOpen
  \bibfield  {author} {\bibinfo {author} {\bibfnamefont {E.}~\bibnamefont
  {Vincent}}, \bibinfo {author} {\bibfnamefont {J.}~\bibnamefont {Hammann}},
  \bibinfo {author} {\bibfnamefont {M.}~\bibnamefont {Ocio}}, \bibinfo {author}
  {\bibfnamefont {J.-P.}\ \bibnamefont {Bouchaud}}, \ and\ \bibinfo {author}
  {\bibfnamefont {L.~F.}\ \bibnamefont {Cugliandolo}},\ }in\ \href@noop {}
  {\emph {\bibinfo {booktitle} {Complex Behaviour of Glassy Systems}}}\
  (\bibinfo  {publisher} {Springer},\ \bibinfo {year} {1997})\ pp.\ \bibinfo
  {pages} {184--219}\BibitemShut {NoStop}%
\bibitem [{\citenamefont {Corberi}\ \emph
  {et~al.}(2019{\natexlab{b}})\citenamefont {Corberi}, \citenamefont {Kumar},
  \citenamefont {Lippiello},\ and\ \citenamefont {Puri}}]{corberi2019effects}%
  \BibitemOpen
  \bibfield  {author} {\bibinfo {author} {\bibfnamefont {F.}~\bibnamefont
  {Corberi}}, \bibinfo {author} {\bibfnamefont {M.}~\bibnamefont {Kumar}},
  \bibinfo {author} {\bibfnamefont {E.}~\bibnamefont {Lippiello}}, \ and\
  \bibinfo {author} {\bibfnamefont {S.}~\bibnamefont {Puri}},\ }\href@noop {}
  {\bibfield  {journal} {\bibinfo  {journal} {Physical Review E}\ }\textbf
  {\bibinfo {volume} {99}},\ \bibinfo {pages} {012131} (\bibinfo {year}
  {2019}{\natexlab{b}})}\BibitemShut {NoStop}%
\bibitem [{\citenamefont {Amoruso}\ \emph {et~al.}(2003)\citenamefont
  {Amoruso}, \citenamefont {Marinari}, \citenamefont {Martin},\ and\
  \citenamefont {Pagnani}}]{amoruso2003scalings}%
  \BibitemOpen
  \bibfield  {author} {\bibinfo {author} {\bibfnamefont {C.}~\bibnamefont
  {Amoruso}}, \bibinfo {author} {\bibfnamefont {E.}~\bibnamefont {Marinari}},
  \bibinfo {author} {\bibfnamefont {O.~C.}\ \bibnamefont {Martin}}, \ and\
  \bibinfo {author} {\bibfnamefont {A.}~\bibnamefont {Pagnani}},\ }\href@noop
  {} {\bibfield  {journal} {\bibinfo  {journal} {Physical review letters}\
  }\textbf {\bibinfo {volume} {91}},\ \bibinfo {pages} {087201} (\bibinfo
  {year} {2003})}\BibitemShut {NoStop}%
\bibitem [{\citenamefont {J{\"o}rg}\ \emph {et~al.}(2006)\citenamefont
  {J{\"o}rg}, \citenamefont {Lukic}, \citenamefont {Marinari},\ and\
  \citenamefont {Martin}}]{jorg2006strong}%
  \BibitemOpen
  \bibfield  {author} {\bibinfo {author} {\bibfnamefont {T.}~\bibnamefont
  {J{\"o}rg}}, \bibinfo {author} {\bibfnamefont {J.}~\bibnamefont {Lukic}},
  \bibinfo {author} {\bibfnamefont {E.}~\bibnamefont {Marinari}}, \ and\
  \bibinfo {author} {\bibfnamefont {O.}~\bibnamefont {Martin}},\ }\href@noop {}
  {\bibfield  {journal} {\bibinfo  {journal} {Physical review letters}\
  }\textbf {\bibinfo {volume} {96}},\ \bibinfo {pages} {237205} (\bibinfo
  {year} {2006})}\BibitemShut {NoStop}%
\bibitem [{\citenamefont {M{\'e}zard}\ \emph {et~al.}(1987)\citenamefont
  {M{\'e}zard}, \citenamefont {Parisi},\ and\ \citenamefont
  {Virasoro}}]{mezard1987spin}%
  \BibitemOpen
  \bibfield  {author} {\bibinfo {author} {\bibfnamefont {M.}~\bibnamefont
  {M{\'e}zard}}, \bibinfo {author} {\bibfnamefont {G.}~\bibnamefont {Parisi}},
  \ and\ \bibinfo {author} {\bibfnamefont {M.}~\bibnamefont {Virasoro}},\
  }\href@noop {} {\emph {\bibinfo {title} {Spin glass theory and beyond: An
  Introduction to the Replica Method and Its Applications}}},\ Vol.~\bibinfo
  {volume} {9}\ (\bibinfo  {publisher} {World Scientific Publishing Company},\
  \bibinfo {year} {1987})\BibitemShut {NoStop}%
\bibitem [{\citenamefont {Cugliandolo}(2002)}]{cugliandolo2002dynamics}%
  \BibitemOpen
  \bibfield  {author} {\bibinfo {author} {\bibfnamefont {L.~F.}\ \bibnamefont
  {Cugliandolo}},\ }\href@noop {} {\bibfield  {journal} {\bibinfo  {journal}
  {arXiv preprint cond-mat/0210312}\ } (\bibinfo {year} {2002})}\BibitemShut
  {NoStop}%
\bibitem [{\citenamefont {Calabrese}\ and\ \citenamefont
  {Gambassi}(2005)}]{calabrese2005ageing}%
  \BibitemOpen
  \bibfield  {author} {\bibinfo {author} {\bibfnamefont {P.}~\bibnamefont
  {Calabrese}}\ and\ \bibinfo {author} {\bibfnamefont {A.}~\bibnamefont
  {Gambassi}},\ }\href@noop {} {\bibfield  {journal} {\bibinfo  {journal}
  {Journal of Physics A: Mathematical and General}\ }\textbf {\bibinfo {volume}
  {38}},\ \bibinfo {pages} {R133} (\bibinfo {year} {2005})}\BibitemShut
  {NoStop}%
\bibitem [{\citenamefont {Bouchaud}\ \emph {et~al.}(1998)\citenamefont
  {Bouchaud}, \citenamefont {Cugliandolo}, \citenamefont {Kurchan},\ and\
  \citenamefont {Mezard}}]{bouchaud1998out}%
  \BibitemOpen
  \bibfield  {author} {\bibinfo {author} {\bibfnamefont {J.-P.}\ \bibnamefont
  {Bouchaud}}, \bibinfo {author} {\bibfnamefont {L.~F.}\ \bibnamefont
  {Cugliandolo}}, \bibinfo {author} {\bibfnamefont {J.}~\bibnamefont
  {Kurchan}}, \ and\ \bibinfo {author} {\bibfnamefont {M.}~\bibnamefont
  {Mezard}},\ }\href@noop {} {\bibfield  {journal} {\bibinfo  {journal} {Spin
  glasses and random fields}\ ,\ \bibinfo {pages} {161}} (\bibinfo {year}
  {1998})}\BibitemShut {NoStop}%
\bibitem [{\citenamefont {Lippiello}\ \emph {et~al.}(2005)\citenamefont
  {Lippiello}, \citenamefont {Corberi},\ and\ \citenamefont
  {Zannetti}}]{lippiello2005off}%
  \BibitemOpen
  \bibfield  {author} {\bibinfo {author} {\bibfnamefont {E.}~\bibnamefont
  {Lippiello}}, \bibinfo {author} {\bibfnamefont {F.}~\bibnamefont {Corberi}},
  \ and\ \bibinfo {author} {\bibfnamefont {M.}~\bibnamefont {Zannetti}},\
  }\href@noop {} {\bibfield  {journal} {\bibinfo  {journal} {Physical Review
  E}\ }\textbf {\bibinfo {volume} {71}},\ \bibinfo {pages} {036104} (\bibinfo
  {year} {2005})}\BibitemShut {NoStop}%
\bibitem [{\citenamefont {Parisi}\ \emph {et~al.}(1999)\citenamefont {Parisi},
  \citenamefont {Ricci-Tersenghi},\ and\ \citenamefont
  {Ruiz-Lorenzo}}]{parisi1999generalized}%
  \BibitemOpen
  \bibfield  {author} {\bibinfo {author} {\bibfnamefont {G.}~\bibnamefont
  {Parisi}}, \bibinfo {author} {\bibfnamefont {F.}~\bibnamefont
  {Ricci-Tersenghi}}, \ and\ \bibinfo {author} {\bibfnamefont {J.~J.}\
  \bibnamefont {Ruiz-Lorenzo}},\ }\href@noop {} {\bibfield  {journal} {\bibinfo
   {journal} {The European Physical Journal B-Condensed Matter and Complex
  Systems}\ }\textbf {\bibinfo {volume} {11}},\ \bibinfo {pages} {317}
  (\bibinfo {year} {1999})}\BibitemShut {NoStop}%
\bibitem [{\citenamefont {Franz}\ \emph {et~al.}(1999)\citenamefont {Franz},
  \citenamefont {Mezard}, \citenamefont {Parisi},\ and\ \citenamefont
  {Peliti}}]{franz1999response}%
  \BibitemOpen
  \bibfield  {author} {\bibinfo {author} {\bibfnamefont {S.}~\bibnamefont
  {Franz}}, \bibinfo {author} {\bibfnamefont {M.}~\bibnamefont {Mezard}},
  \bibinfo {author} {\bibfnamefont {G.}~\bibnamefont {Parisi}}, \ and\ \bibinfo
  {author} {\bibfnamefont {L.}~\bibnamefont {Peliti}},\ }\href@noop {}
  {\bibfield  {journal} {\bibinfo  {journal} {Journal of statistical physics}\
  }\textbf {\bibinfo {volume} {97}},\ \bibinfo {pages} {459} (\bibinfo {year}
  {1999})}\BibitemShut {NoStop}%
\bibitem [{\citenamefont {Corberi}\ \emph
  {et~al.}(2004{\natexlab{a}})\citenamefont {Corberi}, \citenamefont
  {Lippiello},\ and\ \citenamefont {Zannetti}}]{corberi2004effective}%
  \BibitemOpen
  \bibfield  {author} {\bibinfo {author} {\bibfnamefont {F.}~\bibnamefont
  {Corberi}}, \bibinfo {author} {\bibfnamefont {E.}~\bibnamefont {Lippiello}},
  \ and\ \bibinfo {author} {\bibfnamefont {M.}~\bibnamefont {Zannetti}},\
  }\href@noop {} {\bibfield  {journal} {\bibinfo  {journal} {Journal of
  Statistical Mechanics: Theory and Experiment}\ }\textbf {\bibinfo {volume}
  {2004}},\ \bibinfo {pages} {P12007} (\bibinfo {year}
  {2004}{\natexlab{a}})}\BibitemShut {NoStop}%
\bibitem [{\citenamefont {Corberi}\ \emph
  {et~al.}(2011{\natexlab{b}})\citenamefont {Corberi}, \citenamefont
  {Cugliandolo},\ and\ \citenamefont {Yoshino}}]{corberi2011growing}%
  \BibitemOpen
  \bibfield  {author} {\bibinfo {author} {\bibfnamefont {F.}~\bibnamefont
  {Corberi}}, \bibinfo {author} {\bibfnamefont {L.~F.}\ \bibnamefont
  {Cugliandolo}}, \ and\ \bibinfo {author} {\bibfnamefont {H.}~\bibnamefont
  {Yoshino}},\ }\href@noop {} {\bibfield  {journal} {\bibinfo  {journal}
  {Dynamical heterogeneities in glasses, colloids, and granular media}\
  }\textbf {\bibinfo {volume} {150}},\ \bibinfo {pages} {370} (\bibinfo {year}
  {2011}{\natexlab{b}})}\BibitemShut {NoStop}%
\bibitem [{\citenamefont {Barrat}(1998)}]{barrat1998monte}%
  \BibitemOpen
  \bibfield  {author} {\bibinfo {author} {\bibfnamefont {A.}~\bibnamefont
  {Barrat}},\ }\href@noop {} {\bibfield  {journal} {\bibinfo  {journal}
  {Physical Review E}\ }\textbf {\bibinfo {volume} {57}},\ \bibinfo {pages}
  {3629} (\bibinfo {year} {1998})}\BibitemShut {NoStop}%
\bibitem [{\citenamefont {Franz}\ \emph {et~al.}(1998)\citenamefont {Franz},
  \citenamefont {M{\'e}zard}, \citenamefont {Parisi},\ and\ \citenamefont
  {Peliti}}]{franz1998measuring}%
  \BibitemOpen
  \bibfield  {author} {\bibinfo {author} {\bibfnamefont {S.}~\bibnamefont
  {Franz}}, \bibinfo {author} {\bibfnamefont {M.}~\bibnamefont {M{\'e}zard}},
  \bibinfo {author} {\bibfnamefont {G.}~\bibnamefont {Parisi}}, \ and\ \bibinfo
  {author} {\bibfnamefont {L.}~\bibnamefont {Peliti}},\ }\href@noop {}
  {\bibfield  {journal} {\bibinfo  {journal} {Physical Review Letters}\
  }\textbf {\bibinfo {volume} {81}},\ \bibinfo {pages} {1758} (\bibinfo {year}
  {1998})}\BibitemShut {NoStop}%
\bibitem [{\citenamefont {Baiesi}\ \emph {et~al.}(2009)\citenamefont {Baiesi},
  \citenamefont {Maes},\ and\ \citenamefont
  {Wynants}}]{baiesi2009fluctuations}%
  \BibitemOpen
  \bibfield  {author} {\bibinfo {author} {\bibfnamefont {M.}~\bibnamefont
  {Baiesi}}, \bibinfo {author} {\bibfnamefont {C.}~\bibnamefont {Maes}}, \ and\
  \bibinfo {author} {\bibfnamefont {B.}~\bibnamefont {Wynants}},\ }\href@noop
  {} {\bibfield  {journal} {\bibinfo  {journal} {Physical review letters}\
  }\textbf {\bibinfo {volume} {103}},\ \bibinfo {pages} {010602} (\bibinfo
  {year} {2009})}\BibitemShut {NoStop}%
\bibitem [{\citenamefont {Corberi}\ \emph {et~al.}(2010)\citenamefont
  {Corberi}, \citenamefont {Lippiello}, \citenamefont {Sarracino},\ and\
  \citenamefont {Zannetti}}]{corberi2010fluctuation}%
  \BibitemOpen
  \bibfield  {author} {\bibinfo {author} {\bibfnamefont {F.}~\bibnamefont
  {Corberi}}, \bibinfo {author} {\bibfnamefont {E.}~\bibnamefont {Lippiello}},
  \bibinfo {author} {\bibfnamefont {A.}~\bibnamefont {Sarracino}}, \ and\
  \bibinfo {author} {\bibfnamefont {M.}~\bibnamefont {Zannetti}},\ }\href@noop
  {} {\bibfield  {journal} {\bibinfo  {journal} {Physical Review E}\ }\textbf
  {\bibinfo {volume} {81}},\ \bibinfo {pages} {011124} (\bibinfo {year}
  {2010})}\BibitemShut {NoStop}%
\bibitem [{\citenamefont {Lippiello}\ \emph
  {et~al.}(2008{\natexlab{a}})\citenamefont {Lippiello}, \citenamefont
  {Corberi}, \citenamefont {Sarracino},\ and\ \citenamefont
  {Zannetti}}]{lippiello2008non}%
  \BibitemOpen
  \bibfield  {author} {\bibinfo {author} {\bibfnamefont {E.}~\bibnamefont
  {Lippiello}}, \bibinfo {author} {\bibfnamefont {F.}~\bibnamefont {Corberi}},
  \bibinfo {author} {\bibfnamefont {A.}~\bibnamefont {Sarracino}}, \ and\
  \bibinfo {author} {\bibfnamefont {M.}~\bibnamefont {Zannetti}},\ }\href@noop
  {} {\bibfield  {journal} {\bibinfo  {journal} {Physical Review B}\ }\textbf
  {\bibinfo {volume} {77}},\ \bibinfo {pages} {212201} (\bibinfo {year}
  {2008}{\natexlab{a}})}\BibitemShut {NoStop}%
\bibitem [{\citenamefont {Lippiello}\ \emph
  {et~al.}(2008{\natexlab{b}})\citenamefont {Lippiello}, \citenamefont
  {Corberi}, \citenamefont {Sarracino},\ and\ \citenamefont
  {Zannetti}}]{lippiello2008nonlinear}%
  \BibitemOpen
  \bibfield  {author} {\bibinfo {author} {\bibfnamefont {E.}~\bibnamefont
  {Lippiello}}, \bibinfo {author} {\bibfnamefont {F.}~\bibnamefont {Corberi}},
  \bibinfo {author} {\bibfnamefont {A.}~\bibnamefont {Sarracino}}, \ and\
  \bibinfo {author} {\bibfnamefont {M.}~\bibnamefont {Zannetti}},\ }\href@noop
  {} {\bibfield  {journal} {\bibinfo  {journal} {Physical Review E}\ }\textbf
  {\bibinfo {volume} {78}},\ \bibinfo {pages} {041120} (\bibinfo {year}
  {2008}{\natexlab{b}})}\BibitemShut {NoStop}%
\bibitem [{\citenamefont {Crisanti}\ and\ \citenamefont
  {Ritort}(2003)}]{crisanti2003violation}%
  \BibitemOpen
  \bibfield  {author} {\bibinfo {author} {\bibfnamefont {A.}~\bibnamefont
  {Crisanti}}\ and\ \bibinfo {author} {\bibfnamefont {F.}~\bibnamefont
  {Ritort}},\ }\href@noop {} {\bibfield  {journal} {\bibinfo  {journal}
  {Journal of Physics A: Mathematical and General}\ }\textbf {\bibinfo {volume}
  {36}},\ \bibinfo {pages} {R181} (\bibinfo {year} {2003})}\BibitemShut
  {NoStop}%
\bibitem [{\citenamefont {Landry}\ and\ \citenamefont
  {Coppersmith}(2002)}]{landry2002ground}%
  \BibitemOpen
  \bibfield  {author} {\bibinfo {author} {\bibfnamefont {J.}~\bibnamefont
  {Landry}}\ and\ \bibinfo {author} {\bibfnamefont {S.}~\bibnamefont
  {Coppersmith}},\ }\href@noop {} {\bibfield  {journal} {\bibinfo  {journal}
  {Physical Review B}\ }\textbf {\bibinfo {volume} {65}},\ \bibinfo {pages}
  {134404} (\bibinfo {year} {2002})}\BibitemShut {NoStop}%
\bibitem [{\citenamefont {Thomas}\ and\ \citenamefont
  {Middleton}(2007)}]{thomas2007matching}%
  \BibitemOpen
  \bibfield  {author} {\bibinfo {author} {\bibfnamefont {C.~K.}\ \bibnamefont
  {Thomas}}\ and\ \bibinfo {author} {\bibfnamefont {A.~A.}\ \bibnamefont
  {Middleton}},\ }\href@noop {} {\bibfield  {journal} {\bibinfo  {journal}
  {Physical Review B}\ }\textbf {\bibinfo {volume} {76}},\ \bibinfo {pages}
  {220406} (\bibinfo {year} {2007})}\BibitemShut {NoStop}%
\bibitem [{\citenamefont {Khoshbakht}\ and\ \citenamefont
  {Weigel}(2018)}]{khoshbakht2018domain}%
  \BibitemOpen
  \bibfield  {author} {\bibinfo {author} {\bibfnamefont {H.}~\bibnamefont
  {Khoshbakht}}\ and\ \bibinfo {author} {\bibfnamefont {M.}~\bibnamefont
  {Weigel}},\ }\href@noop {} {\bibfield  {journal} {\bibinfo  {journal}
  {Physical Review B}\ }\textbf {\bibinfo {volume} {97}},\ \bibinfo {pages}
  {064410} (\bibinfo {year} {2018})}\BibitemShut {NoStop}%
\bibitem [{\citenamefont {Rieger}\ \emph {et~al.}(1994)\citenamefont {Rieger},
  \citenamefont {Steckemetz},\ and\ \citenamefont
  {Schreckenberg}}]{rieger1994aging}%
  \BibitemOpen
  \bibfield  {author} {\bibinfo {author} {\bibfnamefont {H.}~\bibnamefont
  {Rieger}}, \bibinfo {author} {\bibfnamefont {B.}~\bibnamefont {Steckemetz}},
  \ and\ \bibinfo {author} {\bibfnamefont {M.}~\bibnamefont {Schreckenberg}},\
  }\href@noop {} {\bibfield  {journal} {\bibinfo  {journal} {EPL (Europhysics
  Letters)}\ }\textbf {\bibinfo {volume} {27}},\ \bibinfo {pages} {485}
  (\bibinfo {year} {1994})}\BibitemShut {NoStop}%
\bibitem [{\citenamefont {Kisker}\ \emph {et~al.}(1996)\citenamefont {Kisker},
  \citenamefont {Santen}, \citenamefont {Schreckenberg},\ and\ \citenamefont
  {Rieger}}]{kisker1996off}%
  \BibitemOpen
  \bibfield  {author} {\bibinfo {author} {\bibfnamefont {J.}~\bibnamefont
  {Kisker}}, \bibinfo {author} {\bibfnamefont {L.}~\bibnamefont {Santen}},
  \bibinfo {author} {\bibfnamefont {M.}~\bibnamefont {Schreckenberg}}, \ and\
  \bibinfo {author} {\bibfnamefont {H.}~\bibnamefont {Rieger}},\ }\href@noop {}
  {\bibfield  {journal} {\bibinfo  {journal} {Physical Review B}\ }\textbf
  {\bibinfo {volume} {53}},\ \bibinfo {pages} {6418} (\bibinfo {year}
  {1996})}\BibitemShut {NoStop}%
\bibitem [{\citenamefont {Franz}\ \emph {et~al.}(2003)\citenamefont {Franz},
  \citenamefont {Lecomte},\ and\ \citenamefont
  {Mulet}}]{franz2003quasiequilibrium}%
  \BibitemOpen
  \bibfield  {author} {\bibinfo {author} {\bibfnamefont {S.}~\bibnamefont
  {Franz}}, \bibinfo {author} {\bibfnamefont {V.}~\bibnamefont {Lecomte}}, \
  and\ \bibinfo {author} {\bibfnamefont {R.}~\bibnamefont {Mulet}},\
  }\href@noop {} {\bibfield  {journal} {\bibinfo  {journal} {Physical Review
  E}\ }\textbf {\bibinfo {volume} {68}},\ \bibinfo {pages} {066128} (\bibinfo
  {year} {2003})}\BibitemShut {NoStop}%
\bibitem [{\citenamefont {Chamon}\ \emph {et~al.}(2011)\citenamefont {Chamon},
  \citenamefont {Corberi},\ and\ \citenamefont
  {Cugliandolo}}]{chamon2011fluctuations}%
  \BibitemOpen
  \bibfield  {author} {\bibinfo {author} {\bibfnamefont {C.}~\bibnamefont
  {Chamon}}, \bibinfo {author} {\bibfnamefont {F.}~\bibnamefont {Corberi}}, \
  and\ \bibinfo {author} {\bibfnamefont {L.~F.}\ \bibnamefont {Cugliandolo}},\
  }\href@noop {} {\bibfield  {journal} {\bibinfo  {journal} {Journal of
  Statistical Mechanics: Theory and Experiment}\ }\textbf {\bibinfo {volume}
  {2011}},\ \bibinfo {pages} {P08015} (\bibinfo {year} {2011})}\BibitemShut
  {NoStop}%
\bibitem [{\citenamefont {Corberi}\ \emph
  {et~al.}(2003{\natexlab{a}})\citenamefont {Corberi}, \citenamefont
  {Lippiello},\ and\ \citenamefont {Zannetti}}]{corberi2003scaling}%
  \BibitemOpen
  \bibfield  {author} {\bibinfo {author} {\bibfnamefont {F.}~\bibnamefont
  {Corberi}}, \bibinfo {author} {\bibfnamefont {E.}~\bibnamefont {Lippiello}},
  \ and\ \bibinfo {author} {\bibfnamefont {M.}~\bibnamefont {Zannetti}},\
  }\href@noop {} {\bibfield  {journal} {\bibinfo  {journal} {Physical Review
  E}\ }\textbf {\bibinfo {volume} {68}},\ \bibinfo {pages} {046131} (\bibinfo
  {year} {2003}{\natexlab{a}})}\BibitemShut {NoStop}%
\bibitem [{\citenamefont {Corberi}\ \emph
  {et~al.}(2001{\natexlab{a}})\citenamefont {Corberi}, \citenamefont
  {Lippiello},\ and\ \citenamefont {Zannetti}}]{corberi2001connection}%
  \BibitemOpen
  \bibfield  {author} {\bibinfo {author} {\bibfnamefont {F.}~\bibnamefont
  {Corberi}}, \bibinfo {author} {\bibfnamefont {E.}~\bibnamefont {Lippiello}},
  \ and\ \bibinfo {author} {\bibfnamefont {M.}~\bibnamefont {Zannetti}},\
  }\href@noop {} {\bibfield  {journal} {\bibinfo  {journal} {The European
  Physical Journal B-Condensed Matter and Complex Systems}\ }\textbf {\bibinfo
  {volume} {24}},\ \bibinfo {pages} {359} (\bibinfo {year}
  {2001}{\natexlab{a}})}\BibitemShut {NoStop}%
\bibitem [{\citenamefont {Hinrichsen}(2008)}]{hinrichsen2008dynamical}%
  \BibitemOpen
  \bibfield  {author} {\bibinfo {author} {\bibfnamefont {H.}~\bibnamefont
  {Hinrichsen}},\ }\href@noop {} {\bibfield  {journal} {\bibinfo  {journal}
  {Journal of Statistical Mechanics: Theory and Experiment}\ }\textbf {\bibinfo
  {volume} {2008}},\ \bibinfo {pages} {P02016} (\bibinfo {year}
  {2008})}\BibitemShut {NoStop}%
\bibitem [{\citenamefont {Henkel}\ and\ \citenamefont
  {Pleimling}(2003)}]{henkel2003local}%
  \BibitemOpen
  \bibfield  {author} {\bibinfo {author} {\bibfnamefont {M.}~\bibnamefont
  {Henkel}}\ and\ \bibinfo {author} {\bibfnamefont {M.}~\bibnamefont
  {Pleimling}},\ }\href@noop {} {\bibfield  {journal} {\bibinfo  {journal}
  {Physical Review E}\ }\textbf {\bibinfo {volume} {68}},\ \bibinfo {pages}
  {065101} (\bibinfo {year} {2003})}\BibitemShut {NoStop}%
\bibitem [{\citenamefont {Henkel}\ \emph {et~al.}(2003)\citenamefont {Henkel},
  \citenamefont {Pae{\ss}ens},\ and\ \citenamefont
  {Pleimling}}]{henkel2003scaling}%
  \BibitemOpen
  \bibfield  {author} {\bibinfo {author} {\bibfnamefont {M.}~\bibnamefont
  {Henkel}}, \bibinfo {author} {\bibfnamefont {M.}~\bibnamefont {Pae{\ss}ens}},
  \ and\ \bibinfo {author} {\bibfnamefont {M.}~\bibnamefont {Pleimling}},\
  }\href@noop {} {\bibfield  {journal} {\bibinfo  {journal} {EPL (Europhysics
  Letters)}\ }\textbf {\bibinfo {volume} {62}},\ \bibinfo {pages} {664}
  (\bibinfo {year} {2003})}\BibitemShut {NoStop}%
\bibitem [{\citenamefont {Henkel}\ \emph {et~al.}(2001)\citenamefont {Henkel},
  \citenamefont {Pleimling}, \citenamefont {Godreche},\ and\ \citenamefont
  {Luck}}]{henkel2001aging}%
  \BibitemOpen
  \bibfield  {author} {\bibinfo {author} {\bibfnamefont {M.}~\bibnamefont
  {Henkel}}, \bibinfo {author} {\bibfnamefont {M.}~\bibnamefont {Pleimling}},
  \bibinfo {author} {\bibfnamefont {C.}~\bibnamefont {Godreche}}, \ and\
  \bibinfo {author} {\bibfnamefont {J.-M.}\ \bibnamefont {Luck}},\ }\href@noop
  {} {\bibfield  {journal} {\bibinfo  {journal} {Physical review letters}\
  }\textbf {\bibinfo {volume} {87}},\ \bibinfo {pages} {265701} (\bibinfo
  {year} {2001})}\BibitemShut {NoStop}%
\bibitem [{\citenamefont {Corberi}\ \emph
  {et~al.}(2003{\natexlab{b}})\citenamefont {Corberi}, \citenamefont
  {Lippiello},\ and\ \citenamefont {Zannetti}}]{corberi2003comment}%
  \BibitemOpen
  \bibfield  {author} {\bibinfo {author} {\bibfnamefont {F.}~\bibnamefont
  {Corberi}}, \bibinfo {author} {\bibfnamefont {E.}~\bibnamefont {Lippiello}},
  \ and\ \bibinfo {author} {\bibfnamefont {M.}~\bibnamefont {Zannetti}},\
  }\href@noop {} {\bibfield  {journal} {\bibinfo  {journal} {Physical review
  letters}\ }\textbf {\bibinfo {volume} {90}},\ \bibinfo {pages} {099601}
  (\bibinfo {year} {2003}{\natexlab{b}})}\BibitemShut {NoStop}%
\bibitem [{\citenamefont {Corberi}\ \emph
  {et~al.}(2001{\natexlab{b}})\citenamefont {Corberi}, \citenamefont
  {Lippiello},\ and\ \citenamefont {Zannetti}}]{corberi2001interface}%
  \BibitemOpen
  \bibfield  {author} {\bibinfo {author} {\bibfnamefont {F.}~\bibnamefont
  {Corberi}}, \bibinfo {author} {\bibfnamefont {E.}~\bibnamefont {Lippiello}},
  \ and\ \bibinfo {author} {\bibfnamefont {M.}~\bibnamefont {Zannetti}},\
  }\href@noop {} {\bibfield  {journal} {\bibinfo  {journal} {Physical Review
  E}\ }\textbf {\bibinfo {volume} {63}},\ \bibinfo {pages} {061506} (\bibinfo
  {year} {2001}{\natexlab{b}})}\BibitemShut {NoStop}%
\bibitem [{\citenamefont {Corberi}\ \emph
  {et~al.}(2004{\natexlab{b}})\citenamefont {Corberi}, \citenamefont
  {Castellano}, \citenamefont {Lippiello},\ and\ \citenamefont
  {Zannetti}}]{corberi2004generic}%
  \BibitemOpen
  \bibfield  {author} {\bibinfo {author} {\bibfnamefont {F.}~\bibnamefont
  {Corberi}}, \bibinfo {author} {\bibfnamefont {C.}~\bibnamefont {Castellano}},
  \bibinfo {author} {\bibfnamefont {E.}~\bibnamefont {Lippiello}}, \ and\
  \bibinfo {author} {\bibfnamefont {M.}~\bibnamefont {Zannetti}},\ }\href@noop
  {} {\bibfield  {journal} {\bibinfo  {journal} {Physical Review E}\ }\textbf
  {\bibinfo {volume} {70}},\ \bibinfo {pages} {017103} (\bibinfo {year}
  {2004}{\natexlab{b}})}\BibitemShut {NoStop}%
\bibitem [{\citenamefont {Corberi}\ \emph
  {et~al.}(2005{\natexlab{a}})\citenamefont {Corberi}, \citenamefont
  {Lippiello},\ and\ \citenamefont {Zannetti}}]{corberi2005correction}%
  \BibitemOpen
  \bibfield  {author} {\bibinfo {author} {\bibfnamefont {F.}~\bibnamefont
  {Corberi}}, \bibinfo {author} {\bibfnamefont {E.}~\bibnamefont {Lippiello}},
  \ and\ \bibinfo {author} {\bibfnamefont {M.}~\bibnamefont {Zannetti}},\
  }\href@noop {} {\bibfield  {journal} {\bibinfo  {journal} {Physical Review
  E}\ }\textbf {\bibinfo {volume} {72}},\ \bibinfo {pages} {056103} (\bibinfo
  {year} {2005}{\natexlab{a}})}\BibitemShut {NoStop}%
\bibitem [{\citenamefont {Henkel}\ \emph {et~al.}(2004)\citenamefont {Henkel},
  \citenamefont {Paessens},\ and\ \citenamefont
  {Pleimling}}]{henkel2004scaling}%
  \BibitemOpen
  \bibfield  {author} {\bibinfo {author} {\bibfnamefont {M.}~\bibnamefont
  {Henkel}}, \bibinfo {author} {\bibfnamefont {M.}~\bibnamefont {Paessens}}, \
  and\ \bibinfo {author} {\bibfnamefont {M.}~\bibnamefont {Pleimling}},\
  }\href@noop {} {\bibfield  {journal} {\bibinfo  {journal} {Physical Review
  E}\ }\textbf {\bibinfo {volume} {69}},\ \bibinfo {pages} {056109} (\bibinfo
  {year} {2004})}\BibitemShut {NoStop}%
\bibitem [{\citenamefont {Corberi}\ \emph
  {et~al.}(2005{\natexlab{b}})\citenamefont {Corberi}, \citenamefont
  {Lippiello},\ and\ \citenamefont {Zannetti}}]{corberi2005comment}%
  \BibitemOpen
  \bibfield  {author} {\bibinfo {author} {\bibfnamefont {F.}~\bibnamefont
  {Corberi}}, \bibinfo {author} {\bibfnamefont {E.}~\bibnamefont {Lippiello}},
  \ and\ \bibinfo {author} {\bibfnamefont {M.}~\bibnamefont {Zannetti}},\
  }\href@noop {} {\bibfield  {journal} {\bibinfo  {journal} {Physical Review
  E}\ }\textbf {\bibinfo {volume} {72}},\ \bibinfo {pages} {028103} (\bibinfo
  {year} {2005}{\natexlab{b}})}\BibitemShut {NoStop}%
\bibitem [{\citenamefont {Henkel}\ and\ \citenamefont
  {Pleimling}(2005)}]{henkel2005reply}%
  \BibitemOpen
  \bibfield  {author} {\bibinfo {author} {\bibfnamefont {M.}~\bibnamefont
  {Henkel}}\ and\ \bibinfo {author} {\bibfnamefont {M.}~\bibnamefont
  {Pleimling}},\ }\href@noop {} {\bibfield  {journal} {\bibinfo  {journal}
  {Physical Review E}\ }\textbf {\bibinfo {volume} {72}},\ \bibinfo {pages}
  {028104} (\bibinfo {year} {2005})}\BibitemShut {NoStop}%
\bibitem [{\citenamefont {Lippiello}\ \emph {et~al.}(2006)\citenamefont
  {Lippiello}, \citenamefont {Corberi},\ and\ \citenamefont
  {Zannetti}}]{lippiello2006test}%
  \BibitemOpen
  \bibfield  {author} {\bibinfo {author} {\bibfnamefont {E.}~\bibnamefont
  {Lippiello}}, \bibinfo {author} {\bibfnamefont {F.}~\bibnamefont {Corberi}},
  \ and\ \bibinfo {author} {\bibfnamefont {M.}~\bibnamefont {Zannetti}},\
  }\href@noop {} {\bibfield  {journal} {\bibinfo  {journal} {Physical Review
  E}\ }\textbf {\bibinfo {volume} {74}},\ \bibinfo {pages} {041113} (\bibinfo
  {year} {2006})}\BibitemShut {NoStop}%
\bibitem [{\citenamefont {Henkel}\ and\ \citenamefont
  {Pleimling}(2011)}]{henkel2011non}%
  \BibitemOpen
  \bibfield  {author} {\bibinfo {author} {\bibfnamefont {M.}~\bibnamefont
  {Henkel}}\ and\ \bibinfo {author} {\bibfnamefont {M.}~\bibnamefont
  {Pleimling}},\ }\href@noop {} {\emph {\bibinfo {title} {Non-Equilibrium Phase
  Transitions: Volume 2: Ageing and Dynamical Scaling Far from Equilibrium}}}\
  (\bibinfo  {publisher} {Springer Science \& Business Media},\ \bibinfo {year}
  {2011})\BibitemShut {NoStop}%
\bibitem [{\citenamefont {Mazenko}\ \emph {et~al.}(1985)\citenamefont
  {Mazenko}, \citenamefont {Valls},\ and\ \citenamefont
  {Zhang}}]{mazenko1985kinetics}%
  \BibitemOpen
  \bibfield  {author} {\bibinfo {author} {\bibfnamefont {G.~F.}\ \bibnamefont
  {Mazenko}}, \bibinfo {author} {\bibfnamefont {O.~T.}\ \bibnamefont {Valls}},
  \ and\ \bibinfo {author} {\bibfnamefont {F.}~\bibnamefont {Zhang}},\
  }\href@noop {} {\bibfield  {journal} {\bibinfo  {journal} {Physical Review
  B}\ }\textbf {\bibinfo {volume} {31}},\ \bibinfo {pages} {4453} (\bibinfo
  {year} {1985})}\BibitemShut {NoStop}%
\bibitem [{\citenamefont {Bray}(1990)}]{bray1990renormalization}%
  \BibitemOpen
  \bibfield  {author} {\bibinfo {author} {\bibfnamefont {A.}~\bibnamefont
  {Bray}},\ }\href@noop {} {\bibfield  {journal} {\bibinfo  {journal} {Physical
  Review B}\ }\textbf {\bibinfo {volume} {41}},\ \bibinfo {pages} {6724}
  (\bibinfo {year} {1990})}\BibitemShut {NoStop}%
\bibitem [{\citenamefont {Godreche}\ and\ \citenamefont
  {Luck}(2002)}]{godreche2002nonequilibrium}%
  \BibitemOpen
  \bibfield  {author} {\bibinfo {author} {\bibfnamefont {C.}~\bibnamefont
  {Godreche}}\ and\ \bibinfo {author} {\bibfnamefont {J.}~\bibnamefont
  {Luck}},\ }\href@noop {} {\bibfield  {journal} {\bibinfo  {journal} {Journal
  of Physics: Condensed Matter}\ }\textbf {\bibinfo {volume} {14}},\ \bibinfo
  {pages} {1589} (\bibinfo {year} {2002})}\BibitemShut {NoStop}%
\bibitem [{\citenamefont {Corberi}\ \emph {et~al.}(2006)\citenamefont
  {Corberi}, \citenamefont {Lippiello},\ and\ \citenamefont
  {Zannetti}}]{corberi2006scaling}%
  \BibitemOpen
  \bibfield  {author} {\bibinfo {author} {\bibfnamefont {F.}~\bibnamefont
  {Corberi}}, \bibinfo {author} {\bibfnamefont {E.}~\bibnamefont {Lippiello}},
  \ and\ \bibinfo {author} {\bibfnamefont {M.}~\bibnamefont {Zannetti}},\
  }\href@noop {} {\bibfield  {journal} {\bibinfo  {journal} {Physical Review
  E}\ }\textbf {\bibinfo {volume} {74}},\ \bibinfo {pages} {041106} (\bibinfo
  {year} {2006})}\BibitemShut {NoStop}%
\bibitem [{\citenamefont {Lippiello}\ and\ \citenamefont
  {Zannetti}(2000)}]{lippiello2000fluctuation}%
  \BibitemOpen
  \bibfield  {author} {\bibinfo {author} {\bibfnamefont {E.}~\bibnamefont
  {Lippiello}}\ and\ \bibinfo {author} {\bibfnamefont {M.}~\bibnamefont
  {Zannetti}},\ }\href@noop {} {\bibfield  {journal} {\bibinfo  {journal}
  {Physical Review E}\ }\textbf {\bibinfo {volume} {61}},\ \bibinfo {pages}
  {3369} (\bibinfo {year} {2000})}\BibitemShut {NoStop}%
\bibitem [{\citenamefont {Godr{\`e}che}\ and\ \citenamefont
  {Luck}(2000)}]{godreche2000response}%
  \BibitemOpen
  \bibfield  {author} {\bibinfo {author} {\bibfnamefont {C.}~\bibnamefont
  {Godr{\`e}che}}\ and\ \bibinfo {author} {\bibfnamefont {J.}~\bibnamefont
  {Luck}},\ }\href@noop {} {\bibfield  {journal} {\bibinfo  {journal} {Journal
  of Physics A: Mathematical and General}\ }\textbf {\bibinfo {volume} {33}},\
  \bibinfo {pages} {1151} (\bibinfo {year} {2000})}\BibitemShut {NoStop}%
\bibitem [{\citenamefont {Corberi}\ \emph
  {et~al.}(2002{\natexlab{a}})\citenamefont {Corberi}, \citenamefont
  {Castellano}, \citenamefont {Lippiello},\ and\ \citenamefont
  {Zannetti}}]{corberi2002universality}%
  \BibitemOpen
  \bibfield  {author} {\bibinfo {author} {\bibfnamefont {F.}~\bibnamefont
  {Corberi}}, \bibinfo {author} {\bibfnamefont {C.}~\bibnamefont {Castellano}},
  \bibinfo {author} {\bibfnamefont {E.}~\bibnamefont {Lippiello}}, \ and\
  \bibinfo {author} {\bibfnamefont {M.}~\bibnamefont {Zannetti}},\ }\href@noop
  {} {\bibfield  {journal} {\bibinfo  {journal} {Physical Review E}\ }\textbf
  {\bibinfo {volume} {65}},\ \bibinfo {pages} {066114} (\bibinfo {year}
  {2002}{\natexlab{a}})}\BibitemShut {NoStop}%
\bibitem [{\citenamefont {Corberi}\ \emph
  {et~al.}(2002{\natexlab{b}})\citenamefont {Corberi}, \citenamefont
  {Lippiello},\ and\ \citenamefont {Zannetti}}]{corberi2002slow}%
  \BibitemOpen
  \bibfield  {author} {\bibinfo {author} {\bibfnamefont {F.}~\bibnamefont
  {Corberi}}, \bibinfo {author} {\bibfnamefont {E.}~\bibnamefont {Lippiello}},
  \ and\ \bibinfo {author} {\bibfnamefont {M.}~\bibnamefont {Zannetti}},\
  }\href@noop {} {\bibfield  {journal} {\bibinfo  {journal} {Physical Review
  E}\ }\textbf {\bibinfo {volume} {65}},\ \bibinfo {pages} {046136} (\bibinfo
  {year} {2002}{\natexlab{b}})}\BibitemShut {NoStop}%
\bibitem [{\citenamefont {Corberi}\ \emph
  {et~al.}(2002{\natexlab{c}})\citenamefont {Corberi}, \citenamefont
  {De~Candia}, \citenamefont {Lippiello},\ and\ \citenamefont
  {Zannetti}}]{corberi2002off}%
  \BibitemOpen
  \bibfield  {author} {\bibinfo {author} {\bibfnamefont {F.}~\bibnamefont
  {Corberi}}, \bibinfo {author} {\bibfnamefont {A.}~\bibnamefont {De~Candia}},
  \bibinfo {author} {\bibfnamefont {E.}~\bibnamefont {Lippiello}}, \ and\
  \bibinfo {author} {\bibfnamefont {M.}~\bibnamefont {Zannetti}},\ }\href@noop
  {} {\bibfield  {journal} {\bibinfo  {journal} {Physical Review E}\ }\textbf
  {\bibinfo {volume} {65}},\ \bibinfo {pages} {046114} (\bibinfo {year}
  {2002}{\natexlab{c}})}\BibitemShut {NoStop}%
\bibitem [{\citenamefont {Burioni}\ \emph {et~al.}(2006)\citenamefont
  {Burioni}, \citenamefont {Cassi}, \citenamefont {Corberi},\ and\
  \citenamefont {Vezzani}}]{burioni2006aging}%
  \BibitemOpen
  \bibfield  {author} {\bibinfo {author} {\bibfnamefont {R.}~\bibnamefont
  {Burioni}}, \bibinfo {author} {\bibfnamefont {D.}~\bibnamefont {Cassi}},
  \bibinfo {author} {\bibfnamefont {F.}~\bibnamefont {Corberi}}, \ and\
  \bibinfo {author} {\bibfnamefont {A.}~\bibnamefont {Vezzani}},\ }\href@noop
  {} {\bibfield  {journal} {\bibinfo  {journal} {Physical review letters}\
  }\textbf {\bibinfo {volume} {96}},\ \bibinfo {pages} {235701} (\bibinfo
  {year} {2006})}\BibitemShut {NoStop}%
\bibitem [{\citenamefont {Cugliandolo}\ and\ \citenamefont
  {Kurchan}(1993)}]{cugliandolo1993analytical}%
  \BibitemOpen
  \bibfield  {author} {\bibinfo {author} {\bibfnamefont {L.~F.}\ \bibnamefont
  {Cugliandolo}}\ and\ \bibinfo {author} {\bibfnamefont {J.}~\bibnamefont
  {Kurchan}},\ }\href@noop {} {\bibfield  {journal} {\bibinfo  {journal}
  {Physical Review Letters}\ }\textbf {\bibinfo {volume} {71}},\ \bibinfo
  {pages} {173} (\bibinfo {year} {1993})}\BibitemShut {NoStop}%
\bibitem [{\citenamefont {Cugliandolo}\ and\ \citenamefont
  {Kurchan}(1995)}]{cugliandolo1995weak}%
  \BibitemOpen
  \bibfield  {author} {\bibinfo {author} {\bibfnamefont {L.~F.}\ \bibnamefont
  {Cugliandolo}}\ and\ \bibinfo {author} {\bibfnamefont {J.}~\bibnamefont
  {Kurchan}},\ }\href@noop {} {\bibfield  {journal} {\bibinfo  {journal}
  {Philosophical Magazine B}\ }\textbf {\bibinfo {volume} {71}},\ \bibinfo
  {pages} {501} (\bibinfo {year} {1995})}\BibitemShut {NoStop}%
\bibitem [{\citenamefont {Cugliandolo}\ and\ \citenamefont
  {Kurchan}(1994)}]{cugliandolo1994out}%
  \BibitemOpen
  \bibfield  {author} {\bibinfo {author} {\bibfnamefont {L.~F.}\ \bibnamefont
  {Cugliandolo}}\ and\ \bibinfo {author} {\bibfnamefont {J.}~\bibnamefont
  {Kurchan}},\ }\href@noop {} {\bibfield  {journal} {\bibinfo  {journal}
  {Journal of Physics A: Mathematical and General}\ }\textbf {\bibinfo {volume}
  {27}},\ \bibinfo {pages} {5749} (\bibinfo {year} {1994})}\BibitemShut
  {NoStop}%
\bibitem [{\citenamefont {Humayun}\ and\ \citenamefont
  {Bray}(1991)}]{humayun1991non}%
  \BibitemOpen
  \bibfield  {author} {\bibinfo {author} {\bibfnamefont {K.}~\bibnamefont
  {Humayun}}\ and\ \bibinfo {author} {\bibfnamefont {A.}~\bibnamefont {Bray}},\
  }\href@noop {} {\bibfield  {journal} {\bibinfo  {journal} {Journal of Physics
  A: Mathematical and General}\ }\textbf {\bibinfo {volume} {24}},\ \bibinfo
  {pages} {1915} (\bibinfo {year} {1991})}\BibitemShut {NoStop}%
\bibitem [{\citenamefont {Liu}\ and\ \citenamefont
  {Mazenko}(1991)}]{liu1991nonequilibrium}%
  \BibitemOpen
  \bibfield  {author} {\bibinfo {author} {\bibfnamefont {F.}~\bibnamefont
  {Liu}}\ and\ \bibinfo {author} {\bibfnamefont {G.~F.}\ \bibnamefont
  {Mazenko}},\ }\href@noop {} {\bibfield  {journal} {\bibinfo  {journal}
  {Physical Review B}\ }\textbf {\bibinfo {volume} {44}},\ \bibinfo {pages}
  {9185} (\bibinfo {year} {1991})}\BibitemShut {NoStop}%
\bibitem [{\citenamefont {Lorenz}\ and\ \citenamefont
  {Janke}(2007)}]{lorenz2007numerical}%
  \BibitemOpen
  \bibfield  {author} {\bibinfo {author} {\bibfnamefont {E.}~\bibnamefont
  {Lorenz}}\ and\ \bibinfo {author} {\bibfnamefont {W.}~\bibnamefont {Janke}},\
  }\href@noop {} {\bibfield  {journal} {\bibinfo  {journal} {EPL (Europhysics
  Letters)}\ }\textbf {\bibinfo {volume} {77}},\ \bibinfo {pages} {10003}
  (\bibinfo {year} {2007})}\BibitemShut {NoStop}%
\bibitem [{\citenamefont {Yeung}\ \emph {et~al.}(1996)\citenamefont {Yeung},
  \citenamefont {Rao},\ and\ \citenamefont {Desai}}]{yeung1996bounds}%
  \BibitemOpen
  \bibfield  {author} {\bibinfo {author} {\bibfnamefont {C.}~\bibnamefont
  {Yeung}}, \bibinfo {author} {\bibfnamefont {M.}~\bibnamefont {Rao}}, \ and\
  \bibinfo {author} {\bibfnamefont {R.~C.}\ \bibnamefont {Desai}},\ }\href@noop
  {} {\bibfield  {journal} {\bibinfo  {journal} {Physical Review E}\ }\textbf
  {\bibinfo {volume} {53}},\ \bibinfo {pages} {3073} (\bibinfo {year}
  {1996})}\BibitemShut {NoStop}%
\bibitem [{\citenamefont {Picone}\ and\ \citenamefont
  {Henkel}(2004)}]{picone2004local}%
  \BibitemOpen
  \bibfield  {author} {\bibinfo {author} {\bibfnamefont {A.}~\bibnamefont
  {Picone}}\ and\ \bibinfo {author} {\bibfnamefont {M.}~\bibnamefont
  {Henkel}},\ }\href@noop {} {\bibfield  {journal} {\bibinfo  {journal}
  {Nuclear Physics B}\ }\textbf {\bibinfo {volume} {688}},\ \bibinfo {pages}
  {217} (\bibinfo {year} {2004})}\BibitemShut {NoStop}%
\bibitem [{\citenamefont {Berthier}\ and\ \citenamefont
  {Bouchaud}(2002)}]{berthier2002geometrical}%
  \BibitemOpen
  \bibfield  {author} {\bibinfo {author} {\bibfnamefont {L.}~\bibnamefont
  {Berthier}}\ and\ \bibinfo {author} {\bibfnamefont {J.-P.}\ \bibnamefont
  {Bouchaud}},\ }\href@noop {} {\bibfield  {journal} {\bibinfo  {journal}
  {Physical Review B}\ }\textbf {\bibinfo {volume} {66}},\ \bibinfo {pages}
  {054404} (\bibinfo {year} {2002})}\BibitemShut {NoStop}%
\bibitem [{\citenamefont {Manssen}\ and\ \citenamefont
  {Hartmann}(2015)}]{manssen2015aging}%
  \BibitemOpen
  \bibfield  {author} {\bibinfo {author} {\bibfnamefont {M.}~\bibnamefont
  {Manssen}}\ and\ \bibinfo {author} {\bibfnamefont {A.~K.}\ \bibnamefont
  {Hartmann}},\ }\href@noop {} {\bibfield  {journal} {\bibinfo  {journal}
  {Physical Review B}\ }\textbf {\bibinfo {volume} {91}},\ \bibinfo {pages}
  {174433} (\bibinfo {year} {2015})}\BibitemShut {NoStop}%
\end{thebibliography}%

\end{document}